\documentclass[3p,12pt]{elsarticle}
\usepackage{booktabs}
\usepackage[fleqn]{amsmath}
\usepackage{cases}
\usepackage{multirow}
\usepackage{xcolor}
\usepackage[normalem]{ulem}
\usepackage{tabularx}
\usepackage{siunitx}
\usepackage{comment}
\usepackage{algorithm}
\usepackage{algpseudocode}
\usepackage{algorithmicx}
\usepackage{grffile}
\usepackage{rotating}
\usepackage{lineno}
\usepackage{hyperref}
\usepackage{graphicx,enumerate,subfigure}
\usepackage{amssymb}
\usepackage{bm,amsmath}
\usepackage{caption,color,xcolor}
\usepackage{subeqnarray}
\usepackage{multirow}
\usepackage{lscape}
\usepackage{rotating}
\usepackage{graphics}
\usepackage{float}
\usepackage{epstopdf,epsfig}
\usepackage{threeparttable}
\usepackage{soul}
\floatname{algorithm}{Algorithm}
\biboptions{numbers,sort&compress} 
\journal{XXX}

\newcounter{bla}

\begin{document}

\begin{frontmatter}
\title{Efficient parallel solver for rarefied gas flow using GSIS}

\author{Yanbing Zhang}
\author{Jianan Zeng}
\author{Ruifeng Yuan}
\author{Wei Liu}
\author{Qi Li\corref{mycorrespondingauthor}}
\author{Lei Wu\corref{mycorrespondingauthor}}
\cortext[mycorrespondingauthor]{Corresponding authors:}
\ead{liq33@sustech.edu.cn; wul@sustech.edu.cn}

\address{Department of Mechanics and Aerospace Engineering, Southern University of Science and Technology, Shenzhen 518055, China}

\begin{abstract}
Recently, the general synthetic iterative scheme (GSIS) has been proposed to find the steady-state solution of the Boltzmann equation in the whole range of gas rarefaction, where its fast-converging and asymptotic-preserving properties lead to the significant reduction of iteration numbers and spatial cells in the near-continuum flow regime.  
However, the efficiency and accuracy of GSIS has only been demonstrated in two-dimensional problems with small numbers of spatial cell and discrete velocities. Here, a large-scale parallel computing strategy is designed to extend the GSIS to three-dimensional  flow problems, including the supersonic flows which are usually difficult to solve by the discrete velocity method. Since the GSIS involves the calculation of the mesoscopic kinetic equation which is defined in six-dimensional phase-space, and the macroscopic high-temperature Navier-Stokes-Fourier equations in three-dimensional physical space, the proper partition of the spatial and velocity spaces, and the allocation of CPU cores to the mesoscopic and macroscopic solvers, are the keys to improving the overall computational efficiency. These factors are systematically tested to achieve optimal performance, up to 100 billion spatial and velocity grids. For hypersonic flows around the Apollo reentry capsule, the X38-like vehicle, and the space station, our parallel solver can obtain the converged solution within one hour. 
\end{abstract}


\end{frontmatter}

\section{Introduction}

Due to the development in space exploration~\cite{votta2013hypersonic,ivanov1998computational}, EUV lithography~\cite{EUV-plasma2021,Teng2023JCProd} and vacuum sysmtes for nuclear fusion~\cite{Tantos2024NuclFusion}, the study of rarefied (non-equilibrium) gas dynamics has become more and more important. From the theoretical perspective, these non-equilibrium flows are governed by the Boltzmann-type kinetic equations~\cite{LeiBook2022} that use the velocity distribution function (VDF) to describe the system state at the mesoscopic level, rather than the Naiver-Stokes-Fourier (NSF) equations in the macroscopic level.  From the computational perspective, the efficient and accurate simulation of the kinetic equation is crucial to emerging technologies in aerospace engineering, where the numerical scheme should be carefully designed as the VDF is defined in the high-dimensional phase space (e.g., for polyatomic gas, it includes the time, the three-dimensional physical space, the three-dimensional molecular velocity space, and the one-dimensional internal energy space). 

The Boltzmann equation can be solved by the stochastic and deterministic methods. Historically, due to the limitation of computer memory, the Boltzmann equation is simulated by the direct simulation Monte Carlo (DSMC) method~\cite{bird1994molecular}. This method uses the simulation particles (each represents a huge number of real gas molecules) to mimic the free streaming and binary collisions of gas molecules. Therefore, the number of simulation particles can be kept small, but the macroscopic quantities in the steady state are obtained by a large number of statistical averaging. In simulating moderate and highly rarefied gas flows, DSMC becomes the prevailing numerical method.  However, when it comes to the simulation of near-continuum flows, due to the splitting of streaming and collision, the spatial cell size and time step in DSMC simulations should be smaller than the mean free path and mean collision time of gas molecules, respectively, hence it is quite inefficient. 

The discrete velocity method is the deterministic method to solve the Boltzmann equation. In addition to spatial discretization, the molecular velocity space is also discretized, by tens of thousands of discrete velocities. Thus, the computer memory requirement can be thousands times of the DSMC. However, due to its deterministic nature, the averaging process is removed, so that it can be faster than DSMC in low-speed simulations~\cite{Ho2019CPC}. The early version of discrete velocity method also handles the streaming and collision separately, therefore suffers from the similar numerical deficiency as DSMC. 

In the past decade, significant progresses have been achieved in both the deterministic and stochastic methods~\cite{Gorji2014JCP, zhu2016implicit,zhu2017multigrid,xu2021unified, Dimarco2021CMAME, Dimarco2018JCP, su2020can, Zeng2023CaF, Fei2020JCP, Fei2023JCP, Feng2023CPC, LiuZhu2020JCP, wei2023}. For instances, the implicit unified gas kinetic scheme (UGKS)~\cite{zhu2016implicit,zhu2017multigrid,xu2021unified} and the general synthetic iterative scheme (GSIS)~\cite{su2020can,Zeng2023CaF} are proposed and applied to challenging multiscale engineering applications. In UGKS, the analytical solution of the Bhatnagar–Gross–Krook (BGK) kinetic equation is used, so that the streaming and collision are handled simultaneously, and the limitation on the spatial cell size is relieved (the asymptotic-preserving property). 
In GSIS, the traditional discrete velocity method is used to solve the Boltzmann equation and its simplified kinetic model equations, together with the macroscopic synthetic equations that facilitate the fast-converging and asymptotic-preserving properties, so that steady-state solutions can be obtained within dozens of iterations in the whole range of gas rarefaction. 
The stochastic numerical methods worth mentioning are the Fokker-Planck solver which is based on the stochastic Langevin process so that the time step is not limited by the mean collision time~\cite{Gorji2014JCP}, the unified stochastic particle method based on the BGK equation~\cite{Fei2020JCP, Feng2023CPC} and the time-relaxed Monte Carlo method for the Boltzmann equation~\cite{Fei2023JCP} where the numerical dissipation induced in the large spatial cell size is compensated by changing the collision term. More recently, the unified gas-kinetic wave-particle method (UGKWP) for the BGK-type equation~\cite{LiuZhu2020JCP, wei2023}, which combines the advantages of both deterministic and stochastic methods, has been proposed to simulate large-scale three-dimensional problems. 

The efficient and accurate simulation of multiscale gas flow problems lies in two factors. The first factor is the remove or relieve of the constraints in spatial cell size and time step, and the second factor is the fast convergence to the steady-state. For the methods introduced in the last paragraph, we see that the first factor is satisfied in most schemes. However, the second is not, since most of the methods lack the global ``information exchange''  to enhance convergence in the whole computational domain. The synthetic equations in GSIS are designed to facilitate quick information exchange process, skipping the intermediate physical-evolving process.
The rigorous mathematical analysis of GSIS shows that it processes the two properties in linear problems~\cite{su2020fast}, while the numerical results show that it processes the two properties in small-scale nonlinear problems~\cite{Zeng2023CaF}. Therefore, it is of great practical meaning  to extend the GSIS  to solver large-scale nonlinear problems. Especially, while it is commonly recognized that the stochastic method is much more efficient than the deterministic methods for high-speed flow simulations, here we are going to show that the GSIS is able to outperform the state-of-the-art stochastic method even in high-speed multiscale flow simulations. 

The remainder of the paper is organized as follows. In Section~\ref{sec:2}, we introduce the high-temperature Navier-Stokes equations used in the continuum flow regime, the modified Boltzmann-Rykov equation valid from the continuum to free-molecular flow regimes, and their relations. In Section~\ref{sec:gsis}, the numerical procedure in solving GSIS is introduced, while in Section~\ref{sec:parallel},  the parallel computing strategy is proposed and the factors that affect the parallel efficiency are analyzed. In Section~\ref{sec:num_example}, the accuracy and efficiency of the parallel computing of GSIS are assessed in several challenging cases. Finally, conclusions are given in Section~\ref{sec:conclusion}.
\section{Governing equations}\label{sec:2}

In the non-equilibrium dynamics of dilute gas, the kinetic model equations have been proposed to describe the evolution of gas VDFs; while the multi-temperature macroscopic equations are usually adopted in the near continuum regime.
Without losing of generality, we consider the molecular gas with 3 translational degrees of freedom and $d_r$ internal degrees of freedom. 

\subsection{High-temperature Navier-Stokes equations}

When thermal non-equilibrium occurs in high-temperature gas, the multi-temperature governing equations for the molecular gas with mass density $\rho$, flow velocity $\bm{u}=(u_1,u_2,u_3)$, translational temperature $T_t$, and internal temperature $T_r$ are given by:
\begin{equation}\label{eq:macroscopic_equation_2}
	\begin{aligned}
		\frac{\partial{\rho}}{\partial{t}} + \nabla\cdot\left(\rho\bm{u}\right)  &= 0, \\
		\frac{\partial}{\partial{t}}\left(\rho\bm{u}\right) + \nabla\cdot\left(\rho\bm{u}\bm{u}\right) + \nabla\cdot\bm{P} &= 0, \\
		\frac{\partial}{\partial{t}}\left(\rho e\right) + \nabla\cdot\left(\rho e\bm{u}\right) + \nabla\cdot\left(\bm{P}\cdot\bm{u}+\bm{q}_t+\bm{q}_r\right) &= 0, \\
        \frac{\partial}{\partial{t}}\left(\rho e_r\right) + \nabla\cdot\left(\rho e_r\bm{u}+\bm{q}_r\right) &= \frac{d_r\rho R}{2}\frac{T-T_r}{Z_r\tau},
	\end{aligned}
\end{equation} 
Here, $t$ is the time and $\bm{x}=(x_1,x_2,x_3)$ is the spatial coordinate; $e_r=d_rRT_r/2$ and $e=(3RT_t+u^2)/2+e_r$ are the specific total and internal energies, respectively; the pressure tensor $\bm{P}$ is given by $\bm{P} = p_t\mathrm{I} + \bm{\sigma}$, with $\bm{\sigma}$ being the shear stress tensor, $p_t=\rho R T_t$ the kinetic pressure, $\mathrm{I}$ the $3\times 3$ identity matrix, and $R$ the gas constant; $\bm{q}_t$ and $\bm{q}_r$ are the translational and internal heat fluxes, respectively; the total temperature $T$ is defined as the equilibrium temperature between the translational and internal modes $T=(3T_t+d_rT_r)/(3+d_r)$. Finally, $Z_r$ is the internal collision number, and $Z_r\tau$ characterizes how fast the internal-translational energy exchange is when compared to the mean collision time $\tau=\mu/p_t$, where $\mu$ is the shear viscosity of the gas. The power-law intermolecular potential is considered, so that the viscosity can be expressed as 
\begin{equation}
\mu(T_t)=\mu(T_0)\left(\frac{T_t}{T_0}\right)^{\omega},
\end{equation}
with $\omega$ the viscosity index and $T_0$ the reference temperature.

In the continuum flow regime, i.e., when the Knudsen number (defined as the ratio of the molecular mean free path $\lambda$ to the characteristic flow length $L_0$)
\begin{equation}
     \text{Kn}=\frac{\lambda}{L_0}\equiv
     \frac{\mu(T_0)}{p_0L}\sqrt{\frac{\pi R T_0}{2}},
 \end{equation}
is small ($\text{Kn}<0.001$), the constitutive relations are given by the Newton law of viscosity and the Fourier law of heat conduction:
\begin{equation}\label{eq:NSF_constitutive}
    \begin{aligned}
        \bm{\sigma}_{\text{NSF}} &= -\mu \left(\nabla\bm{u}+\nabla\bm{u}^{\mathrm{T}}-\frac{2}{3}\nabla\cdot\bm{u}\mathrm{I}\right),\\
        \bm{q}_{t,\text{NSF}} &= -\kappa_t\nabla T_t,\\
        \bm{q}_{r,\text{NSF}} &= -\kappa_r\nabla T_r.
    \end{aligned}
\end{equation}
where $\kappa_t$ and $\kappa_r$ are the transitional and internal thermal conductivities, respectively, and the superscript $\mathrm{T}$ is the matrix transpose.

\subsection{Gas kinetic equations}

Kinetic model equations simplified from the Wang-Chang \& Uhlenbeck equation \cite{wangcs1951transport} are usually adopted in numerical simulations to describe the dynamics of molecular gas in the whole range of gas rarefaction. The model equation applied in this work is initially developed by Rykov~\cite{rykov1975model} and recently extended~\cite{LeiJFM2015,li2021uncertainty,Li2023JFM} to reflect the proper relaxations of energy and heat-flux exchanges between translational and internal modes. Two VDFs, $f_0(t, \bm{x}, \bm{\xi})$ and $f_1(t, \bm{x}, \bm{\xi})$, are used to describe the translational and internal states of gas molecules, where $\bm{\xi}=(\xi_1,\xi_2,\xi_3)$ is the molecular velocity. The macroscopic quantities are obtained by taking moments of VDFs $f_0$ and $f_1$:
\begin{equation}\label{eq:getmoment}
    \begin{aligned}
        \left(\rho,~\rho\bm{u},~\bm{\sigma},~\frac{3}{2}\rho RT_t,~\bm{q}_{t}\right)&=\int\left(1,~\bm{\xi},~\bm{c}\bm{c}-\frac{c^2}{3}\mathrm{I},~\frac{c^2}{2},~\frac{c^2}{2}\bm{c}
        \right) f_0 \mathrm{d}\bm{\xi},\\
        \left(\frac{d_r}{2}\rho RT_r,~\bm{q}_{r}\right)&=\int\left(1,~\bm{c}\right)f_1\mathrm{d}\bm{\xi},
    \end{aligned}
\end{equation}
where $\bm{c}=\bm{\xi}-\bm{u}$ is the peculiar (thermal) velocity. The pressure related to the translational motion is $p_t=\rho R T_t$, while the total pressure is $p=\rho RT$.

In the absence of an external force, the evolution of VDFs is governed by the following kinetic equations:
\begin{equation}\label{general_model}
    \begin{aligned}
        &\frac{\partial f_0}{\partial t}+\bm{\xi}\cdot \nabla f_0 = \frac{g_{0t}-f_0}{\tau}+\frac{g_{0r}-g_{0t}}{Z_r\tau}, \\
        & \frac{\partial f_1}{\partial t}+\bm{\xi}\cdot \nabla f_1 = \frac{g_{1t}-f_1}{\tau}+\frac{g_{1r}-g_{1t}}{Z_r\tau}, 
    \end{aligned}
\end{equation}
where the reference distribution functions are given by:
\begin{equation}
    \begin{aligned}
        g_{0t}&= \rho\left(\frac{1}{2\pi RT_t}\right)^{3/2}\exp\left(-\frac{c^2}{2RT_t}\right)\left[1+\frac{2\bm{q}_{t}\cdot\bm{c}}{15RT_tp_t}\left(\frac{c^2}{2RT_t}-\frac{5}{2}\right)\right],\\
        g_{0r}&= \rho\left(\frac{1}{2\pi RT}\right)^{3/2}\exp\left(-\frac{c^2}{2RT}\right)\left[1+\frac{2\bm{q}_{0}\cdot\bm{c}}{15RTp}\left(\frac{c^2}{2RT}-\frac{5}{2}\right)\right],\\
        g_{1t}&=\frac{d_r}{2}RT_rg_{0t} + \left(\frac{1}{2\pi RT_t}\right)^{3/2}\frac{\bm{q}_{r}\cdot\bm{c}}{RT_t}\exp\left(-\frac{c^2}{2RT_t}\right), \\
        g_{1r}&=\frac{d_r}{2}RTg_{0r} + \left(\frac{1}{2\pi RT}\right)^{3/2}\frac{\bm{q}_{1}\cdot\bm{c}}{RT}\exp\left(-\frac{c^2}{2RT}\right),
    \end{aligned}
\end{equation}
with $\bm{q}_{0},~\bm{q}_{1}$ being linear combinations of translational and internal heat fluxes \cite{li2021uncertainty}:
\begin{equation}
    \begin{bmatrix} 
        \bm{q}_{0} \\ \bm{q}_{1} 
    \end{bmatrix}
    =
    \begin{bmatrix}		
        (2-3A_{tt})Z_r+1 & -3A_{tr}Z_r  \\		
        -A_{rt}Z_r & -A_{rr}Z_r+1 \\ 
    \end{bmatrix}
    \begin{bmatrix} 
    \bm{q}_{t} \\ \bm{q}_{r} 
    \end{bmatrix},
\end{equation}
where $\bm{A}=[A_{tt},A_{tr},A_{rt},A_{rr}]$ is determined by the relaxation rates of heat flux. 

\subsection{Relation between the mesoscopic and macroscopic descriptions}

Here the relation between the mesoscopic and macroscopic equations is introduced. First, 
Eq.~\eqref{eq:macroscopic_equation_2} is obtained by taking moments of the kinetic equations \eqref{general_model}. Note that, at this stage, the pressure tensor and heat flux are determined by the VDFs via Eq.~\eqref{eq:getmoment} rather than \eqref{eq:NSF_constitutive}, making the macroscopic equations valid in all flow regimes but not closed. 

The Chapman-Enskog expansion method~\cite{chapman1990mathematical} is used to close the macroscopic equations at Euler and Navier-Stokes levels. The VDFs $f_l~(l=0,1)$ are expansions in the form of an infinite series of $\text{Kn}$, $f_l = f_l^{(0)} + \text{Kn}f_l^{(1)} + \cdots $. By substituting the expansions into Eq.~\eqref{general_model} with the assumption $Z_r\sim O(\text{Kn}^{-1})$, the zero-order distribution functions $f_l^{(0)}$ can be obtained immediately as the local equilibrium distribution functions with the temperatures $T_t,T_r$ of respective
modes:
\begin{equation}
    \begin{aligned}[b]
        f_0^{(0)}&= \rho\left(\frac{1}{2\pi RT_t}\right)^{3/2}\exp\left(-\frac{c^2}{2RT_t}\right),\quad 
        f_1^{(0)}&=\frac{d_r}{2}RT_rf_0^{(0)}.
    \end{aligned}
\end{equation}
Then the zero-order pressure $\bm{P}^{(0)}$ and heat fluxes $\bm{q}^{(0)}_t,\bm{q}^{(0)}_r$ can be obtained by taking moments of $f_l^{(0)}$, and gives the constitutive relations at Euler approximation:
$\bm{P}_{\text{Euler}} = \bm{P}^{(0)}= \rho RT_t\bm{\mathrm{I}}$ and  
$\left(\bm{q}_{t,\text{Euler}},\bm{q}_{r,\text{Euler}}\right) = \left(\bm{q}^{(0)}_t,\bm{q}^{(0)}_r\right) = \left(0,0\right)$. 
Next, the first-order correction $f_l^{(1)}$ is solved from the equations:
\begin{equation}\label{eq:1st_equation}
	\begin{aligned}[b]
		\mathcal{D}^{(1)}f_l = -\frac{f_l^{(1)}}{\tau} + \frac{g_{lr}-f_l^{(0)}}{Z_r\tau},
	\end{aligned}
\end{equation}
where $\mathcal{D}^{(1)}f_l$ can be explicitly evaluated by ${\partial{f_l^{(0)}}}/{\partial{t}}+\bm{\xi} \cdot {\partial{f_l^{(0)}}}/{\partial{\bm{x}}}$. Thus the first-order distribution functions $f_l^{(1)}$ are given by:
\begin{equation}
    \begin{aligned}
        f_0^{(1)}&=\frac{g_{0r}-f_0^{(0)}}{Z_r}- \tau f_0^{(0)}\left[\left(\frac{c^{2}}{2RT_{t}}-\frac{5}{2}\right)\bm{c}\cdot\nabla\ln{T_{t}} +\frac{d_r}{3T_{t}}\left(\frac{c^{2}}{2RT_{t}}-\frac{3}{2}\right) \frac{T-T_r}{Z_r\tau} \right. \\
		&\left. +\frac{1}{RT_{t}}\left(\bm{c}\bm{c}-\frac{1}{3}c^2\mathrm{I}\right):\nabla\bm{u} \right], \\
        f_1^{(1)}&=\frac{g_{1r}-f_1^{(0)}}{Z_r}- \tau f_1^{(0)}\left[\left(\frac{c^{2}}{2RT_{t}}-\frac{5}{2}\right)\bm{c}\cdot\nabla\ln{T_{t}} +\bm{c}\cdot\nabla\ln{T_{r}} \right. \\
		&\left. +\left(\frac{d_r}{3T_{t}}\left(\frac{c^{2}}{2RT_{t}}-\frac{3}{2}\right) -\frac{1}{T_r}\right)\frac{T-T_r}{Z_r\tau} +\frac{1}{RT_{t}}\left(\bm{c}\bm{c}-\frac{1}{3}c^2\mathrm{I}\right):\nabla\bm{u} \right] .
    \end{aligned}
\end{equation}
Substituting the approximation $f_l^{(0)} + \text{Kn}f_l^{(1)}$ into the definitions of the pressure tensor and heat fluxes, the constitutive relations at the NSF level read,
\begin{equation}\label{eq:NSF_constitutive0}
    \begin{aligned}
        \bm{P}_{\text{NSF}} &= \bm{P}^{(0)}+\bm{P}^{(1)} = \rho RT_t\bm{\mathrm{I}} -\mu \left(\nabla\bm{u}+\nabla\bm{u}^{\mathrm{T}}-\frac{2}{3}\nabla\cdot\bm{u}\mathrm{I}\right),\\
        \left(\bm{q}_{t,\text{NSF}},\bm{q}_{r,\text{NSF}}\right) &= \left(\bm{q}^{(0)}+\bm{q}^{(1)}_t,\bm{q}^{(0)}_r+\bm{q}^{(1)}_r\right) = -\left(\kappa_t\nabla T_t,\kappa_r\nabla T_r\right). 
    \end{aligned}
\end{equation}
where the shear viscosity is $\mu = \rho RT_t\tau$ and thermal conductivities $\kappa_t,\kappa_r$ are given by,
\begin{equation}\label{eq:mu_kappa}
	\begin{aligned}[b]
        \left[ 
            \begin{array}{ccc} 
              \kappa_t \\ \kappa_r 
            \end{array}
          \right]
          &= \frac{\mu R}{2}
          \left[ 
            \begin{array}{ccc} 
              A_{tt} & A_{tr} \\ A_{rt} & A_{rr} 
            \end{array}
          \right]^{-1}
          \left[ 
            \begin{array}{ccc} 
              5 \\ d_r 
            \end{array}
          \right].
	\end{aligned}
\end{equation} 
It shows that each component of the heat flux is related to the corresponding temperature gradient of
its own mode, due to the slow translational-internal energy relaxation.

\section{General Synthetic Iterative Scheme}\label{sec:gsis}

There are two versions of GSIS, where the difference lies in the macroscopic synthetic equations~\cite{Zeng2023CiCP}: in GSIS-I~\cite{su2020can,su2020fast,su2021multiscale} the synthetic equations include the evolution equations for the mass, momentum, energy, stress and heat flux, while in GSIS-II~\cite{zhu2021general,Zeng2023CaF} only the evolution equations for the mass, momentum and energy are considered. Therefore, the asymptotic-preserving property of GSIS-I is better, but meanwhile, the numerical solving of high-order macroscopic equations is more difficult. Here we choose the GSIS-II, since the sophisticated numerical techniques in computational fluid dynamics can be directly used; indeed, anyone who can write program to solve the NSF equations can implement the GSIS-II without any difficulties. 

We adopt the finite volume scheme with second order of accuracy to solve the kinetic equations and macroscopic equations. We only show the major steps here, leaving the details in Ref.~\cite{zhu2021general,Zeng2023CaF}. 

\begin{figure}[t]
    \centering
    \subfigure[]{
        \label{fig:3D_Ma5_AoA30_X_unstructure_Volume}
        {\includegraphics[width=0.46\textwidth]{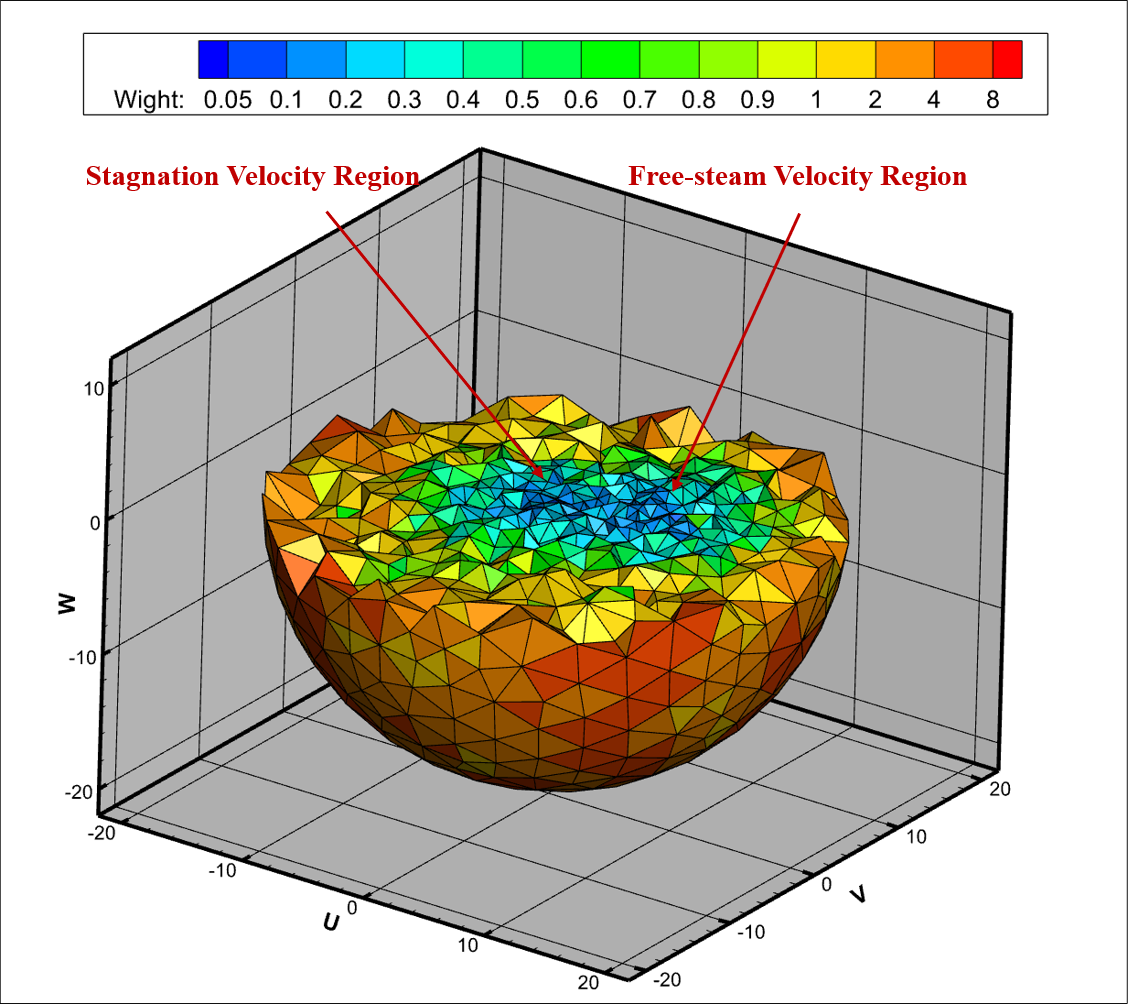}}}
         \subfigure[]{
        \label{fig:3D_Ma5_AoA30_X_hybrid_Volume}
        {\includegraphics[width=0.46\textwidth]{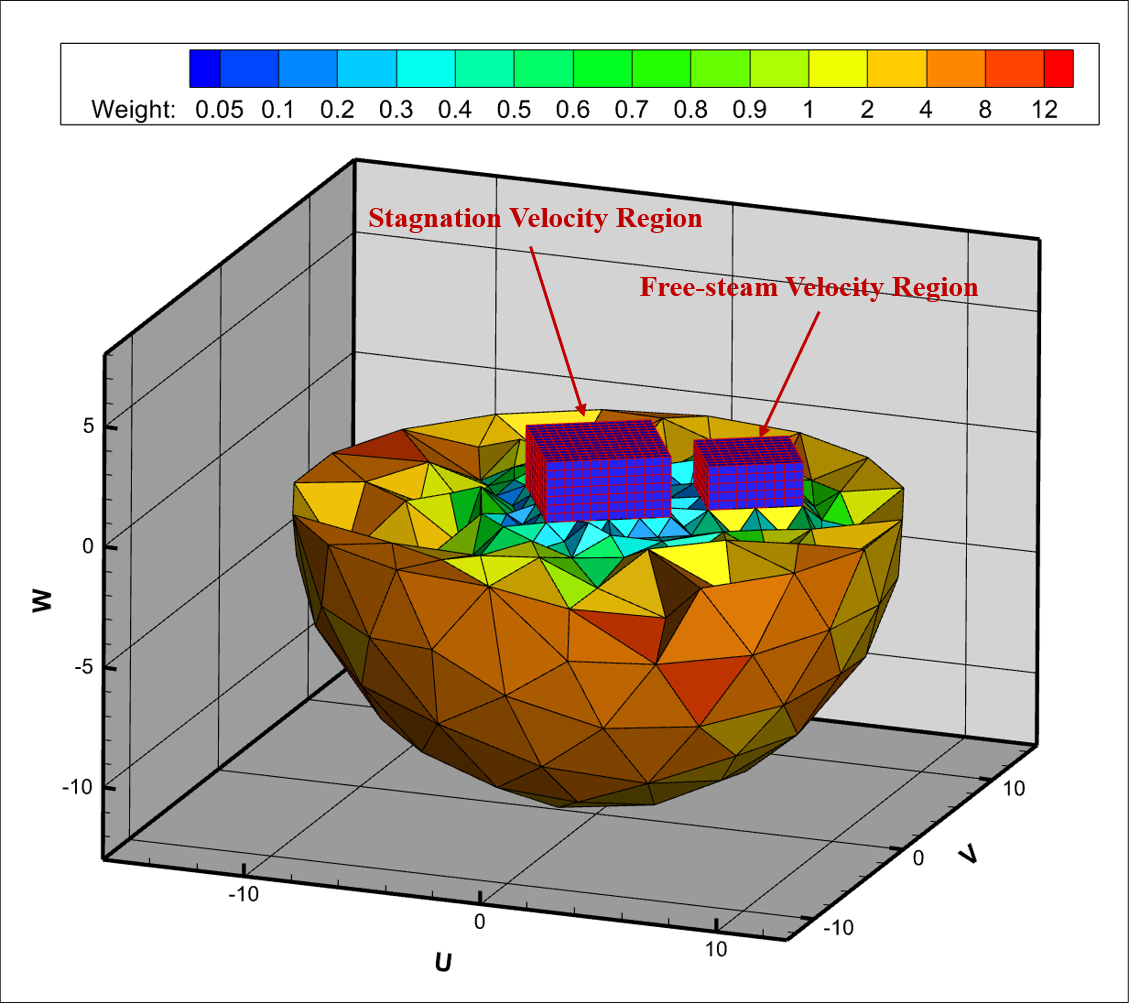}}}
    \caption{The velocity domain is respectively discretized for a hypersonic flow at $\text{Ma}=5$ and $\text{AoA}=30^\circ$ with (a) $27,704$ unstructured tetrahedral cells and (b) $8,166$ structure-unstructured hybrid cells.
    Cells are refined around the stagnation and free stream velocities~\cite{yuanPhd,zhang2023conservative,zhang4724172implicit}. 
    The color indicates the volume of the discretized velocity cells, which is nondimensionalized by $(RT_0)^{3/2}$.}
    \label{fig:Ma5_AOA30_X_velocity_mesh}
\end{figure}

\subsection{Unstructured velocity space discretization}

In the DVM, the continuous velocity space should be discretized first. Normally, for low-speed flows, it can be discretized by the Cartesian grid, both uniformly or non-uniformly~\cite{Wu2014JFM}. In the simulation of hypersonic flows, the velocity space is usually discretized by the unstructured mesh, see Fig.~\ref{fig:3D_Ma5_AoA30_X_unstructure_Volume}.
However, a huge number of velocity grid is required. Recently, it is found that the local refinement of velocity space in front of shock wave, after shock wave and at the wall surface can effectively reduce the number of velocity discretization without affecting the integral accuracy. ~\cite{yuanPhd,zhang2023conservative,zhang4724172implicit}. According to the $3\sigma$ and $5\sigma$ criteria of the standard normal distribution, the probability of the sample falling within the corresponding interval is 99.73\% and 99.99\%, respectively.
In the three-dimensional example, the whole velocity space is truncated into a sphere centered at $\bm{U}_s=\bm{U}_\infty/ \left[\left(\gamma + 1\right)\text{Ma}^2 / \left(2 + \left(\gamma - 1 \right)\text{Ma}^2\right) \right]$ with radius $5\sqrt{RT_s}$, where $T_s=\max\left(T_m, T_w, T_\infty \right)$, $T_m=T_\infty \left[1 + \left(\gamma - 1\right)\text{Ma}^2 / 2\right]$, $\gamma$ is the specific heat ratio, $T_w$ is the wall temperature, $\bm{U}_\infty$ and $T_\infty$ are the velocity and temperature of the free flow. 
Then the refinement discretization is performed at the centers $\bm{U}_\infty$ and $\bm{0}$ with the lengths of $3\sqrt{RT_\infty}$ and $3\sqrt{RT_w}$, respectively. The refinement range can be a sphere or a cube, and the rest velocity domain is filled with unstructured tetrahedra, see Fig.~\ref{fig:3D_Ma5_AoA30_X_hybrid_Volume}. If the refinement range is rectangular and uniform orthogonal discrete, high-precision integration format can be used to improve the integration accuracy. 

\subsection{The kinetic solver}


Since $f_0$ and $f_1$ share the same form, in the following they are represented by $f$ for clarity of presentation. Given the gas information at the $n$-th iteration step, the discretized velocity distribution at the next intermediate (the VDFs will be further modified according to the solution of the synthetic equations) iteration step $n+1/2$ is calculated as
\begin{equation}
    \begin{aligned}[b]
        \frac{f_i^{n+1/2}-f_i^n}{\Delta t} + \frac{1}{V_i}\sum_{j\in N(i)} \xi_nf_{ij}^{n+1/2}S_{ij}=\frac{g^{n}_{i}-f^{n+1/2}_{i}}{\tau^{n}_i},
    \end{aligned}
\end{equation}
where 
\begin{equation}
        g = \left(1-\frac{1}{Z_r}\right)g_t+\frac{1}{Z_r}g_r.
\end{equation}
In the above two equations, $\Delta t $ is the time step;
The subscripts $i,~j$ are the indices of the control cells, and the subscript $ij$ denotes the interface between the adjacent cells $i$ and $j$, with
$S_{ij}$ and $V_i$ being the area of interface $ij$ and the volume of cell $i$, respectively. $\xi_n=\bm{\xi}\cdot\bm{n}$ is the molecular velocity component along normal direction $\bm{n}=\bm{S}/|\bm{S}|$ pointing from cell $i$ to cell $j$; the sum of fluxes $\xi_n f_{ij}$ is taken over all the faces of a cell $N(i)$.

To apply a simple matrix-free implicit solving of the discretized equations, the incremental variable $\Delta f_i=f_i^{n+1/2}-f_i^{n}$ is introduced. Therefore, the delta-form discretized kinetic equation for $\Delta f_i$ is given by:
\begin{equation}\label{eq:delta_equations}
    \left( \frac{1}{\Delta t} + \frac{1}{\tau_i^n}\right)\Delta f_i + \frac{1}{V_i}\sum_{j\in N(i)} \xi_n\Delta f_{ij}S_{ij} = \underbrace{ \frac{g^{n}_{i}-f^{n}_{i}}{\tau^{n}_i}-\frac{1}{V_i}\sum_{j\in N(i)} \xi_nf_{ij}^{n}S_{ij}}_{r_i^n}.
\end{equation}

The interface fluxes $f_{ij}^n$ in the right-hand-side of Eq.~\eqref{eq:delta_equations} are reconstructed using a second-order upwind scheme. Specifically, we have $f_{ij}=[\xi_n^+(f_i+\phi \nabla f_i\cdot \bm{x})+\xi_n^-(f_j + \phi \nabla f_j\cdot \bm{x})]/2$, where $\xi_n^{\pm}=[1\pm \text{sign}(\xi_n)]$ denotes the interface sign directions with respect to the cell center value, and $\phi$ is the Venkatakrishnan limiter. The derivative information is obtained via the least squares method. On the other hand, the increment fluxes $\Delta f_{ij}^n$ in the left-hand-side of Eq.~\eqref{eq:delta_equations} are constructed using a first-order upwind scheme: $\Delta f_{ij}=(\xi_n^+\Delta f_i+\xi_n^-\Delta f_j)/2$.  Finally, Eq.~\eqref{eq:delta_equations} can be rewritten as:
\begin{equation}
\left(\frac{1}{\Delta t} + \frac{1}{\tau_i^n} + \frac{1}{2V_i}\sum_{j\in N(i)}{\xi}_n^+ {\xi}_n S_{ij} \right) \Delta f_i 
+ \sum_{j\in N(i)} \left(\frac{1}{2V_i}{\xi}_n^- {\xi}_n S_{ij} \right) \Delta f_j=r_i^n,
\end{equation}
which can be solved using the standard Lower-Upper Symmetric Gauss-Seidel (LU-SGS) technique. When $\Delta f_i$ is solved, the VDF at the intermediate step is given by 
\begin{equation}
   f_i^{n+1/2}= f_i^{n}+\Delta f_i.
\end{equation}


\subsection{The macroscopic solver}

To describe the rarefaction effects, the constitutive relation in the macroscopic equations~\eqref{eq:macroscopic_equation_2} should not only contain the Newton and Fourier laws of viscosity and heat conduction, but also contain the high-order rarefaction effects. In GSIS-II, the stress and heat fluxes are constructed in the following manner:
\begin{equation}\label{eq:full_constitutive}
    \begin{aligned}
        \bm{\sigma}^{n+1} &= \bm{\sigma}^{n+1}_{\text{NSF}}  + 
        \underbrace{ \int \left(\bm{c}\bm{c}-\frac{c^2}{3}\mathrm{I}\right)f^{n+1/2}_0 \mathrm{d}\bm{v} -\bm{\sigma}^{n+1/2}_{\text{NSF}} }_{\text{HoT}_{\bm{\sigma}}},
        \\
        \bm{q}^{n+1}_{t} &=  \bm{q}^{n+1}_{t,\text{NSF}} + \underbrace{ {\frac{1}{2}}\int \bm{c}c^2f^{n+1/2}_0 \mathrm{d}\bm{v} -\bm{q}^{n+1/2}_{t,\text{NSF}} }_{\text{HoT}_{\bm{q}_{t}}},
        \\
        \bm{q}^{n+1}_{r} &= \bm{q}^{n+1}_{r,\text{NSF}} + \underbrace { \int \bm{c}f^{n+1/2}_1 \mathrm{d}\bm{v}  -\bm{q}^{n+1/2}_{r,\text{NSF}} }_{\text{HoT}_{\bm{q}_{r}}},
    \end{aligned}
\end{equation}
such that when substituting Eq.~\eqref{eq:full_constitutive} into Eq.~\eqref{eq:macroscopic_equation_2}, the traditional NSF equations with source terms coming from the high-order constitutive relations are obtained, where variables without the superscripts are all solved in the $(n+1)$-th step.

The discretized form of the governing equation \eqref{eq:macroscopic_equation_2} for the macroscopic properties with the backward Euler method can be written as:
\begin{equation}\label{eq:macro}
    \frac{\bm{W}_i^{n+1}-\bm{W}_i^n}{\Delta t} + \frac{1}{V_i}\sum_{j\in N(i)}\bm{F}_{ij}^{n+1}S_{ij}=\bm{Q}^{n+1}_i,
\end{equation}
where the detailed expressions of macroscopic variables $\bm{W}$, the fluxes $\bm{F}$ including both convective and viscous parts, and the source terms $\bm{Q}$ are given in the appendix in~\cite{Zeng2023CaF}. Introduce the incremental variables $\Delta \bm{W}_i^m=\bm{W}_i^{m+1}-\bm{W}_i^{m}$ with $m$ being the inner iteration index in solving macroscopic equations, Eq.~\eqref{eq:macro} is converted to
\begin{equation}\label{eq:delta_Macroscopic_equations}
    \begin{aligned}[b]
        \left[\frac{1}{\Delta t_i}- \left(\frac{\partial \bm{Q}}{\partial \bm{W}}\right)_{i}^{m}\right]\Delta \bm{W}_i^{m} + \frac{1}{V_i} \sum_{j\in N(i)} \Delta\bm{F}_{ij}^{m}S_{ij}=
        \underbrace{-\frac{1}{V_i}\sum_{j\in N(i)} \bm{F}_{ij}^{m}S_{ij}+\bm{Q}_i^{m}}_{R_i^m}.
    \end{aligned}
\end{equation}
The general form of the macroscopic fluxes can be expressed as $F_{ij} = F(W_L, W_R, S_{ij})$, where $W_{L,R}$ represents the reconstructed values of the left and right sides of the interface, respectively, and can be further written as $W_{L/R}=W_{i/j}+\phi \nabla W_{i/j}\cdot \bm{x}$. For the reconstruction of the macroscopic flux, the Rusanov scheme~\cite{mohamed2021modified} is applied, while the gradient and the limiter are chosen to be consistent with the mesoscopic equations.


To obtain a matrix-free form, the implicit fluxes in the macroscopic system \eqref{eq:delta_Macroscopic_equations} are approximated by the Euler-type fluxes: $\Delta \bm{F}_{ij}^m = \frac{1}{2}\left[\Delta \bm{F}_i^m + \Delta \bm{F}_j^m+\Gamma_{ij}(\Delta\bm{W}_i^m - \Delta\bm{W}_j^m)\right]$, with $\Gamma_{ij} = |u_n| + c_s + \frac{2\mu}{\rho|\bm{n}_{ij}\cdot(\bm{x}_j-\bm{x}_i)|}$.
Since the control volume satisfies the geometric conservation law, the interface fluxes through the cell accumulate to  $\sum_{j\in N(i)}\bm{F}_{i}S_{ij}=0$. While the flux can be directly represented by the convective one $\bm{F}=\bm{F}_c$, the flux of subscript $j$ can be written as a matrix-free form $\Delta \bm{F}_j^m=\bm{F}(\bm{W}_j^m + \Delta \bm{W}_j^m) - \bm{F}(\bm{W}_j^m)$. 
Substituting this into Eqs.~\eqref{eq:delta_Macroscopic_equations}, the implicit governing equations for macroscopic variables become
\begin{equation}\label{eq:res_macro}
    D_i\Delta \bm{W}_i^{m} + \frac{1}{2V_i}\sum_{j\in N(i)} \left(\Delta\bm{F}_{j}^{m} - \Gamma_{ij}\Delta \bm{W}_j^m\right)S_{ij}=\bm{R}_i^m,
\end{equation}
where $D_i = \frac{1}{\Delta t_i}+\frac{1}{2V_i}\sum_{j\in N(i)}\Gamma_{ij}S_{ij}- \left(\frac{\partial \bm{Q}}{\partial \bm{W}}\right)^m_i$.

The macroscopic solver needs boundary conditions (note that in rarefied gas dynamics, the no-velocity-slip and no-temperature-jump conditions do not hold anymore). In the initial work of the GSIS for nonlinear flows~\cite{zhu2021general}, the macroscopic synthetic equations were solved in the inner domain, excluding four cell layers adjacent to solid walls. Thus, although the total iteration number for the kinetic solver (which is the most time-consuming part) can be greatly reduced when compared to the traditional implicit discrete velocity method, it still needs several hundreds of iterations. This problem is partially fixed in our recent paper~\cite{Zeng2023CaF}, where the boundary flux is modified by the physical quantity increment of the boundary element, in a similar manner as the Roe scheme. 
Very recently, we further proposed a generalized macroscopic boundary treatment to achieve super-accelerated convergence in GSIS~\cite{liu2024further}, where the conservative variables in the NSF solver, and the high-order constitutive relations for stress and heat flux in the kinetic solver, are used to construct the VDFs similar to that used in the Grad 13 moment method, and hence providing the boundary flux for the macroscopic solver in each step $m$. 
The velocity distribution function on both sides $L,R$ of the interface $ij$ is reconstructed based on macroscopic quantity information:
\begin{equation} \label{eq:G13_VDF}
    \begin{aligned}
        f^{L,R}_{0, ij}&=f^{eq}\left[1+\frac{\bm{\sigma}\cdot \bm{cc}}{2\rho T^2} + \frac{\bm{q}_t\cdot\bm{c}}{\rho T^2}\left(\frac{\bm{c}^2}{5T_t}-1\right)\right], \\
        f^{L,R}_{1,ij}&=\left(\frac{d_r}{2} T_r\right) f^{L,R}_{0,ij}\left[1+\frac{\bm{\sigma} \cdot \bm{c c}}{2 \rho T_t^2}+\frac{\bm{q}_t \cdot \bm{c}}{\rho T_t^2}\left(\frac{\bm{c}^2}{5 T_t}-1\right)\right]+f^{L,R}_{0,ij} \frac{\bm{q}_r  \cdot \bm{c}}{\rho T_t},
    \end{aligned}
\end{equation}
where the conservative variables in the NSF solver, and the high-order constitutive relations for stress and heat flux in the kinetic solver, are used to construct the VDFs similar to that used in the Grad 13 moment method,
and then the macroscopic flux at the interface $ij$ could be obtained:
\begin{equation} \label{eq:GSIS-DVM-flux}
    {{\bm{F}}_{ij}} = S\left[ {\begin{array}{*{20}{c}}
        {\rho {u_n}}\\
        {\rho {\bm{u}}{u_n}}\\
        {\rho E{u_n}}\\
        {\rho {E_r}{u_n}}
        \end{array}} \right] = S\left[ {\begin{array}{*{20}{c}}
        {\int_{\xi_n>0} {{\xi_n}{f^L_{0,ij}}} \mathrm{d}\bm{\xi} + \int_{\xi_n<0} {{\xi_n}{f^R_{0,ij}}} \mathrm{d}\bm{\xi}}\\
        {\int_{\xi_n>0} {{\xi_n}{f^L_{0,ij}}{\bm{\xi}}} \mathrm{d}\bm{\xi} + \int_{\xi_n<0} {{\xi_n}{f^R_{0,ij}}{\bm{\xi}}} \mathrm{d}\bm{\xi}}\\
        {\int_{\xi_n>0} {{\xi_n}\left( {\frac{1}{2}{\bm{\xi}^2}{f^L_{0,ij}} + {f^L_{1,ij}}} \right)} \mathrm{d}\bm{\xi} + \int_{\xi_n<0} {{\xi_n}\left( {\frac{1}{2}{\bm{\xi}^2}{f^R_{0,ij}} + {f^R_{1,ij}}} \right)} \mathrm{d}\bm{\xi}} \\
        {\int_{\xi_n>0} {{\xi_n}{f^L_{1,ij}}} \mathrm{d}\bm{\xi} + \int_{\xi_n<0} {{\xi_n}{f^R_{1,ij}}} \mathrm{d}\bm{\xi}}
        \end{array}} \right]
\end{equation}
and hence providing the boundary flux for the macroscopic solver in each step $m$. 
Since the distribution function in Eq.~\eqref{eq:G13_VDF} is an equilibrium distribution, the above macroscopic flux can be expressed and calculated explicitly.
Details are presented in Ref.~\cite{liu2024further} since it involves complicated mathematics; also, the boundary condition affects only the iteration number $n$ but not the parallel efficiency; the latter is the major focus of the present paper.


When the macroscopic conservative variables $\bm{W}^{n+1}$ are solved, they are used to update the VDF. That is, the non-equilibrium part is kept while the equilibrium is modified: 
  \begin{equation}\label{eq:updatef}
      f^{n+1} = f^{n+1/2} + [f_{eq}(\bm{W}^{n+1})-f_{eq}(\bm{W}^{n+1/2})].
  \end{equation}

\begin{algorithm}[t]
    \caption{Overall flowchart of GSIS} 
    \label{GSIS_procedure}
    \begin{algorithmic}[1]
        \Require
            Initial macroscopic quantities $\bm{W}$;
        \Ensure
            Macroscopic quantities $\bm{W}$;
        \State Getting initial field by calculating 1000 steps of the macroscopic solver (see Algorithm~\ref{NS_precedure} below) with Euler constitutive relations and 10 steps of the kinetic solver (see Algorithm~\ref{DVM_procedure} below) in general;
        \State set $steps = 0,\ error = 1$;
        \While {$steps \le \text{MaxSteps} \parallel error \ge 1e-6$}
            \State $steps ++$;
            \State Update velocity distribution function using macroscopic quantities;
            \State Kinetic solver evolves once, see Algorithm~\ref{DVM_procedure};
            \State Calculate the macroscopic quantities $\bm{W}$, high-order terms $\text{HoT}_{\bm{q}}$ and $\text{HoT}_{\bm{\sigma}}$ defined in Eq.~\eqref{eq:full_constitutive}, boundary flux $F_{ij}$;
            \State Macroscopic solver evolves multiple ($300 \sim 400$) times, see Algorithm~\ref{NS_precedure};
            \State Calculate $error$;
        \EndWhile
    \end{algorithmic}
\end{algorithm}

\section{Parallel implementation of GSIS}\label{sec:parallel}


To meet the requirement of solving non-equilibrium flows on complex configurations, the parallel implementation of a solver needs to be carefully designed to achieve high performance. Although the parallelism model on processors with shared memory architectures has the advantage of simplicity, the scalable performance is limited to tens of processors. Thus, the parallelism model works on distributed memory architectures using the Message Passing Interface (MPI) library is usually utilized in large-scale practical simulations.

The overview of GSIS is shown in Algorithm~\ref{GSIS_procedure}. Every single iteration step in GSIS invokes the kinetic solver once and the macroscopic solver several tens or hundreds of times. Considering the significant difference in computing time and memory requirement between the two solvers, their parallel strategies are designed separately to achieve an optimized usage of computational resources. Note that in order to increase the stability of the algorithm, pre-conditioning of the macroscopic and kinetic equations is adopted. Namely, we first run the macroscopic solver with Euler constitutive relations for 1000 steps, then the kinetic solver for 10 steps, before calling the GSIS.

\begin{algorithm}[tb]
    \caption{Macroscopic solver: Spatial domain decomposition to solve macroscopic equations in parallel. The {\color{blue} MPI\_Startall()} and {\color{blue} MPI\_Waitall()} are non-blocking communication subroutines in MPI, which correspond to one-to-one invoking.} 
    \label{NS_precedure}
    \begin{algorithmic}[1]
        \Require
            Macroscopic quantities $\bm{W}$, high-order terms $\text{HoT}_{\bm{q}}$ and $\text{HoT}_{\bm{\sigma}}$, and boundary flux $\bm{F}_{ij}$ obtained from the previous iteration or the initial conditions;
        \Ensure
            Macroscopic quantities $\bm{W}$ in the next iteration step of kinetic solver;
        \State set $steps = 0,\ error = 1,\ n = 4$;
        \While {$steps \le \text{MaxSteps} \parallel error \ge 10^{-6}$}
            \State $steps ++$;
            \State {\color{blue} MPI\_Startall()}, send data of macroscopic quantities on ghost cells between spatial partitions;
            \State Update boundary conditions;
            \State {\color{blue} MPI\_Waitall()}, wait for data reception to complete;
            \State Interpolate the cell-centered macroscopic quantities to the interfaces and calculate the fluxes in each spatial subdomain;
            \State {\color{blue} MPI\_Startall()}, send data of unilateral fluxes on interfaces between subdomains;
            \State Calculate cell-based source terms and time steps;
            \State {\color{blue} MPI\_Waitall()}, wait for interface flux reception to complete;
            \State Calculate total fluxes on interfaces between subdomains;
            \For{$i = 0$; $i<n$; $i++$}
                \State {\color{blue} MPI\_Startall()}, send data of increment of macroscopic quantities $\Delta \bm{W}$;
                \State {\color{blue} MPI\_Waitall()}, wait for $\Delta \bm{W}$ reception to complete;
                \State LU-SGS iteration;
            \EndFor
            \State Boundary flux modification;
            \State Calculate $error$;
        \EndWhile
    \end{algorithmic}
\end{algorithm}

\subsection{Parallel computing strategy}

For the macroscopic solver, a natural parallel implementation on unstructured grids uses spatial domain decomposition to partition the grids across processors, where each processor performs the computations for its assigned grids. In the implicit scheme, the information communicated between processors includes the macroscopic variables and their fluxes at the interfaces between subdomains, as well as at neighboring grid cells outside each subdomain. Thus, additional layers of adjacent grid cells around a subdomain are attached to the associated processor as ghost cells (one layer of ghost cells is sufficient for the numerical schemes up to second order). Algorithm~\ref{NS_precedure} shows the parallel computing strategy for solving macroscopic equations. The non-blocking version of the MPI send/receive subroutines is used to simultaneously execute the computation and MPI communication.

\begin{algorithm}[tb]
    \caption{Kinetic solver: Hybrid parallelization of spatial domain and velocity domain decomposition to calculate mesoscopic equations. The {\color{red} MPI\_Allreduce()} is reduction subroutine in MPI.}
    \label{DVM_procedure}
    \begin{algorithmic}[1]
        \Require
           Distribution function $f$ from the previous iteration or the initial conditions;
        \Ensure
            Macroscopic quantities $\bm{W}$, high order terms $\text{HoT}_{\bm{q}}$ and $\text{HoT}_{\bm{\sigma}}$, boundary flux $F_{ij}$;
        \State set $steps = 0,\ error = 1,\ n = 4$;
        \State The velocity domain decomposition; 
        \While {$steps \le \text{MaxSteps} \parallel error \ge 10^{-6}$}
            \State $steps ++$;
            \State {\color{blue} MPI\_Startall()},  send data of velocity distribution function on ghost cells between spatial partitions;
            \State Calculate collision terms;
            \State {\color{blue} MPI\_Waitall()}, wait for data reception to complete;
            \State Interpolate the cell-centered velocity distribution function to the interfaces and calculate the fluxes in each spatial subdomain;
            \State {\color{blue} MPI\_Startall()}, send data of unilateral fluxes on interfaces between subdomains;
            \State Calculate boundary fluxes;
            \State {\color{blue} MPI\_Waitall()}, wait for interface flux reception to complete;
            \State Calculate total fluxes on interfaces between subdomains;
            \For{$i = 0$; $i<n$; $i++$}
                \State {\color{blue} MPI\_Startall()}, send data of increment of velocity distribution functions $\Delta f$;
                \State {\color{blue} MPI\_Waitall()}, wait for $\Delta f$ reception to complete;
                \State LU-SGS iteration;
            \EndFor
            \State {\color{red} MPI\_Allreduce()},  calculate macroscopic quantities based on Eq.~\eqref{eq:getmoment}; 
            \State Calculate $error$;
        \EndWhile
    \end{algorithmic}
\end{algorithm}

For the kinetic solver, the implementation of spatial domain decomposition can be the same as that for the macroscopic solver. However, the information of entire velocity grids needs to be stored on each computing core. Thus, it can easily exceed the memory limitations of a single core for hypersonic flow simulations, where a large number of discrete velocity grids are required. Besides, an enormous amount of distribution functions needs to be communicated between processors, which may lead to a significant reduction in parallel efficiency when the number of subdomains is large. Therefore, the velocity domain decomposition, as the second level parallelism in the kinetic solver, is also inevitably required. Note that the distribution functions on discrete velocity points are independent of each other in the calculations of streaming and collision terms, while the data dependency and information communications are required only for calculating macroscopic quantities. It is simple to implement velocity domain decomposition and achieve high-efficiency parallelism, as shown in Algorithm~\ref{DVM_procedure}. In addition to the send/receive subroutines, the MPI reduction subroutine is also used to calculate Eq.~\eqref{eq:getmoment}.

\begin{figure}[t]
    \centering
    {\includegraphics[scale=0.6,clip = true]{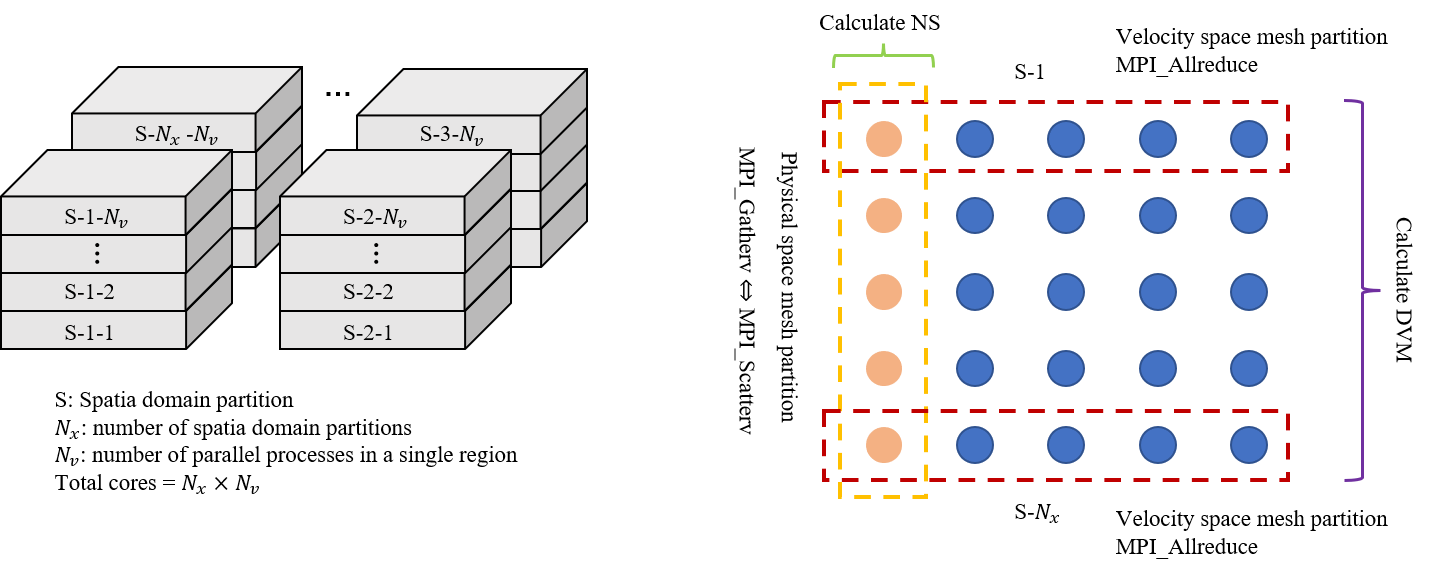}}
    \caption{Allocation and usage of the computing cores in the parallel strategy of GSIS for large-scale simulations. The processor S-m-n performs the computations for grids in the $m$-th spatial subdomain (m=1,2,...,$N_x$); for the kinetic solver, n=1,2,...,$N_v$ represents the second level parallelism for velocity subdomain, while for the macroscopic solver, n=1 is used in each spatial cell partition. }
    \label{fig:HybridParallel}
\end{figure}

The parallel strategy of GSIS is sketched in Fig.~\ref{fig:HybridParallel}: (i) as the first level parallelism for both macroscopic and kinetic solvers, the entire spatial domain is decomposed into $N_x$ subdomains by graph partitioning techniques to achieve optimized load balance and time cost on associated message passing across processors, e.g., see Fig.~\ref{fig:Apollo_gridConfiguration}(c); (ii) for each spatial subdomain, the entire discrete velocity cells are uniformly distributed over $N_v$ processors, as the second level parallelism only for the kinetic solver; (iii) In total, $N_c = N_x\times N_v$ cores, labeled as S-m-n ($m=1,2,...,N_x$, $n=1,2,...,N_v$), are required and used in the kinetic solver, while S-m-1 ($m=1,2,...,N_x$) among those are utilized in macroscopic solver with others waited.

\subsection{Parallel computing efficiency}

\begin{figure}[t]
    \centering
    \subfigure[]{\includegraphics[scale=0.35,clip = true]{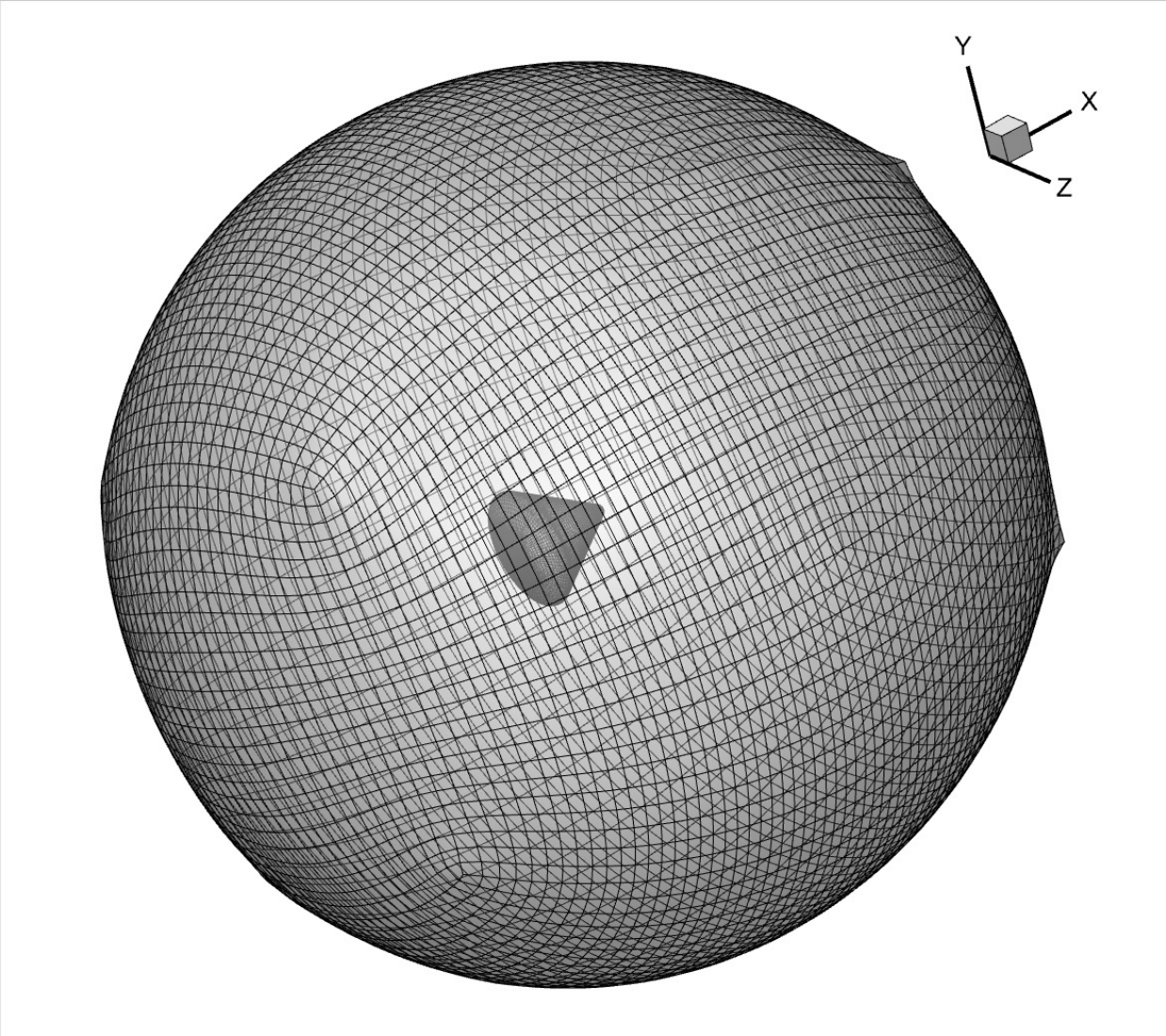}} 
    \subfigure[]{\includegraphics[scale=0.35,clip = true]{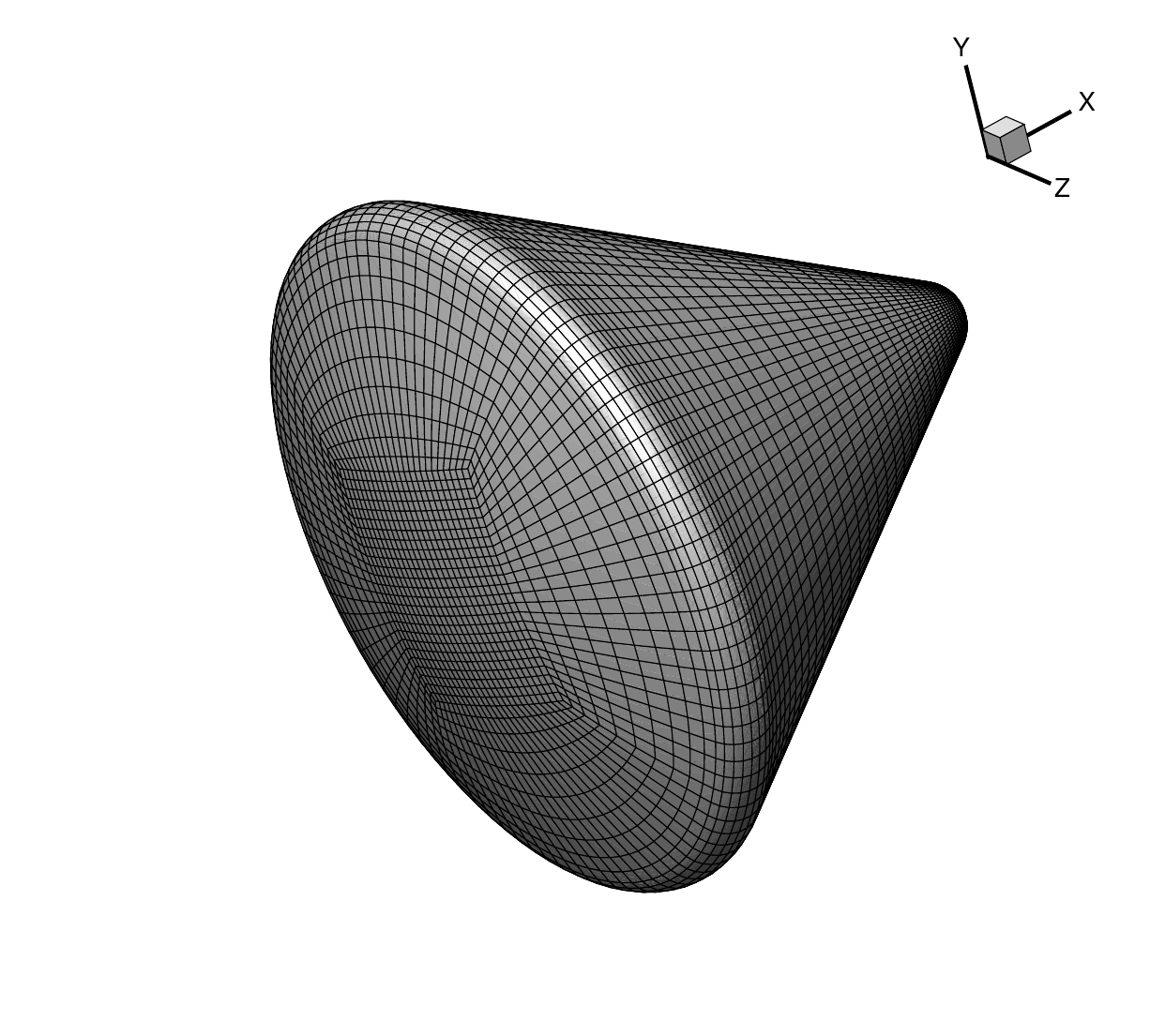}} \\
     \subfigure[] {\includegraphics[scale=0.5,clip = true]{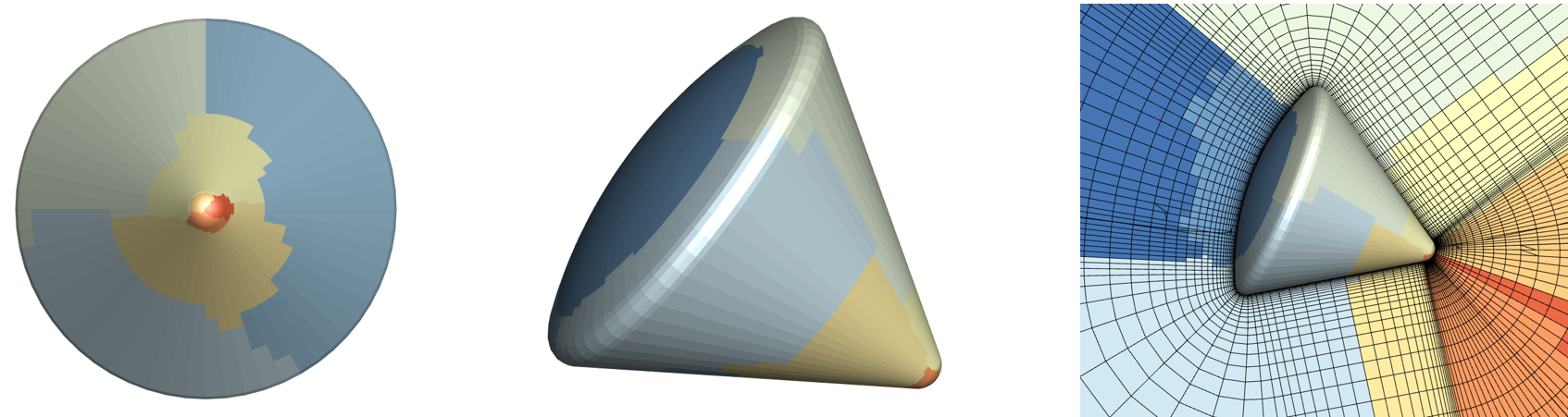}}
    \caption{Spatial domain of the hypersonic flow around Apollo is discretized by 372,500 hexahedral cells. (a) Global view of the simulation domain, (b) a detailed view of the meshes on the wall surface.
    (c) Schematic of partitioning of spatial cells with 10 subdomains indicated by different colors.}
    \label{fig:Apollo_gridConfiguration}
\end{figure}

The parallel computing efficiency of the macroscopic solver, the kinetic solver, as well as the overall GSIS algorithm is assessed individually in the hypersonic flow around the re-entry capsule Apollo at $\text{Ma}=5$, $\text{Kn}=0.0012$, and the angle of attack $\text{AoA}=30^\circ$. The spatial domain consisting of 372,500 hexahedral cells is illustrated in Fig.~\ref{fig:Apollo_gridConfiguration}(a,b) with a detailed view of the mesh on the wall surface, and the velocity domain is discretized by 27,704 tetrahedral cells. The open-source graph partitioning program METIS ~\cite{karypis1998fast} is used to facilitate spatial cell decomposition. For example, Fig.~\ref{fig:Apollo_gridConfiguration}(c) demonstrates a partitioning with 10 subdomains indicated by different colors. All the simulations are conducted on a parallel computer with Inter(R) Core(TM) i7-9700 CPU@3.2GHz.

%

The parallel efficiency of the macroscopic solver for a fixed interval of 2000 iterations is tested, where the spatial partitioning number $N_x$ changes from 1 to 480 by using $N_c=N_x\times 1$ cores. The wall clock time and corresponding parallel efficiency (based on the time cost of a serial solver with $N_x=1$) are shown in Table~\ref{tab:ns_partitions}. It is found that, although the parallel efficiency falls below 90\% when $N_x$ is larger than 20, it keeps fluctuating around 73-88\% as $N_x$ increases from 20 to 400, which is a fair performance in large-scale parallelism. Further increase of $N_x$ leads to a significant reduction of parallel efficiency, due to the chock of message passing between processors. It is noted that $N_x=400$ corresponds to approximate 930 spatial cells assigned to each processor, which can be regarded as a lower limit of cell number per subdomain in spatial partitioning to keep a scalable parallelism in this configuration. In other words, if the total cell number is increased, then using more than 400 spatial partitions will also have a parallel efficiency of around 80\%.

\begin{table}[t]
    \centering
    \caption{Wall clock time and parallel computing efficiency of the macroscopic solver with 2000 iterations in the hypersonic flow passing over Apollo. The number of spatial partitions $N_x$ varies from 1 to 480, and the total number of cores used is $N_c=N_x\times 1$.}
    \begin{tabular}{cccc}\hline
        {Partitions $N_x$} & {Wall time (s)} & {Actual speedup} & {Parallel efficiency} \\ \hline
        1   & 4356.96 & 1.00   & 100.00\% \\ 
        2   & 2330.98 & 1.87   & 93.46\%  \\ 
        5   & 889.01  & 4.90   & 98.02\%  \\ 
        10  & 436.55  & 9.98   & 99.80\%  \\ 
        20  & 246.74  & 17.66  & 88.29\%  \\ 
        80  & 69.37   & 62.81  & 78.51\%  \\ 
        160 & 33.45   & 130.25 & 81.41\%  \\ 
        320 & 17.03   & 255.84 & 79.95\%  \\ 
        360 & 16.42   & 265.34 & 73.71\%  \\ 
        400 & 13.78   & 316.18 & 79.04\%  \\ 
        440 & 15.9    & 274.02 & 62.28\%  \\ 
        480 & 16.2    & 268.95 & 56.03\%  \\ \hline
    \end{tabular}
    \label{tab:ns_partitions}
\end{table}

\begin{figure}[t]
    \centering
    \subfigure[]{
        \label{dvm_Partition}
        {\includegraphics[scale=0.5,clip = true]{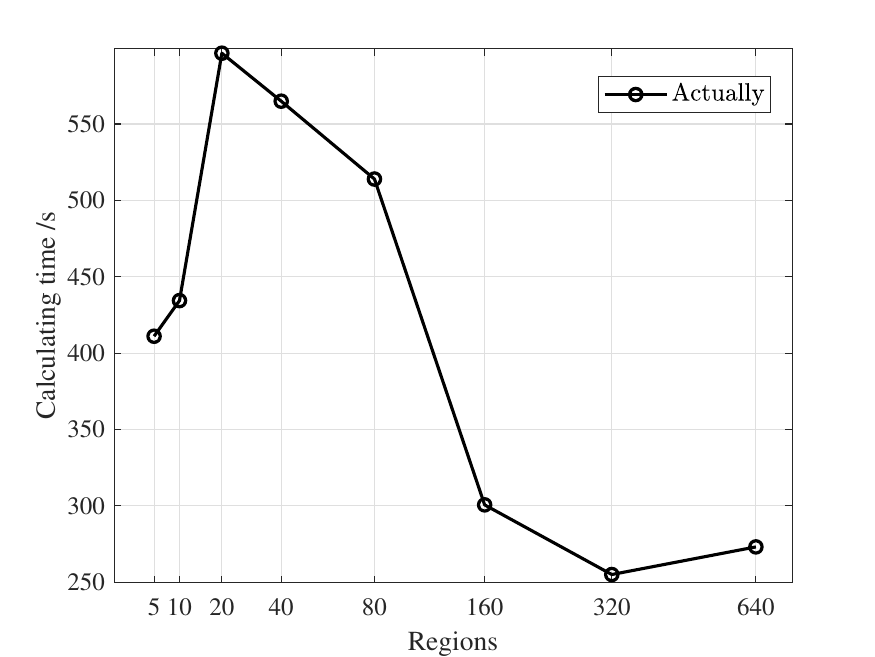}}}
    \subfigure[]{
        \label{dvm_Speedup}
        {\includegraphics[scale=0.5,clip = true]{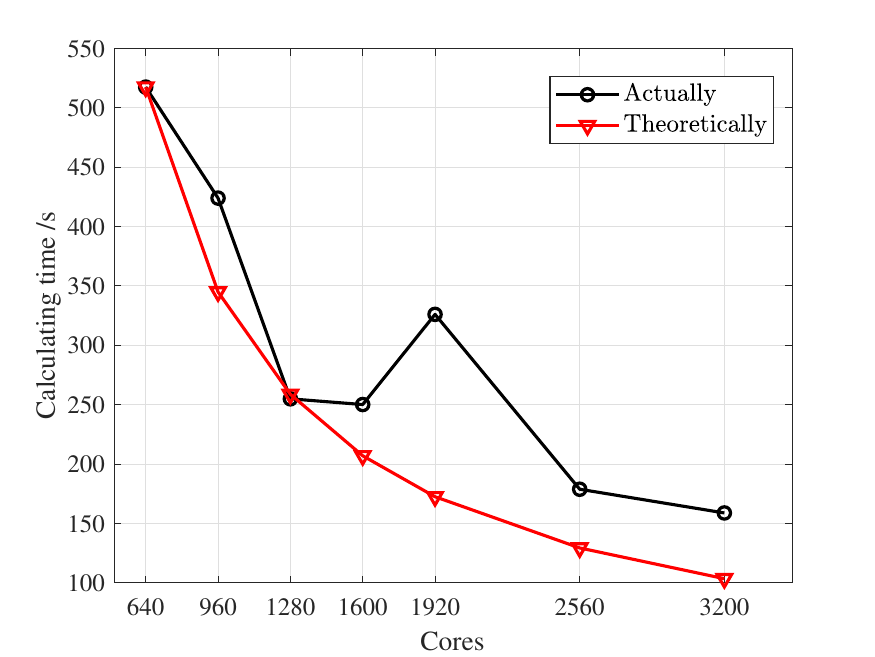}}}
    \caption{The comparison of wall clock time of the kinetic solver with 20 iterations in the hypersonic flow around Apollo by (a) changing the number of partitions $N_x$ with a fixed number of total cores $N_c=1280$ and (b) changing the total core number $N_c$ with a fixed spatial partitioning number $N_x=320$, where the ideal time cost is calculated based on the reference one with $N_c=640$.}
    \label{fig:dvm_Effency}
\end{figure}

To measure the performance of the kinetic solver with the two-level parallel strategy, the computational costs are compared by (i) changing $N_x$ and $N_v$ with $N_c=N_x\times N_v=1280$ fixed; (ii) changing $N_c$ with $N_x=320$ fixed. For the case (i), as shown in Fig.~\ref{dvm_Partition}, once the total core number $N_c$ is fixed, the high efficiency of parallelism (less wall clock time) occurs when using either a small number of spatial subdomains ($N_x\leq10$), or a small number of velocity partitioning ($N_v=N_c/N_x=2\sim8$). The reason lies in the competition in the amount of data transfer between neighboring subdomains and the communication efficiency within the associated processors, both of which reduce when $N_x$ increases. Considering that there are $N_x\times (N_v-1)$ cores waiting in idle for the macroscopic solver, as well as the fact that the number of spatial cells is usually much larger than that of the velocity grids in 3D flow problems, a small number of velocity partitioning will be a practical choice on a fixed number of total cores. For the case (ii), Fig.~\ref{dvm_Speedup} compares the wall clock time by increasing the total core number $N_c$ with $N_x=320$ fixed, where the ideal computational cost is calculated based on the reference one with $N_c=640$. It is found that a high parallel efficiency above 82\% can be guaranteed when $N_c\leq 1600$ ($N_v\leq 5$ correspondingly). Further increase of $N_c$ leads to a significantly larger portion of time cost on the message passing, and thus reduces the efficiency.

Table~\ref{tab:gsis_Speedup} shows the wall clock time and the corresponding parallel efficiency of the overall GSIS solver when running 20 iteration steps, each of which includes one step of kinetic solver and 400 steps of macroscopic solver. The numbers of spatial and velocity partitioning are $N_x=160,320,400$ and $N_v=2,4$, respectively. The parallel computing efficiency is measured based on the time cost of the case with $N_x\times N_v=160 \times 2$ (a serial solver with $N_x\times N_v=1 \times 1$ is too time-consuming to be applied in this problem), and good performance is achieved. 

\begin{table}[h]
    \centering
    \caption{Computational time and parallel computing efficiency of GSIS with 20 iteration steps, each of which includes one step kinetic solver and 400 steps macroscopic solver.  The numbers of spatial and velocity partitioning are $N_x=160,320,400$ and $N_v=2,4$, respectively.}
    \begin{tabular}{ccccc}\hline
        {$N_x\times N_v$} & {Wall time (s)} & {Ideal speedup} & {Actual speedup} & {Parallel efficiency} \\ \hline
        160 $\times$ 2 & 1256.2 & 1   & 1.00    & 100.0\% \\
        320 $\times$ 2 & 638.3  & 2   & 1.97 & 98.4\%  \\
        400 $\times$ 2 & 550.4  & 2.5 & 2.28 & 91.3\%  \\
        160 $\times$ 4 & 708.6  & 2   & 1.77 & 88.6\%  \\
        320 $\times$ 4 & 445.9  & 4   & 2.82 & 70.4\%  \\
        400 $\times$ 4 & 405.5  & 5   & 3.10  & 62.0\%   \\ 
        \hline
    \end{tabular}
    \label{tab:gsis_Speedup}
\end{table}



\section{Numerical results}\label{sec:num_example}



In this section, the parallel GSIS solver is assessed in a 2D lid-driven cavity flow and in hypersonic flows (Apollo reentry module, space vehicle like x38, and space station) with complex 3D configurations. Detailed flow fields are compared with the available data from the DSMC simulations \cite{li2021kinetic} and the latest AUGKWP \cite{wei2023}.

Nitrogen gas with rotational degrees of freedom $d_r=2$, collision number $Z_r=3.5$ and viscosity index $\omega=0.74$ is employed in the following simulations, unless otherwise noted. The thermal relaxation rates are \cite{LeiJFM2015}: $A_{tt}=0.786$, $A_{tr}=-0.201$, $A_{rt}=-0.059$ and $A_{rr}=0.842$, and hence the Eucken factors (corresponding to thermal conductivities) of translational and rotational degrees of freedom are determined as $f_t=2.365, ~f_r=1.435$, respectively.

The convergence criterion of the simulations is that the volume-weighted relative error $\varepsilon$ between two consecutive iterations 
\begin{equation}\label{convergence_critertion}
    \varepsilon=\max \left(\sqrt{\frac{\sum_{i}^{} \left ( W_i^{n}-W_i^{n-1} \right )^2 \varOmega _i }{\sum_{i}^{} \left ( W_i^{n-1} \right )^2 \varOmega _i } }  \right)
\end{equation}
is less than $10^{-6}$, where $W\in \{\rho,\bm{u},T_t\}$. This criterion is used to determine the wall clock time spent in the following simulations. Nevertheless, due to the fast-converging property of GSIS~\cite{su2020fast}, $\varepsilon<10^{-4}$ is sufficient to obtain the converged solutions of critical macroscopic properties, see the results in Fig.~\ref{fig:Apollo_Ma5_Kn00012_ErrorEvolution} for an example.


\begin{figure}[!t]
    \centering
    \subfigure[Spatial discretization]{\includegraphics[width=0.42\textwidth,clip = true]{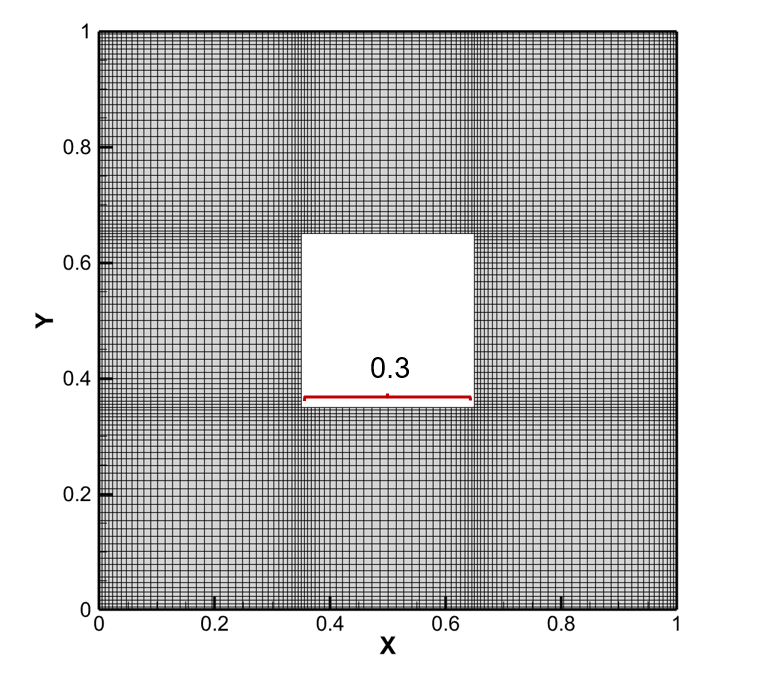}} 
    \subfigure[$\text{Kn}=0.5$]{
        {\includegraphics[width=0.45\textwidth,clip = true]{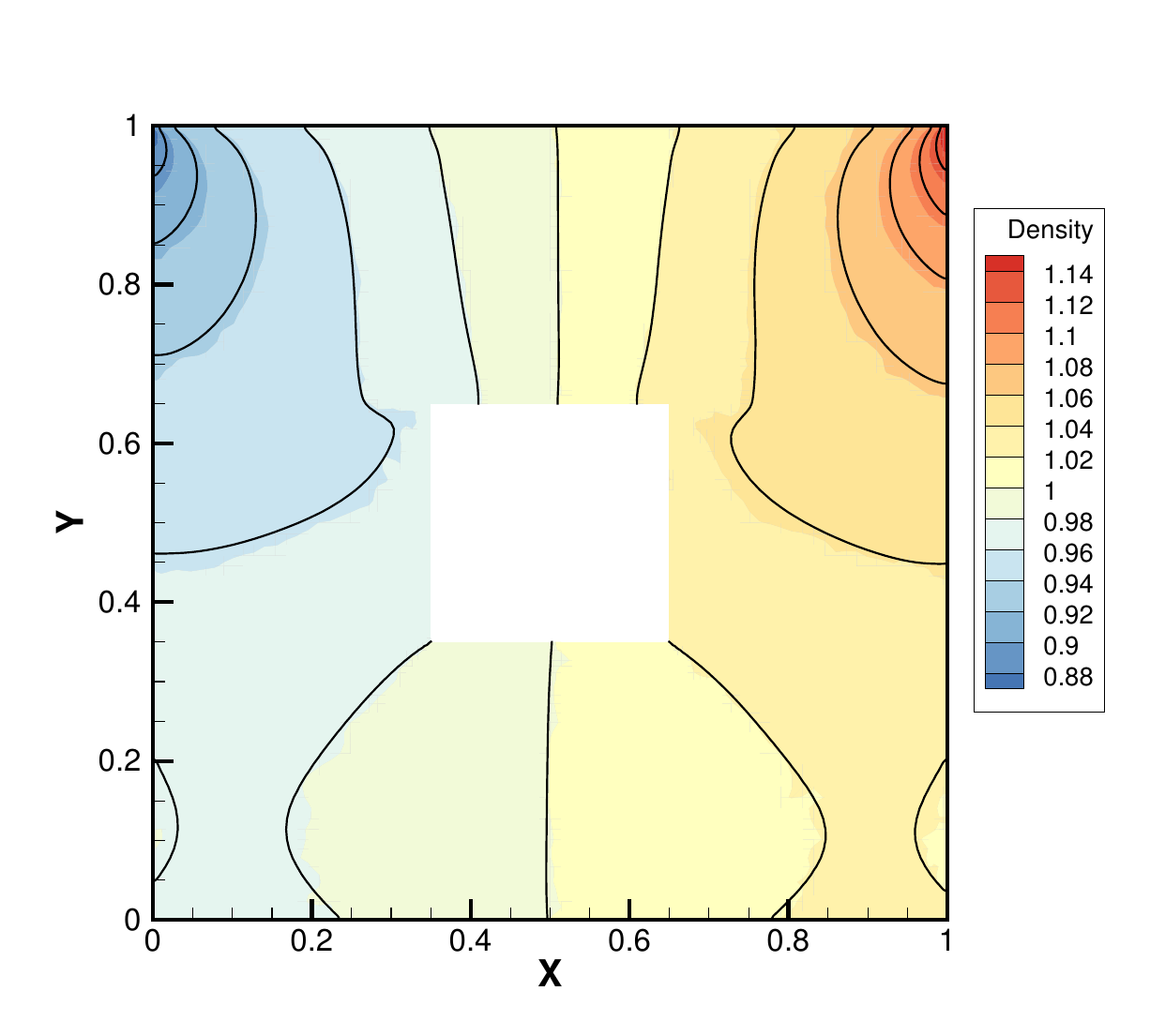}}}
    \subfigure[$\text{Kn}=0.05$]{
        {\includegraphics[width=0.45\textwidth,clip = true]{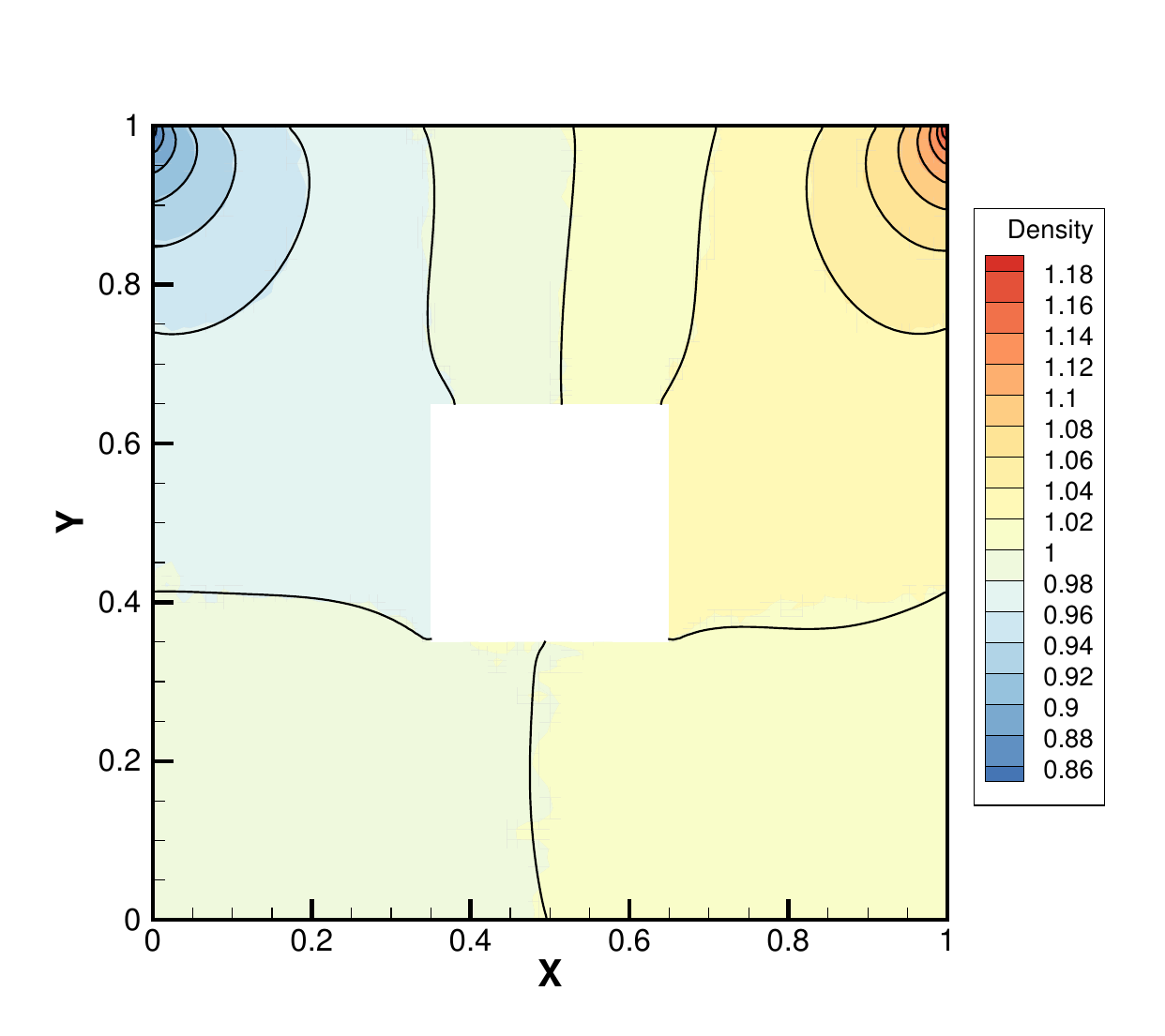}}}
    \subfigure[$\text{Kn}=0.005$]{
        {\includegraphics[width=0.45\textwidth,clip = true]{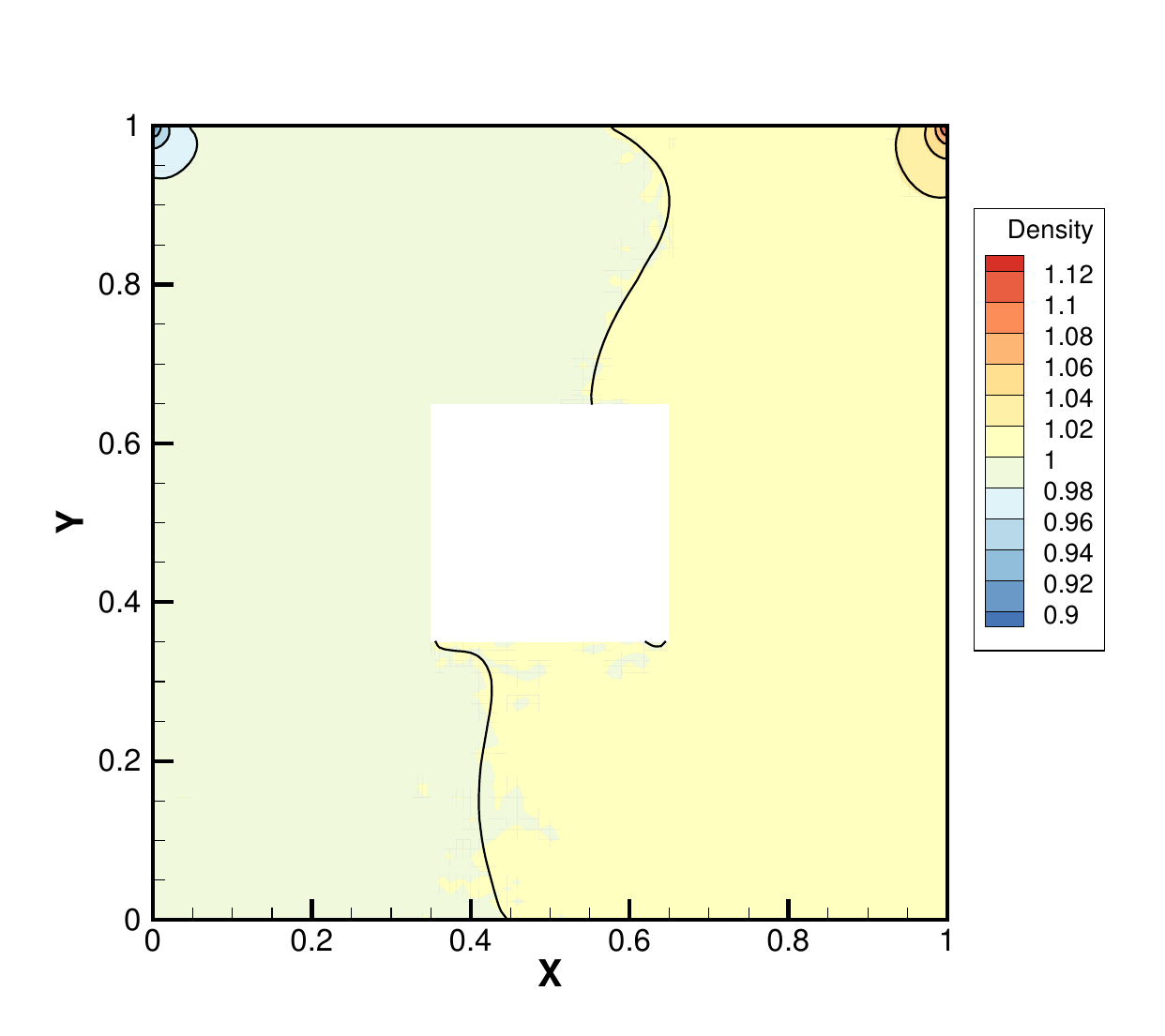}}}
    \caption{Comparison of density distribution in cavity under different Knudsen numbers. Background contour: DSMC; Black solid lines: GSIS.}
    \label{fig:2DCavityPorous_gsis_dsmc_rho_cmp}
\end{figure}

\subsection{2D lid-driven flow in a square cavity} \label{subsec:cavity}

In order to test the acceleration effect of GSIS in low speed flow, a 2D lid-driven cavity flow is tested. The grid structure is shown in Fig.~\ref{fig:2DCavityPorous_gsis_dsmc_rho_cmp}(a), with a square barrier of 0.3 length in the center. The square cavity lid moves forward to the X-axis at a speed of $50\ \text{m/s}$, and all wall temperatures are set to $T_w=273.15\ \text{K}$. The simulated gas is nitrogen, the initial temperature is $T_0=273.15\ \text{K}$, the Knudsen numbers are $\text{Kn}$ = 0.5, 0.05 and 0.005, where the side length of the square cavity is chosen as the reference length. Non-uniform structure spatial grid is used, where the thickness of the first layer mesh near the wall is $0.005$, and the total number of physical space grids is $9100$. The velocity space adopts the uniform orthogonal discretization, the range of the two velocity directions is $[-5\sqrt{RT_0},\ 5\sqrt{RT_0}] \times [-4\sqrt{RT_0},\ 4\sqrt{RT_0}]$, and the discrete number is set to $50 \times 40$.

\begin{figure}[!t]
    \centering
    \subfigure[$\text{Kn}=0.5$]{
        {\includegraphics[width=0.45\textwidth,clip = true]{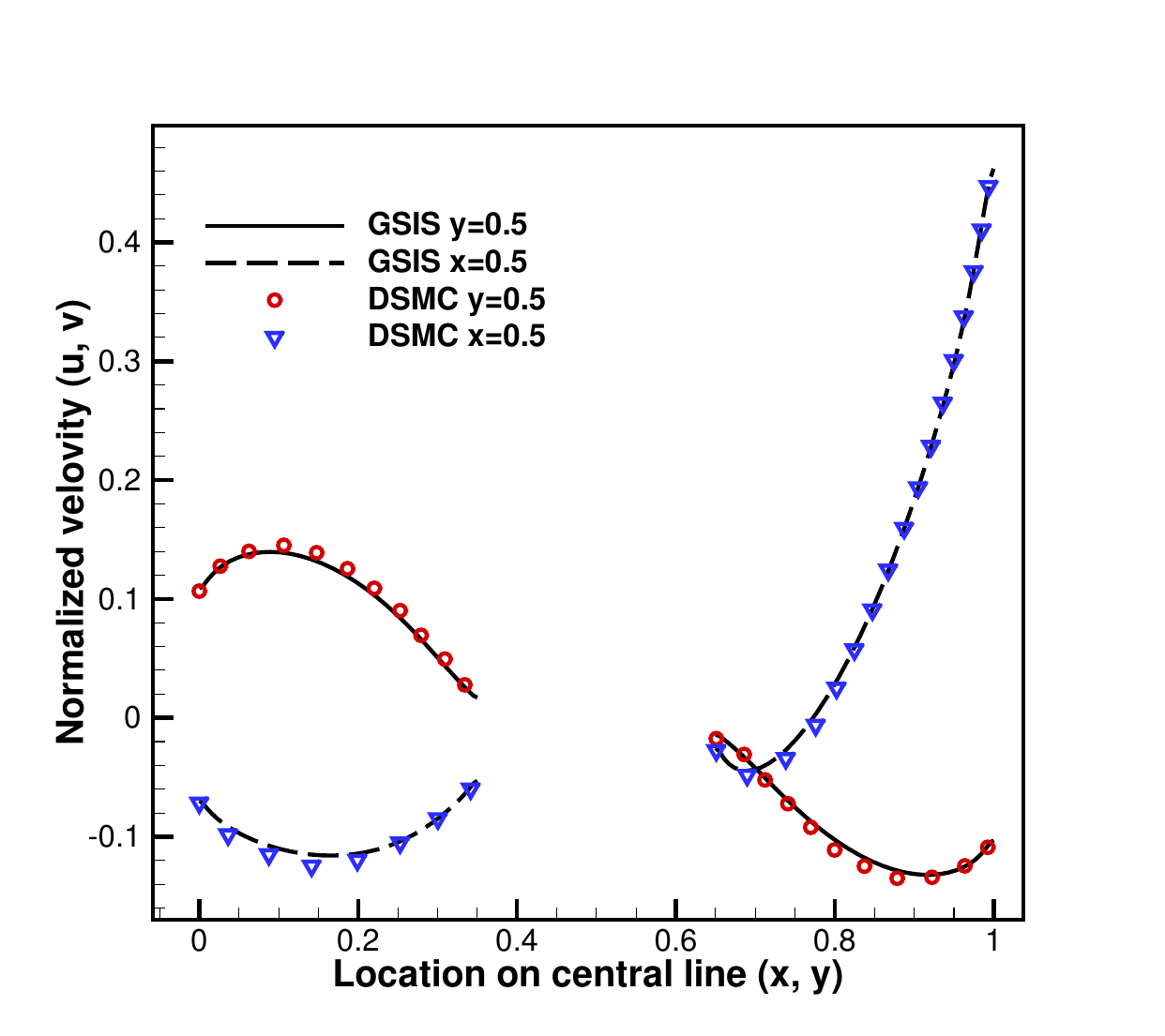}} }
    \subfigure[$\text{Kn}=0.05$]{
        {\includegraphics[width=0.45\textwidth,clip = true]{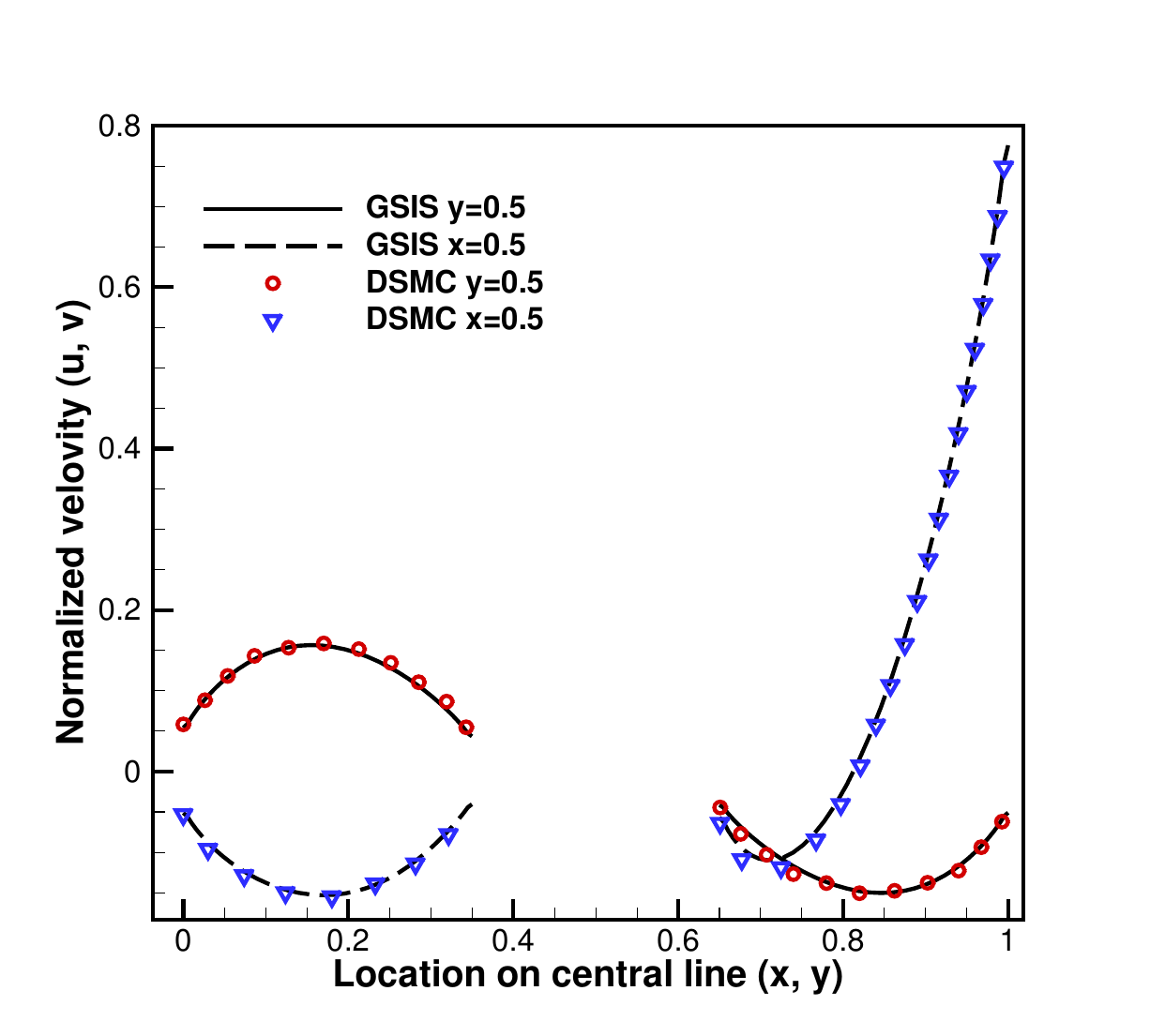}} }
    \subfigure[$\text{Kn}=0.005$]{
        {\includegraphics[width=0.45\textwidth,clip = true]{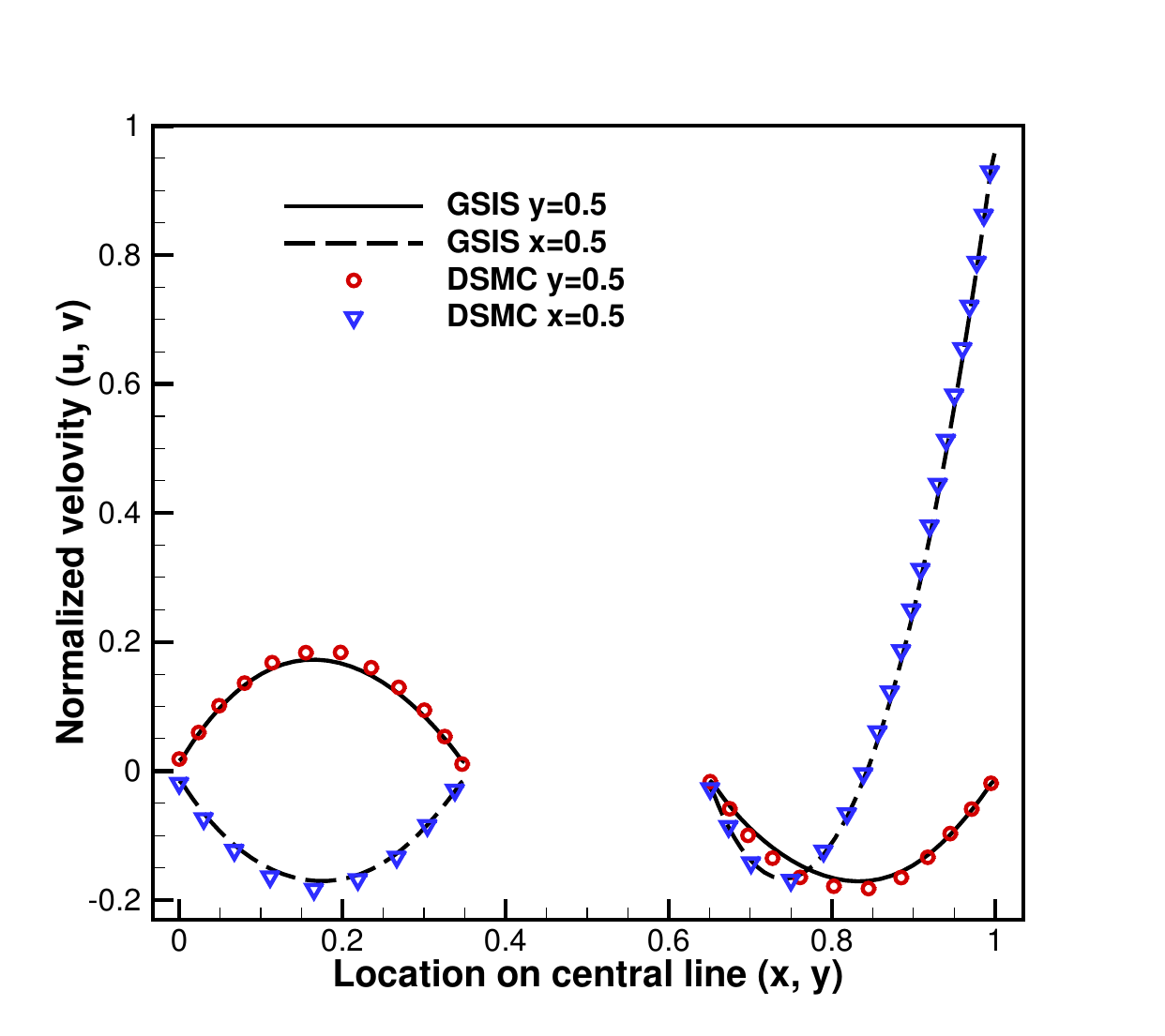}} }
    
    \caption{Comparison of velocity distribution along the central axis of the cavity under different Knudsen numbers.}
    \label{fig:2DCavityPorous_gsis_dsmc_u_cmp}
\end{figure}

Figure~\ref{fig:2DCavityPorous_gsis_dsmc_rho_cmp} compares the density between the GSIS and DSMC under different Knudsen numbers. Meanwhile, the velocity profiles on the central axis in the horizontal and vertical directions of the cavity under different Knudsen numbers are compared in Fig.~\ref{fig:2DCavityPorous_gsis_dsmc_u_cmp}. It can be seen that the calculated results of density and velocity agree well with each other. 

The simulation time of DSMC and the traditional discrete velocity method (DVM) are compared in Table~\ref{tab:2DCavity_gsis_Effency}. It can be found that the GSIS has a considerable advantage over the DSMC and traditional DVM methods in computing time in the low speed flow of small Knudsen numbers. Note that since DVM has no asymptotic preserving property, it cannot calculate the same result as DSMC and GSIS under this physical grid at $\text{Kn}=0.005$. It is necessary to refined the grid at the wall to get the correct result.

\begin{table}[!t]
	\centering
	\caption{The computational costs for the simulations of the lid-driven flow in a square cavity. The results of DSMC are calculated by the open source software SPARTA~\cite{plimpton2015sparta}  using 320 cores, while DVM and GSIS use 8 cores; CPU times = Cores $\times$ Wall times. }
	\begin{threeparttable}
		\begin{tabular}{ccccccccc}\hline
			\multirow{2}{*}{Kn} & ~ & {DSMC}   & ~ &\multicolumn{2}{c}{DVM} & ~ & \multicolumn{2}{c}{GSIS}  \\
			\cline{3-3}  \cline{5-6} \cline{8-9}
			~       & ~ & CPU times (h)   & ~ &  Steps & CPU times (h)   & ~  & Steps & CPU times (h) \\ \hline
			0.5     & ~ & 85       & ~ & 133     & 0.18     & ~                     & 75 & 0.22  \\ 
			0.05    & ~ & 135      & ~ & 370     & 0.44     & ~                     & 61 & 0.18  \\ 
			0.005   & ~ & 2491     & ~ & 3660    & 3.95     & ~                     & 19 & 0.07  \\ \hline
		\end{tabular}
	\end{threeparttable}
	\label{tab:2DCavity_gsis_Effency}
\end{table}

\subsection{Hypersonic flow passing Apollo}\label{subsec:Apollo}


A hypersonic flow passing Apollo at $\text{Ma}=5$ and $\text{AoA}=30^\circ$ is simulated for $\text{Kn}=0.0012, 0.01, 0.1, 1$, which are defined in terms of the reference length $L_0=3.912$ m and temperature $T_0=T_{\infty}=142.2$ K with $T_{\infty}$ being the free stream temperature. The isothermal surface with $T_w=300$ K and fully diffuse gas-wall interaction is adopted. The simulation configuration has been presented in Section~\ref{sec:parallel}, and the spatial domain is discretized by 372,500 hexahedral cells as shown in Fig.~\ref{fig:Apollo_gridConfiguration}.

The velocity domain is truncated to a sphere with radius $12.2\sqrt{RT_0}$, centered at \\($0.866\sqrt{RT_0}$, $0.5\sqrt{RT_0}$, 0). Structured meshes with refinement around the stagnation and free stream velocity points are used, with centers of (0, 0, 0) and ($5.12\sqrt{RT_0}$, $2.96\sqrt{RT_0}$, 0) and lengths of $5\sqrt{RT_0}$ and $3.6\sqrt{RT_0}$, respectively. The rest of the velocity space is partitioned into 8,166 tetrahedral and hexahedral cells, see Fig.~\ref{fig:3D_Ma5_AoA30_X_hybrid_Volume}. Fig.~\ref{fig:3D_Ma5_AoA30_X_unstructure_Volume} shows that the velocity domain is truncated to a sphere centered at (0, 0, 0) with radius $20\sqrt{RT_0}$. In the same way, unstructured meshes with refinement around the stagnation and free stream velocity points are used, which result in 27,704 tetrahedral cells in the velocity domain discretization.
In this paper, the velocity space discretization in Fig.~\ref{fig:3D_Ma5_AoA30_X_hybrid_Volume} is adopted, and the calculated results are consistent with those in Fig.~\ref{fig:3D_Ma5_AoA30_X_unstructure_Volume}. The computational resources required in the simulations include $N_x\times N_v=128\times 1$ cores and 487 GB RAM.

\begin{figure}[p]
	    \centering
	    {\includegraphics[scale=0.25,clip = true]{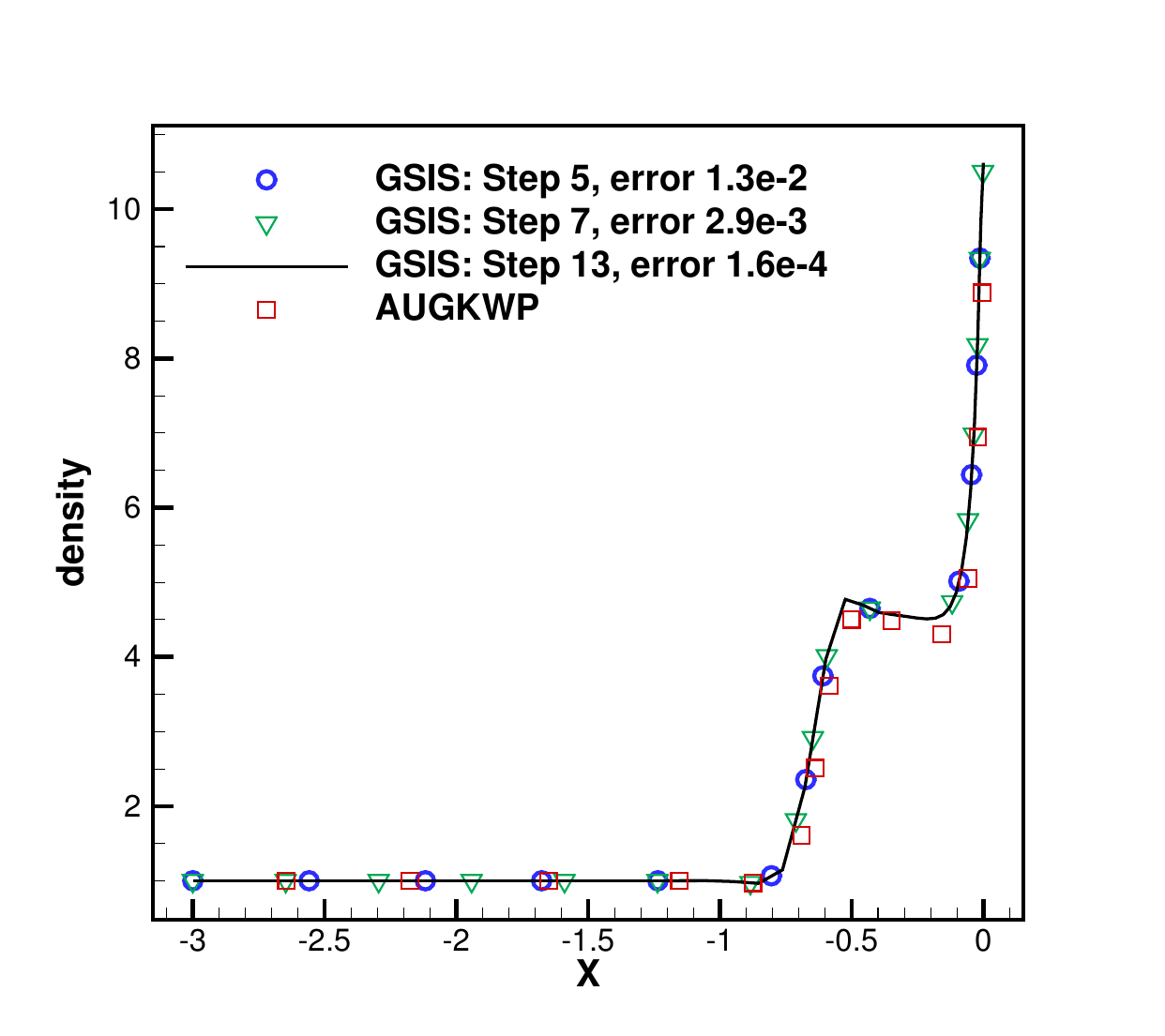}}
	    {\includegraphics[scale=0.25,clip = true]{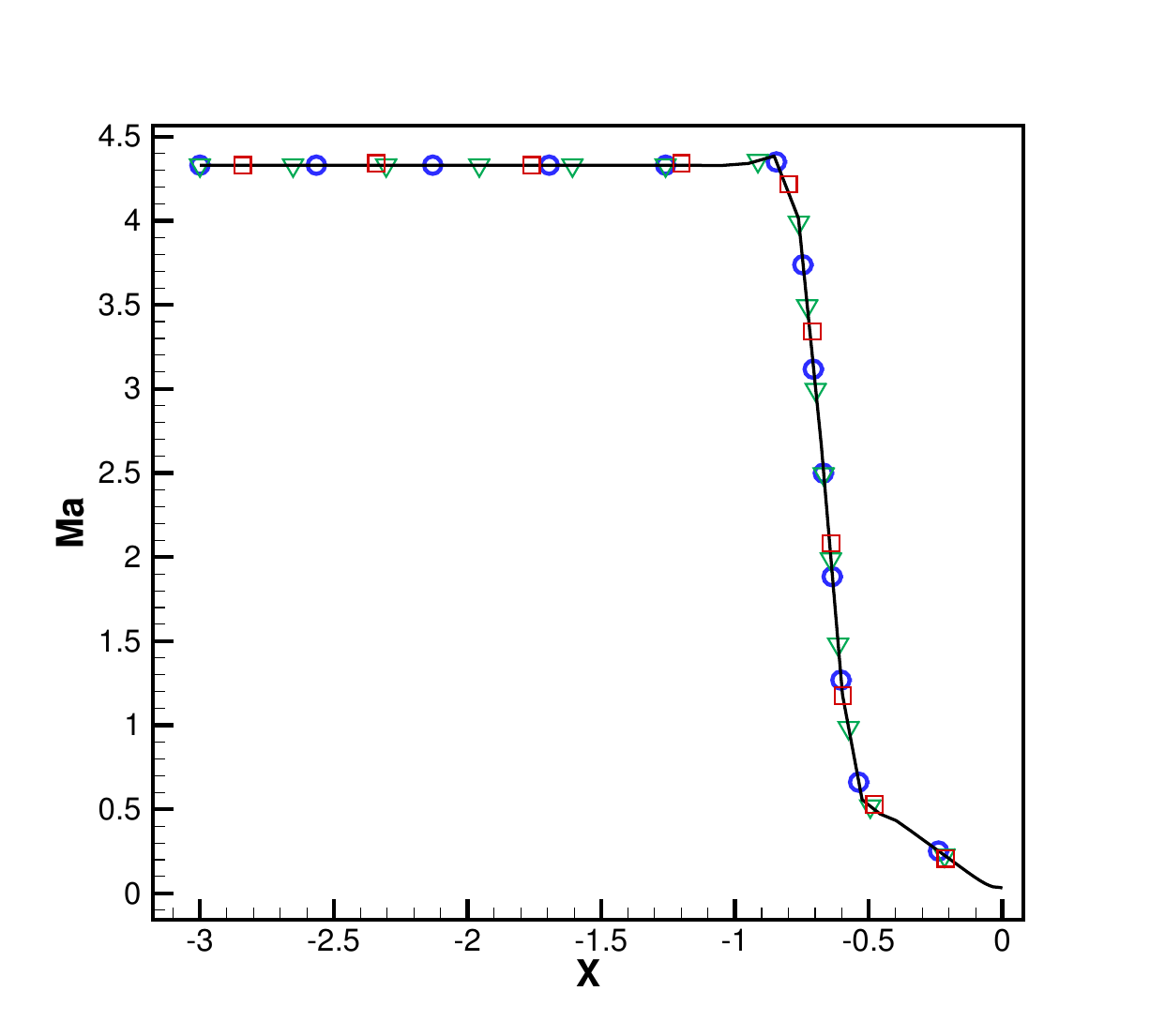}}
	    {\includegraphics[scale=0.25,clip = true]{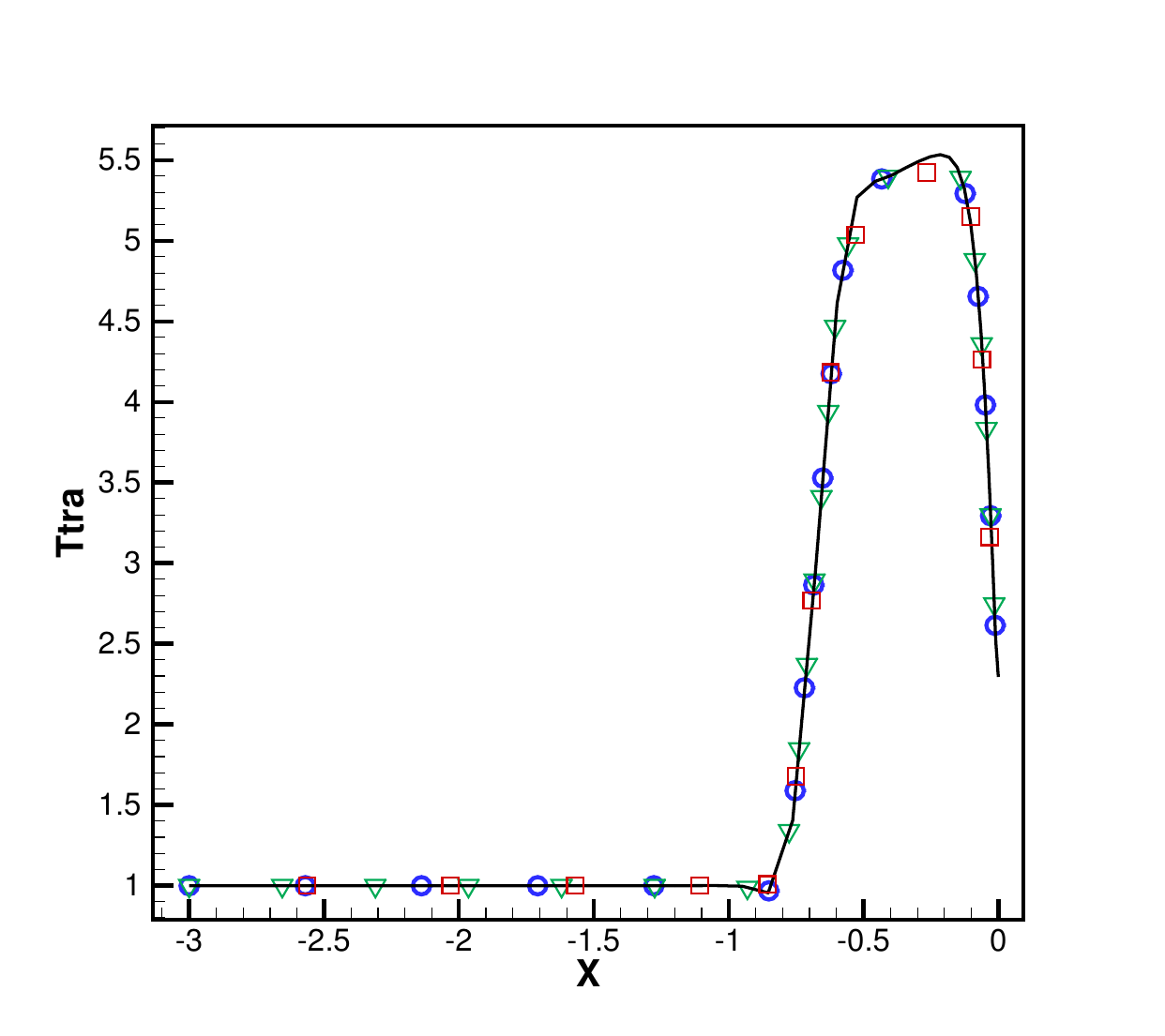}} \\
        {\includegraphics[scale=0.25,clip = true]{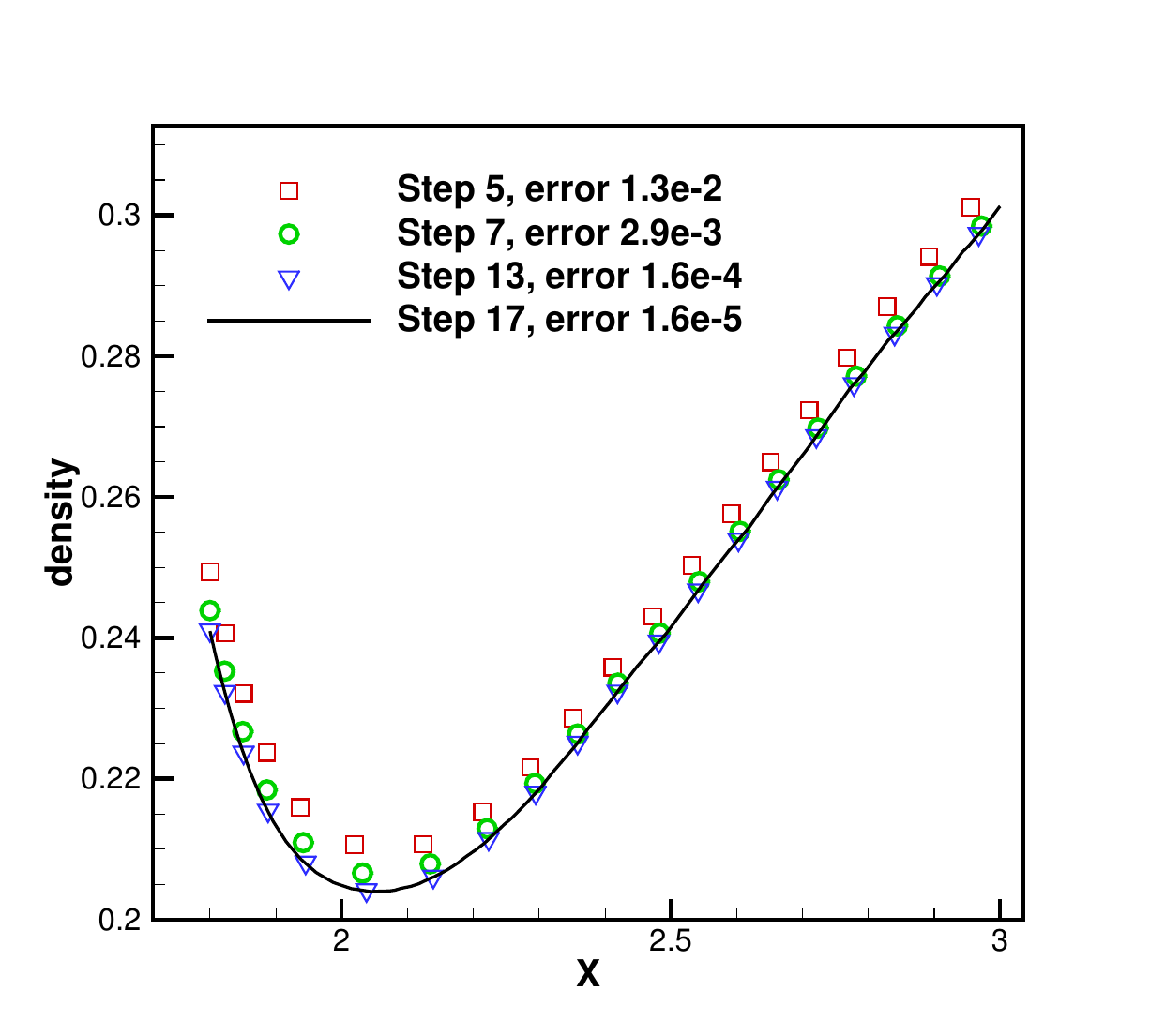}}    
        {\includegraphics[scale=0.25,clip = true]{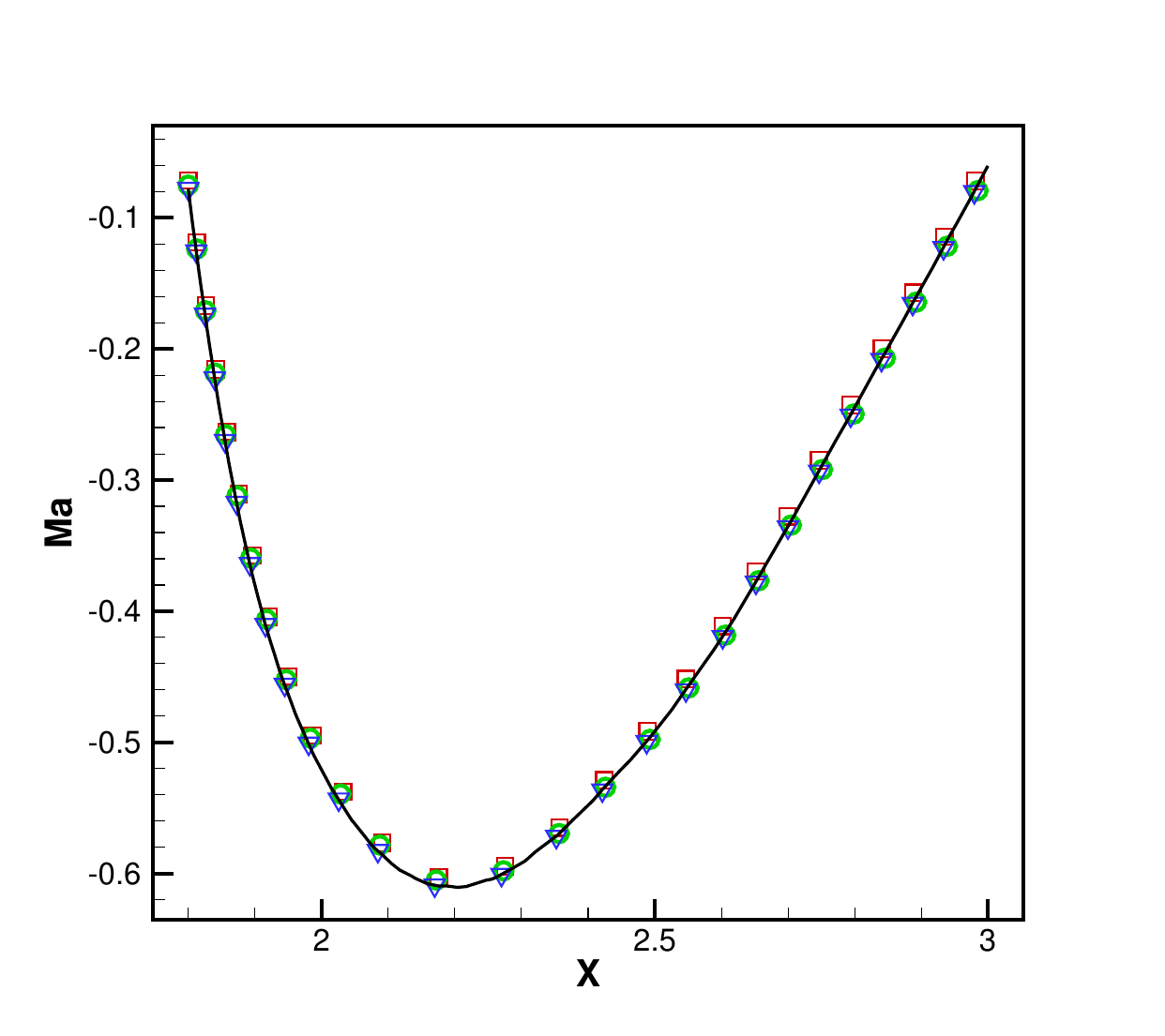}}   
        {\includegraphics[scale=0.25,clip = true]{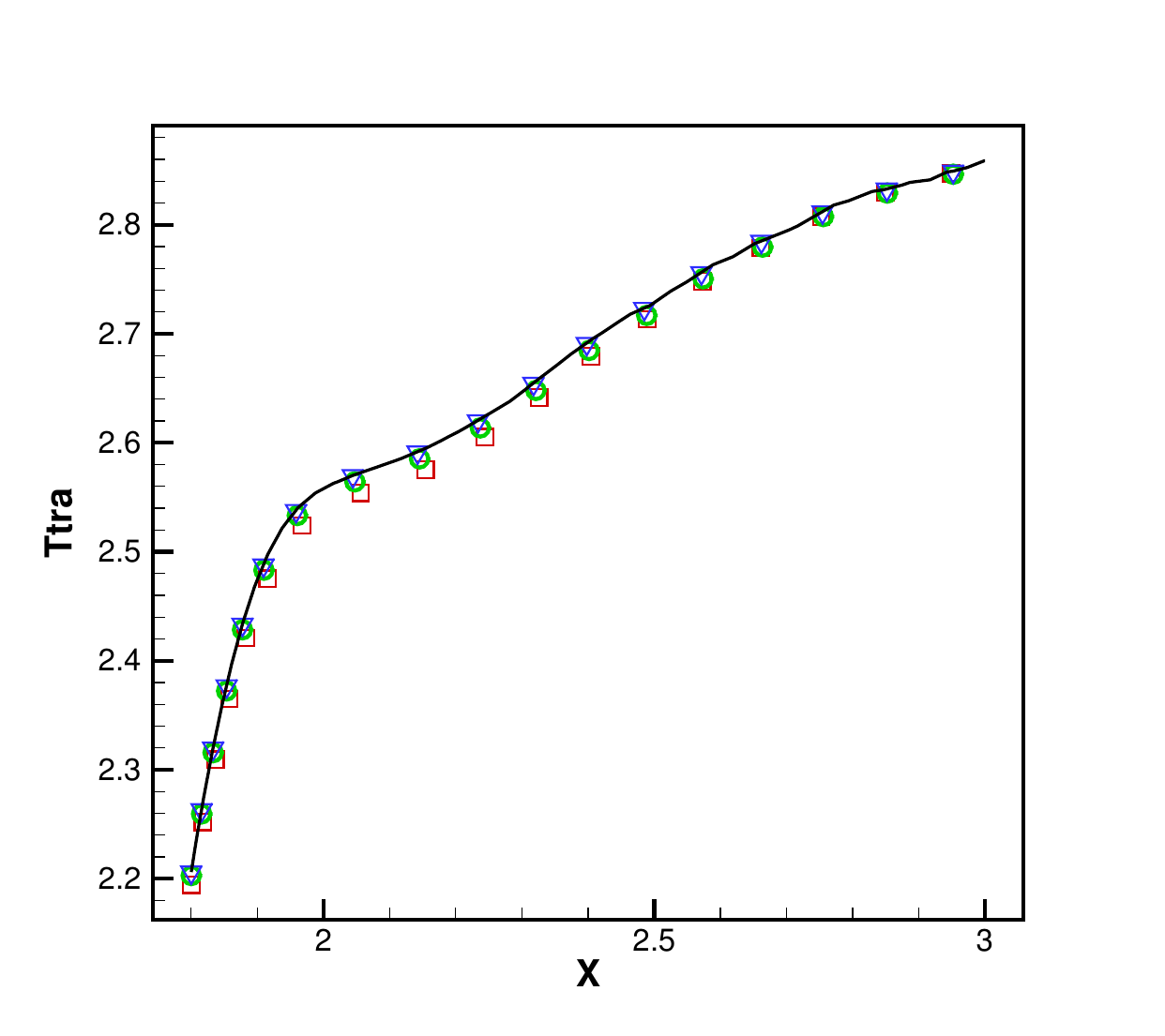}} \\
        {\includegraphics[scale=0.25,clip = true]{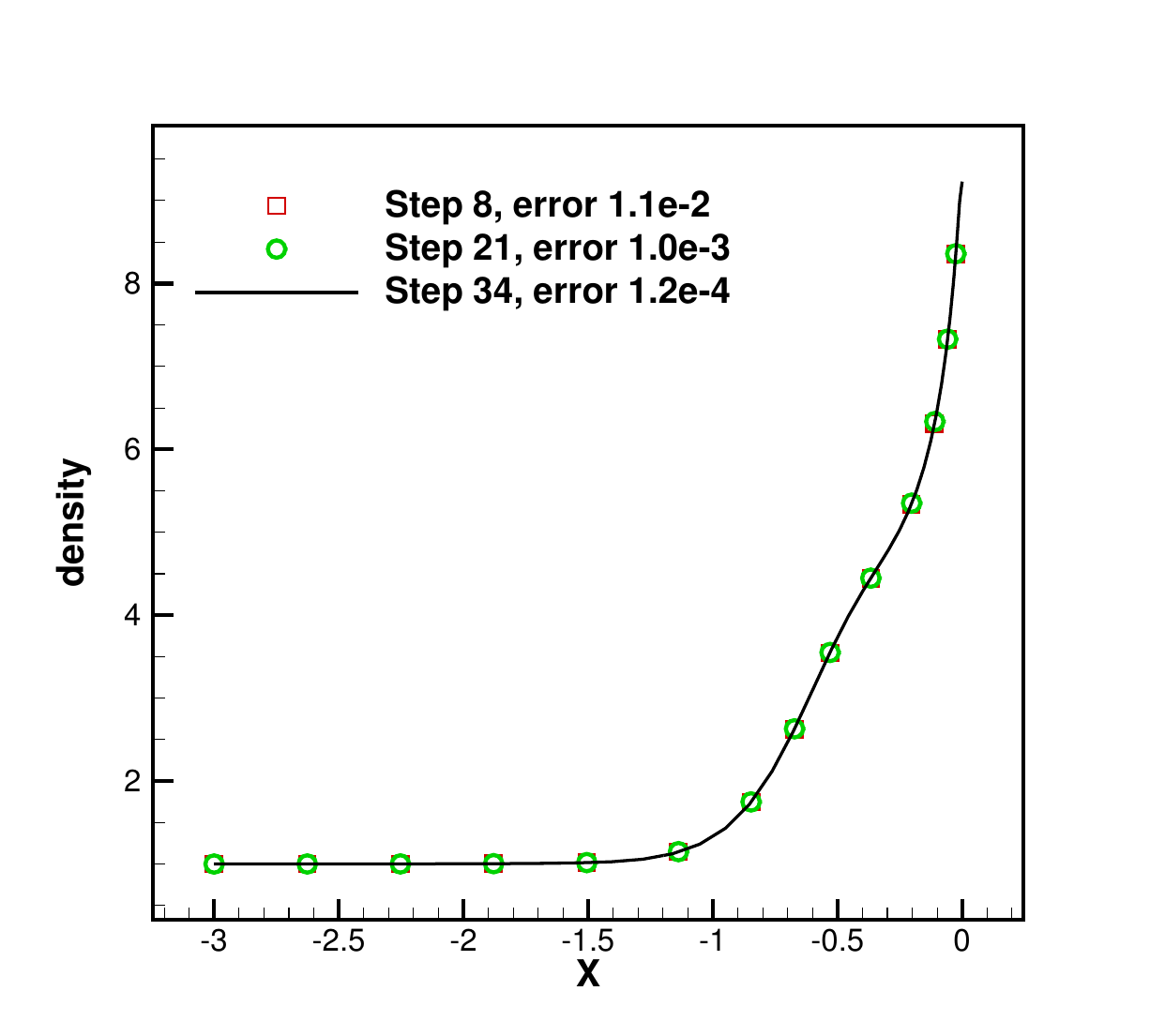}}
        {\includegraphics[scale=0.25,clip = true]{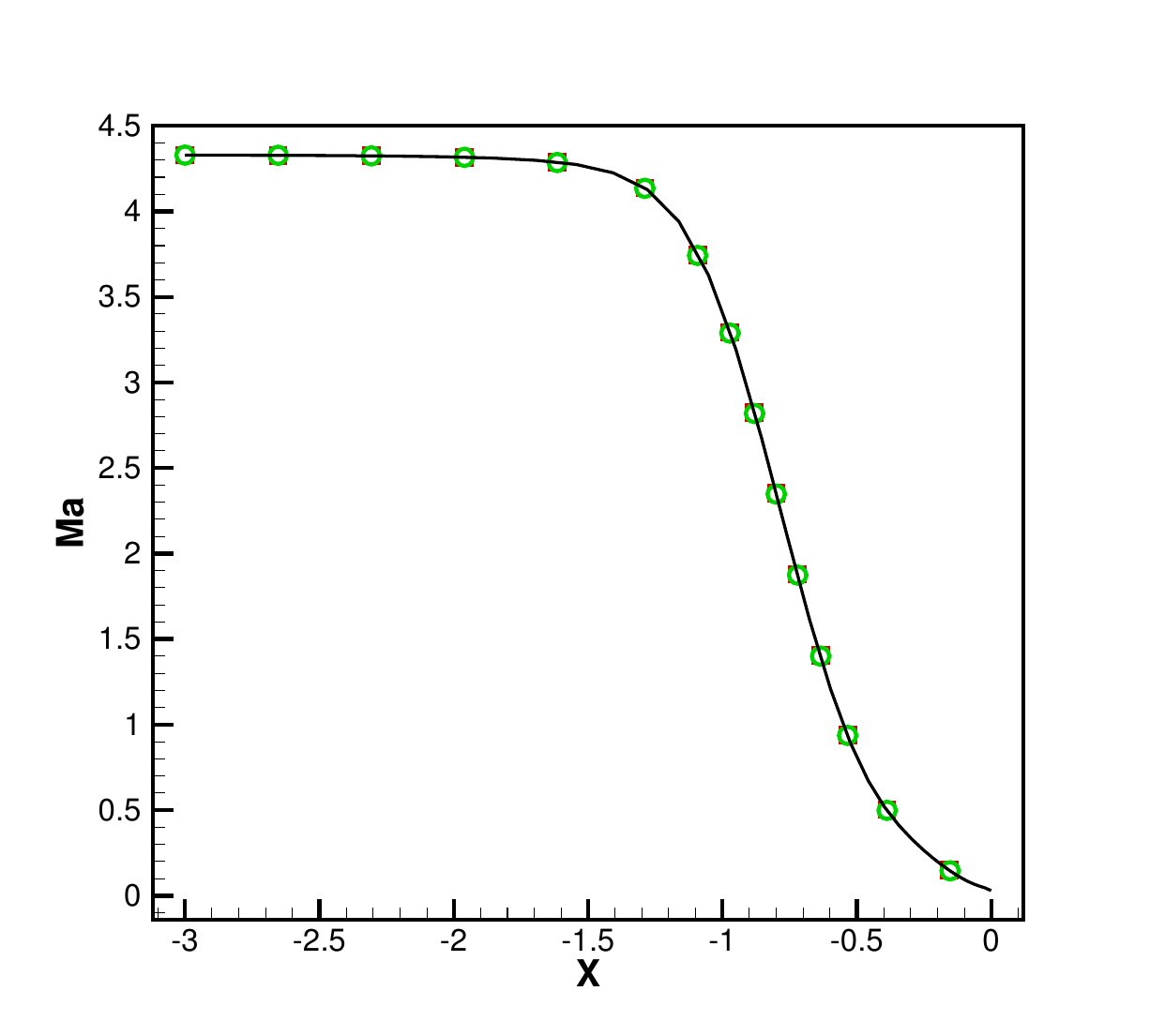}}
        {\includegraphics[scale=0.25,clip = true]{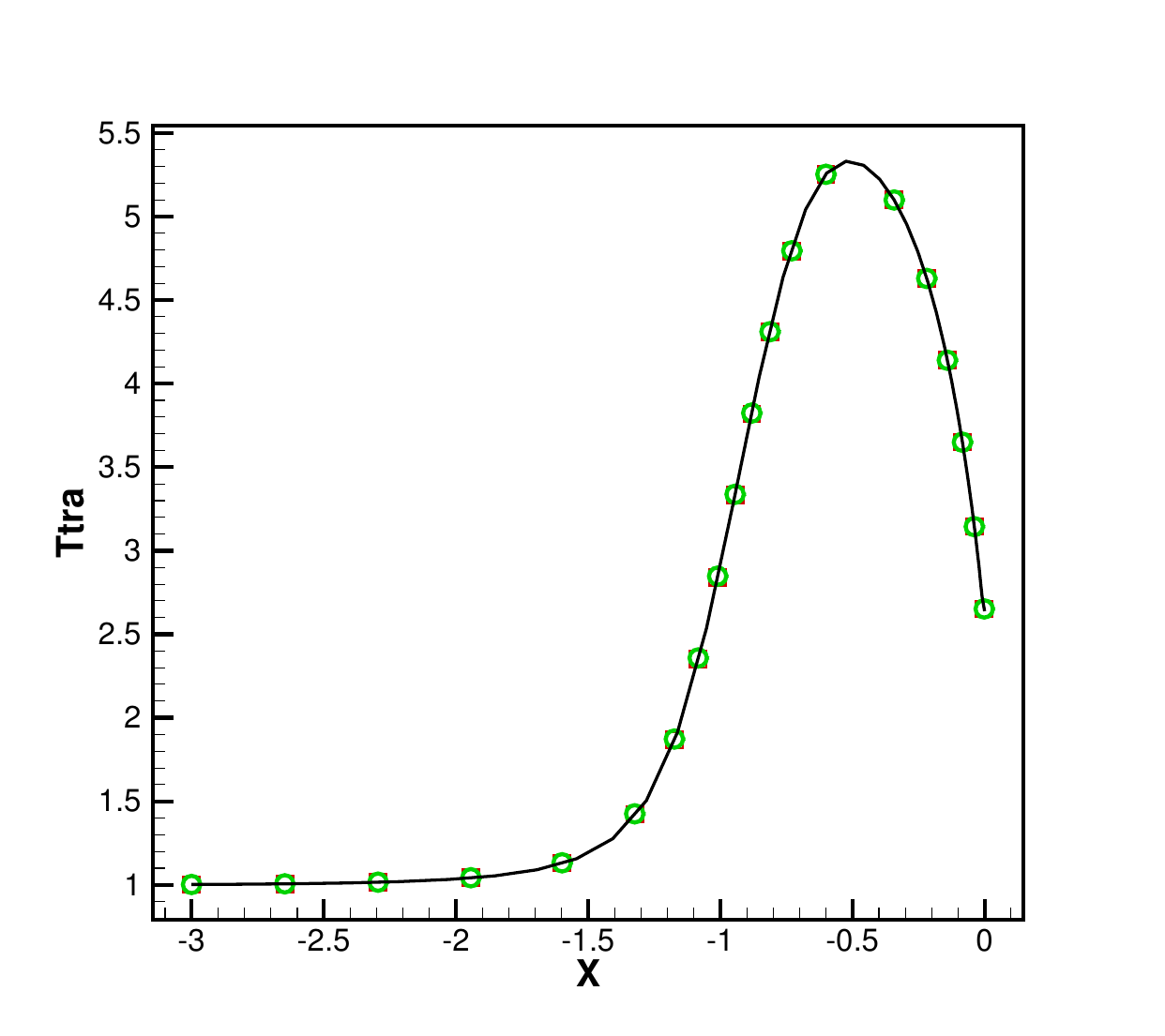}} \\
        {\includegraphics[scale=0.25,clip = true]{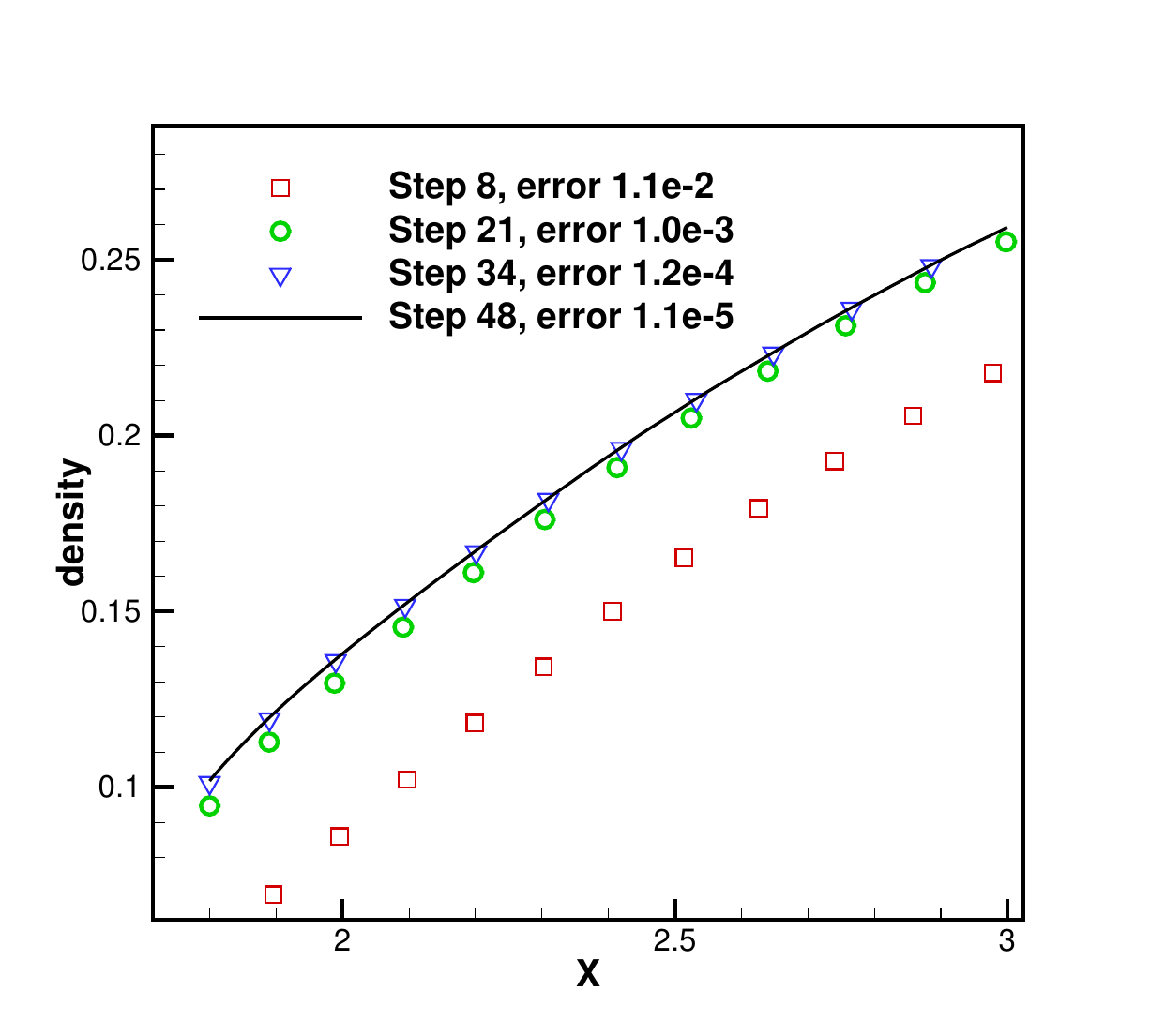}}    
        {\includegraphics[scale=0.25,clip = true]{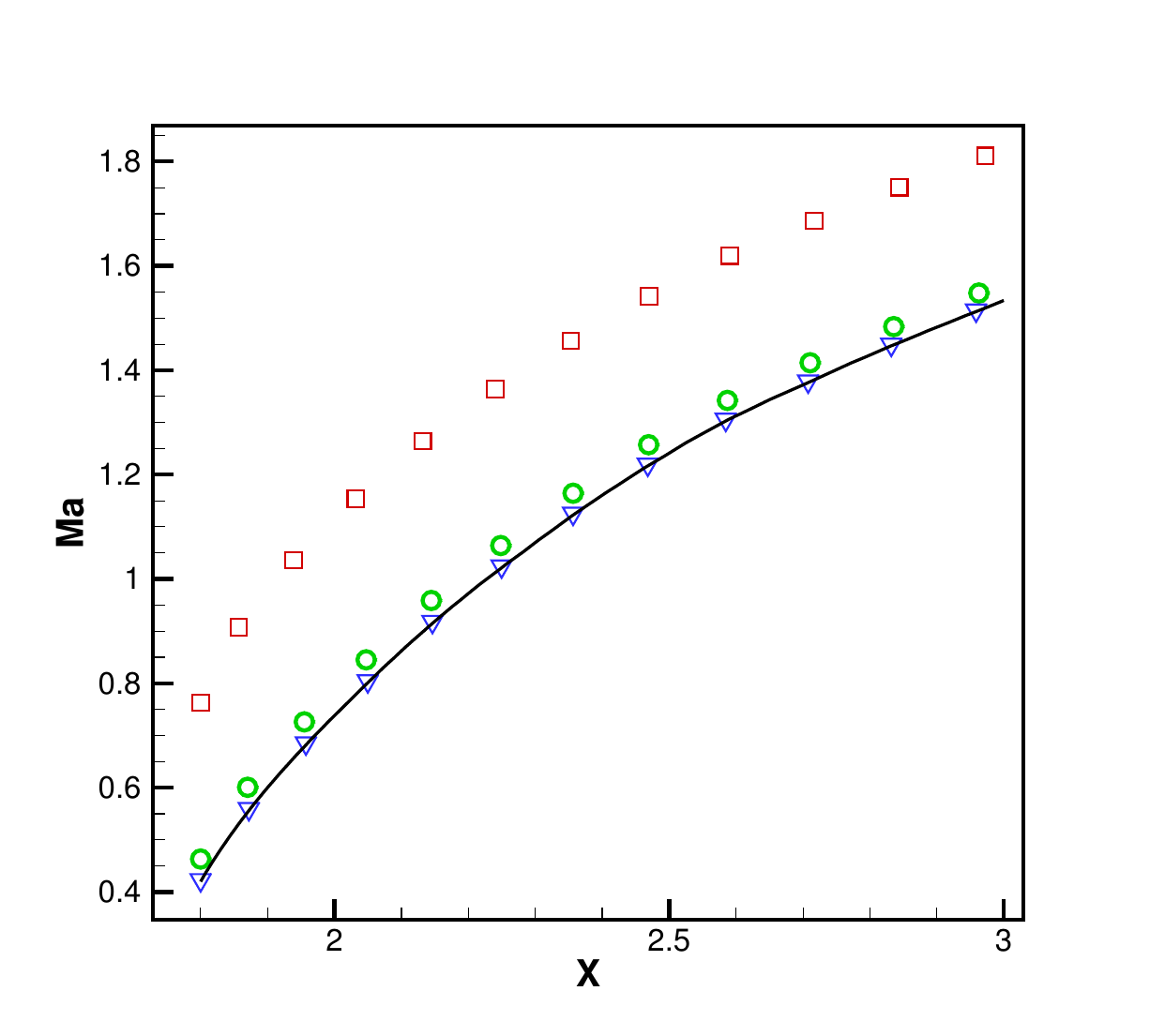}}   
        {\includegraphics[scale=0.25,clip = true]{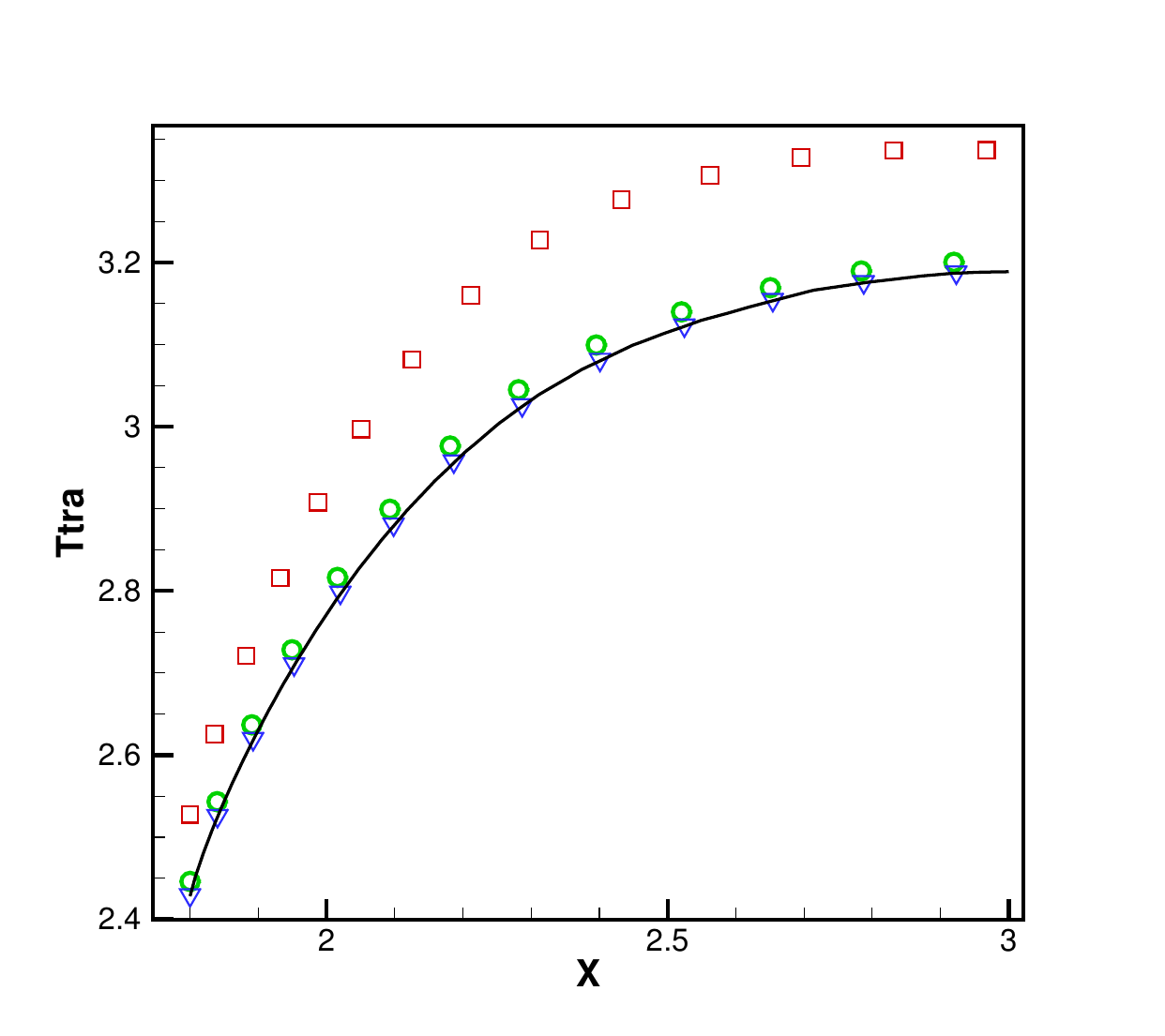}}
     \caption{
            Comparison of the dimensionless macroscopic quantities along the symmetry axis (z-axis) obtained under different convergence criteria, when $\text{Ma}=5$. (1st and 3rd rows) windward region from (-3.0, 0, 0) to (0, 0, 0) and (2nd and 4th rows) leeward region from  (1.8, 1.2, 0) to (3.0, 1.8, 0). The Knudsen number in the first and last two rows are $\text{Kn}=0.0012$ and 0.1, respectively. 
         }
     \label{fig:Apollo_Ma5_Kn00012_ErrorEvolution}
 \end{figure}

We first analyze how the convergence criteria~\eqref{convergence_critertion} affect the final solution. In the previous paper~\cite{su2020fast}, we have rigorously analyzed that the GSIS has the fast-convergence property, which means that the solution converges even when $\epsilon$ in Eq.~\eqref{convergence_critertion} is large. This is indeed supported in the dimensionless macroscopic quantities along the symmetry axis in Fig.~\ref{fig:Apollo_Ma5_Kn00012_ErrorEvolution}. In the near-continuum flow regime ($\text{Kn}=0.0012$), in the windward region, it is seen that the solutions of density, flow velocity and translational temperature converge, when the maximum relative error is as low as $\varepsilon=10^{-2}$. In the leeward region, the solutions converge even when the maximum relative error is around $\varepsilon=10^{-3} \sim 10^{-4}$, which corresponds to only 13 iteration steps. In the transition regime ($\text{Kn}=0.1$),  the solutions can be seen converged after 34 iterations, when the maximum relative error is again around $\varepsilon=10^{-4}$.

\begin{table}[!t]
	\centering
	\caption{The computational costs for the simulations of the hypersonic flow passing Apollo ($\text{Ma}=5$) at different Knudsen numbers. The same spatial grid cells are adopted in both GSIS and AUGKWP. The total wall clock time in GSIS including those spent on initial flow field calculation (1000 steps of macroscopic Euler solver followed by 10 steps of kinetic solver) and GSIS iterations.}
	\begin{threeparttable}
		\begin{tabular}{cccccccc}\hline
			\multirow{2}{*}{Kn}   & ~ &\multicolumn{2}{c}{AUGKWP~\cite{wei2023}} & ~ & \multicolumn{3}{c}{GSIS}  \\
			\cline{3-4}  \cline{6-8}
			~       & ~ & Cores & Wall time (h)   & ~  & Cores & Steps & Wall times (h) \\ \hline
			1       & ~ & -     & -     & ~ & \multirow{4}{*}{128} & 50 & 0.48  \\
			0.1     & ~ & -     & -     & ~ & ~                    & 66 & 0.55  \\ 
			0.01    & ~ & -     & -     & ~ & ~                    & 47 & 0.36  \\ 
			0.0012  & ~ & 120   & 6.82  & ~ & ~                    & 22 & 0.24  \\ \hline
		\end{tabular}
	\end{threeparttable}
	\label{tab:Apollo_gsis_Effency}
\end{table}

\begin{figure}[!t]
    \centering
    \subfigure[]{
        \label{gsis_Ma5_Kn01_density}
        {\includegraphics[scale=0.3,clip = true]{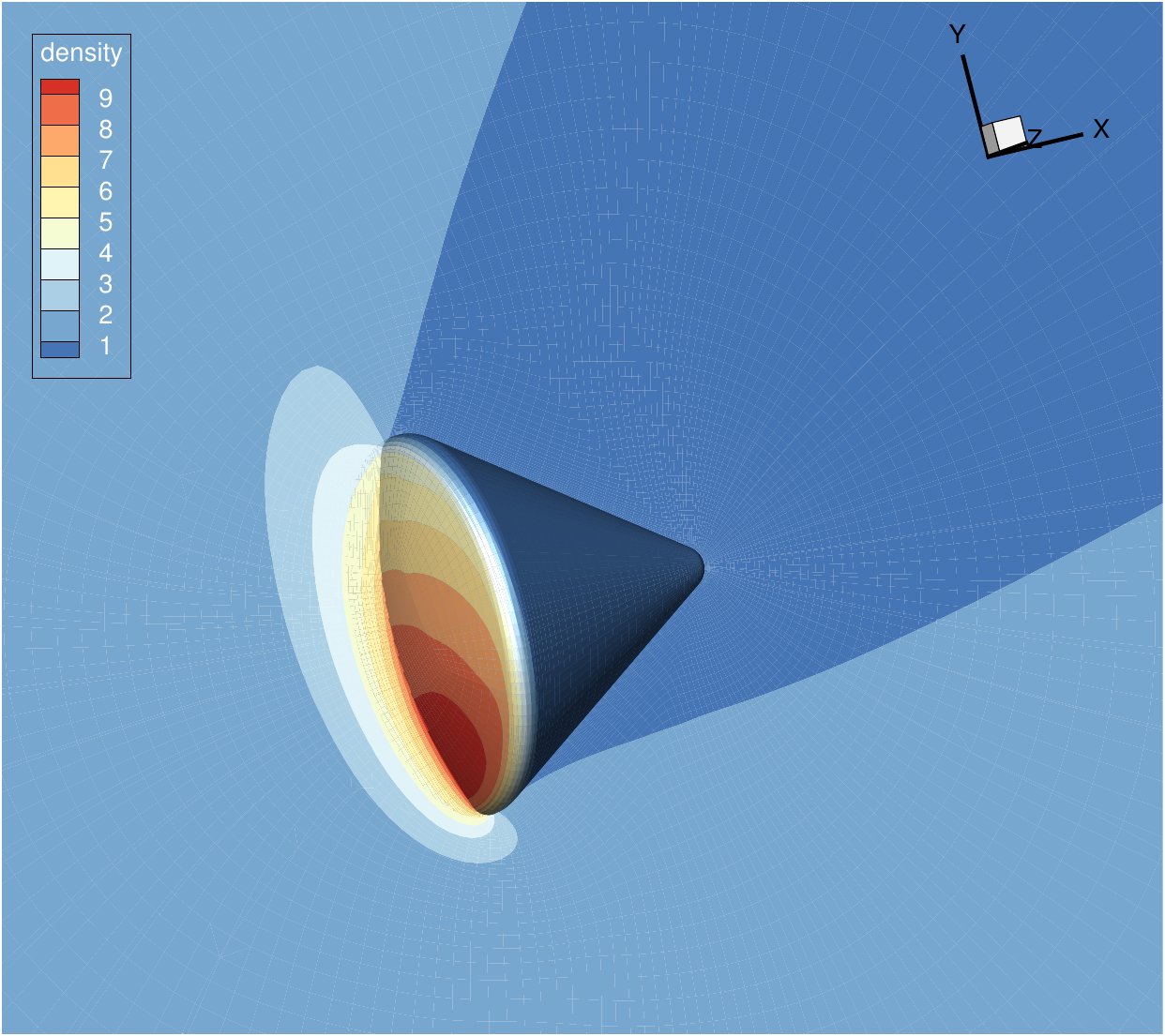}}}
    \subfigure[]{
        \label{gsis_Ma5_Kn01_Ma}
        {\includegraphics[scale=0.3,clip = true]{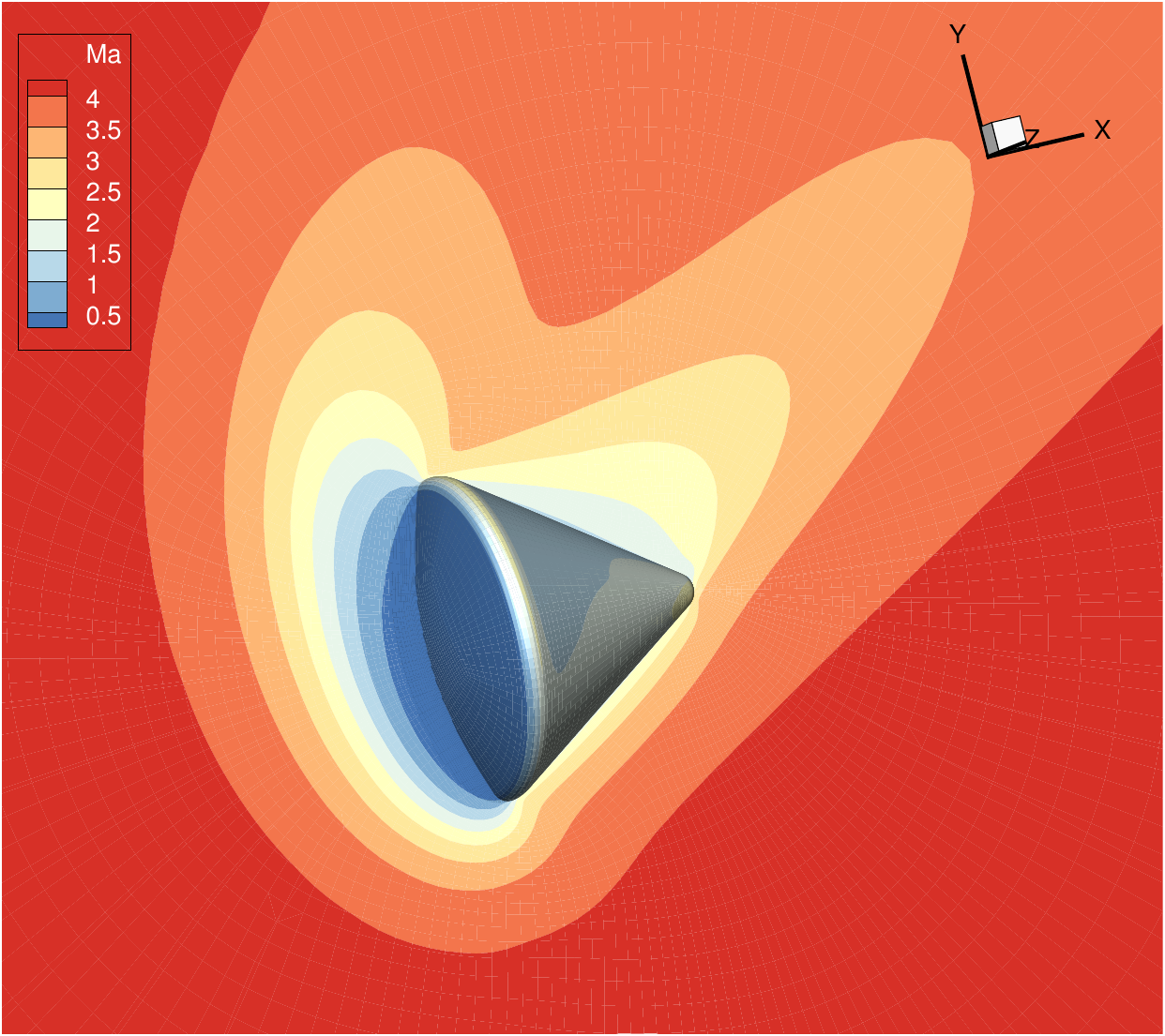}}}
    \\
    \subfigure[]{
        \label{gsis_Ma5_Kn01_Ttra}
        {\includegraphics[scale=0.3,clip = true]{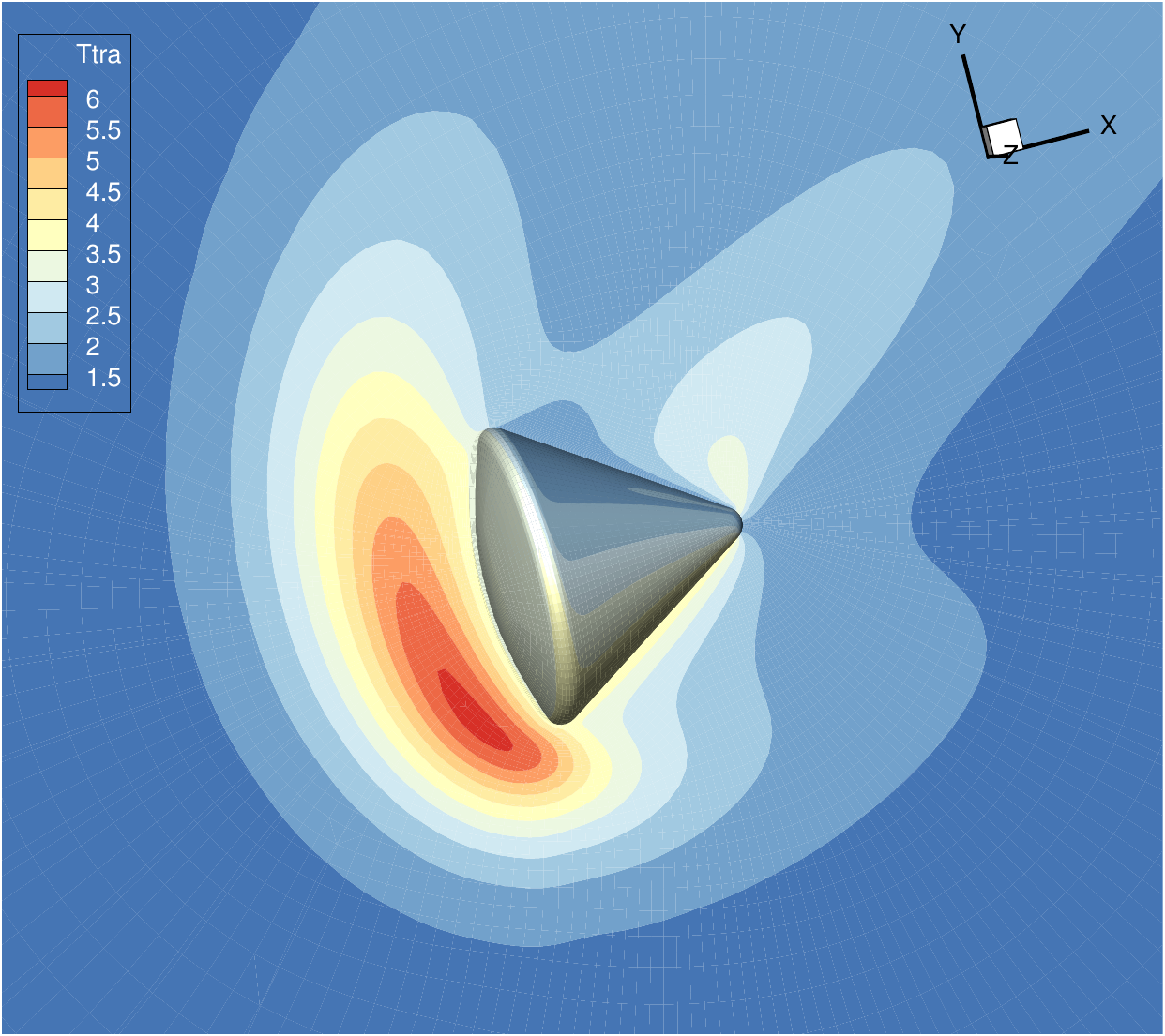}}}
    \subfigure[]{
        \label{gsis_Ma5_Kn01_Trot}
        {\includegraphics[scale=0.3,clip = true]{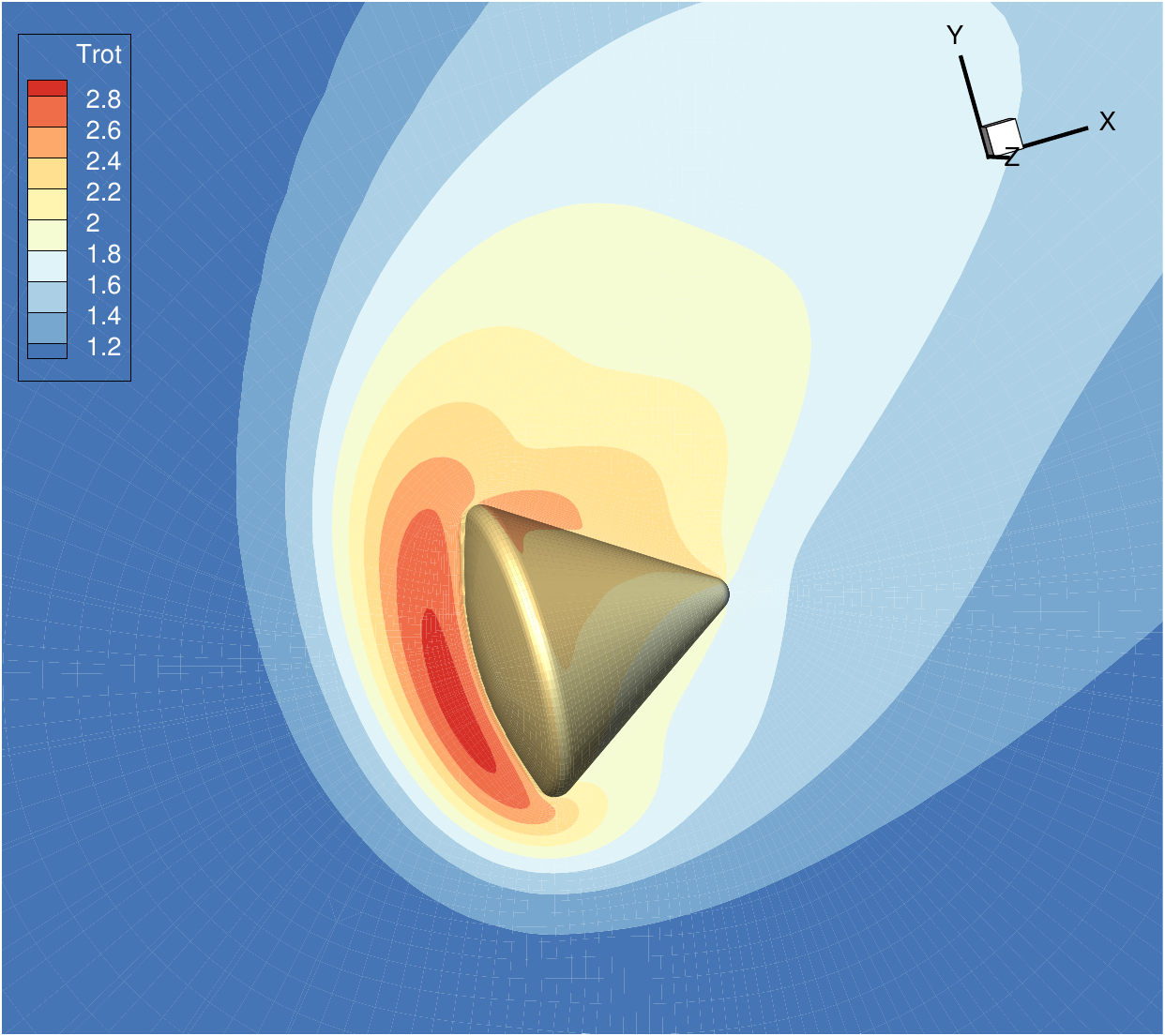}}}
    \caption{The distributions of dimensionless (a) density, (b) local Mach number, (c) translational and (d) rotational temperatures calculated by GSIS solver for the hypersonic flow passing Apollo, when $\text{Kn}=1$ and $\text{Ma}=5$, and $\text{AoA}=30^\circ$.}
    \label{fig:gsis_Ma5_Kn01_macro}
\end{figure}

\begin{figure}[H]
    \centering
    \subfigure[]{
        \label{gsis_Ma5_Kn00012_density}
        {\includegraphics[scale=0.3,clip = true]{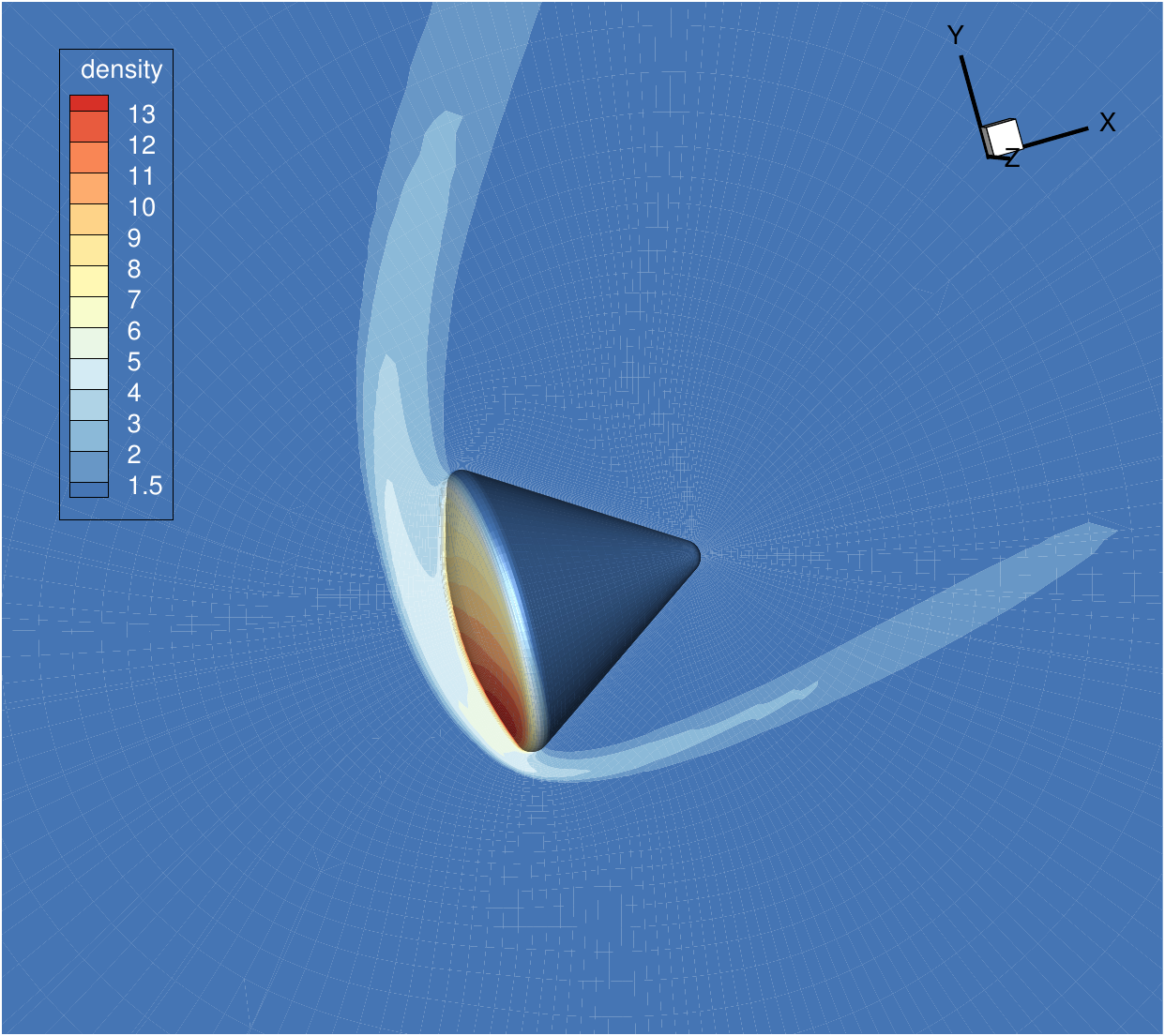}}}
    \subfigure[]{
        \label{gsis_Ma5_Kn00012_Ma}
        {\includegraphics[scale=0.3,clip = true]{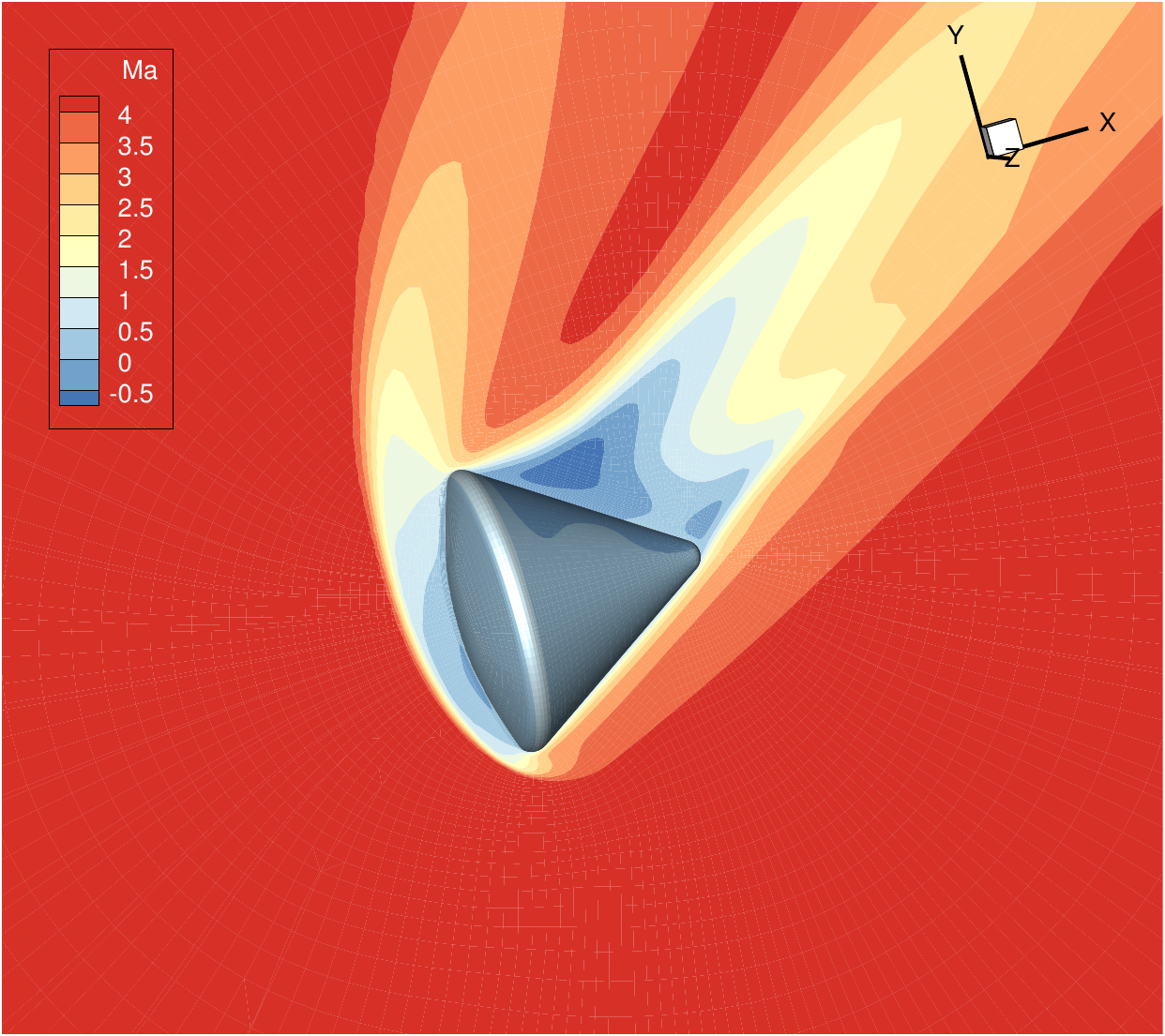}}}
    \\
    \subfigure[]{
        \label{gsis_Ma5_Kn00012_Ttra}
        {\includegraphics[scale=0.3,clip = true]{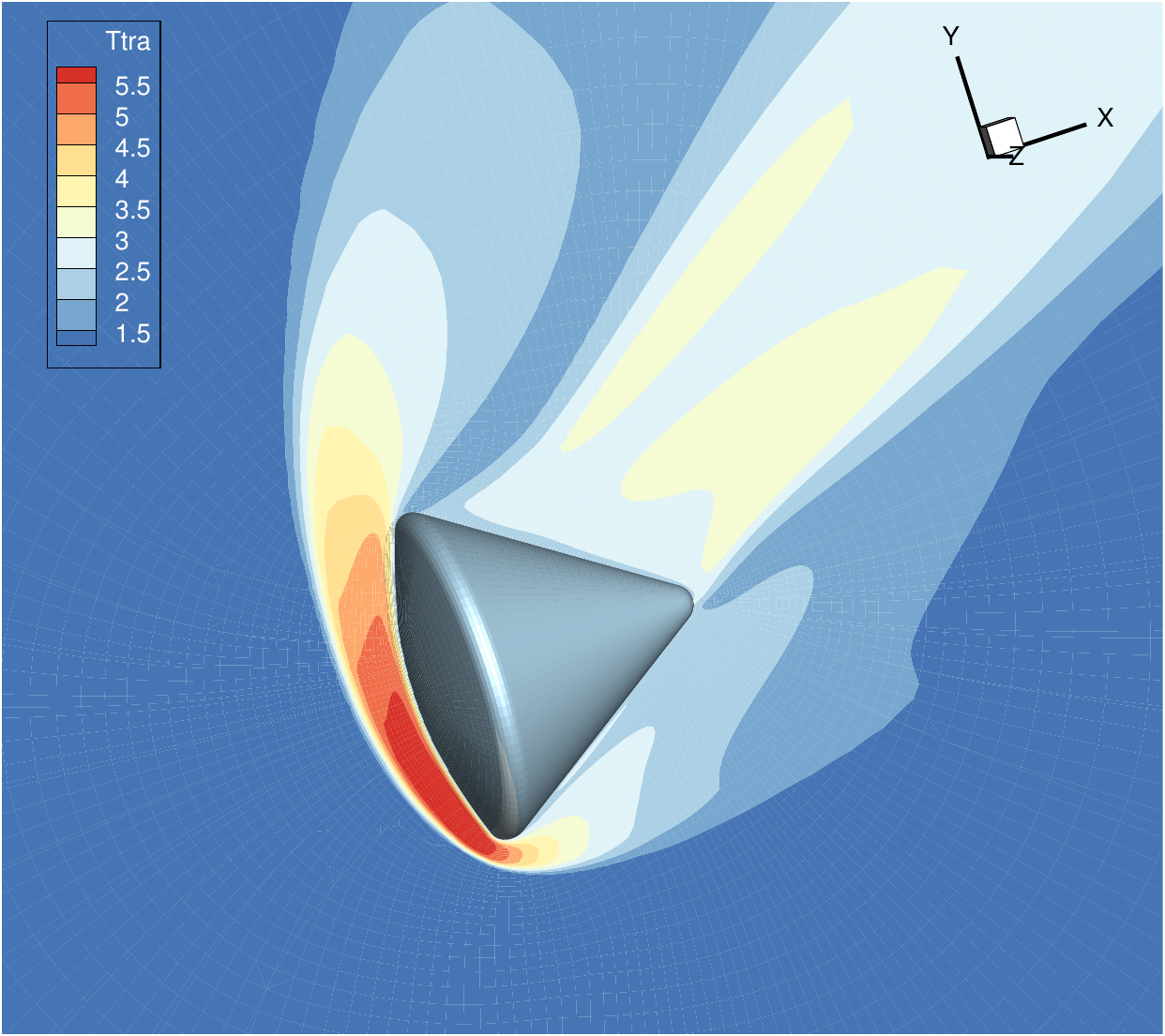}}}
    \subfigure[]{
        \label{gsis_Ma5_Kn00012_Trot}
        {\includegraphics[scale=0.3,clip = true]{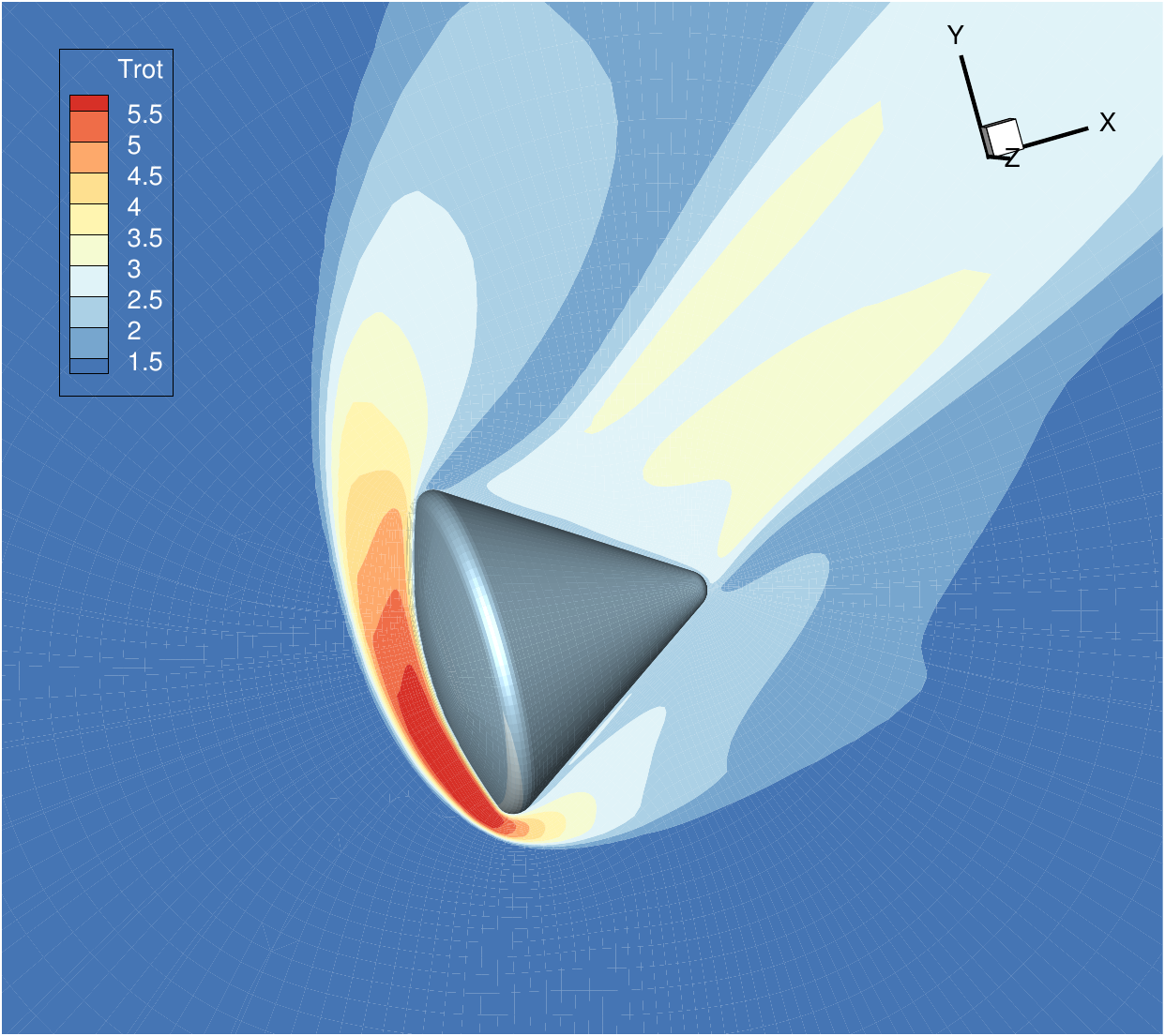}}}
    \caption{The distributions of dimensionless (a) density, (b) local Mach number, (c) translational and (d) rotational temperatures calculated by GSIS solver for the hypersonic flow passing Apollo, when $\text{Kn}=0.0012$ and $\text{Ma}=5$, and $\text{AoA}=30^\circ$.}
    \label{fig:gsis_Ma5_Kn00012_macro}
\end{figure}

The first row in Figure~\ref{fig:Apollo_Ma5_Kn00012_ErrorEvolution} compares our results of the windward side density, flow velocity and total temperature along the symmetry axis with those solved by AUGKWP \cite{wei2023} when $\text{Kn}=0.0012$. The good agreement between the two methods proofs the accuracy of GSIS. It should be noted that the Knudsen number in Ref.~\cite{wei2023} is 0.001 due to a different definition. Therefore, we show the distributions of dimensionless density, local Mach number, translational and rotational temperatures calculated by the GSIS solver in
Figs.~\ref{fig:gsis_Ma5_Kn01_macro} and~\ref{fig:gsis_Ma5_Kn00012_macro}, when $\text{Kn}=1$ and $0.0012$, respectively. As the Knudsen number increases, the shock thickness at the windward region increases significantly; meanwhile, the thermal non-equilibrium grows stronger and a distinguishable difference between the translational and rotational temperatures is observed. 

Table~\ref{tab:Apollo_gsis_Effency} summarizes the computational costs at different Knudsen numbers, and also a comparison with that of AUGKWP~\cite{wei2023} simulation when $\text{Kn}=0.0012$. It is seen that the GSIS is efficient across the different degrees of rarefaction, with a converged solution
found within dozens of iterations. Particularly, it shows a significant advantage in the near-continuous flow regime, which outperforms the AUGKWP solver by 26 times, in terms of the total CPU hours. 

\subsection{Hypersonic flow passing an X38-like space vehicle}

\begin{figure}[th]
    \centering
    {\includegraphics[scale=0.55,clip = true]{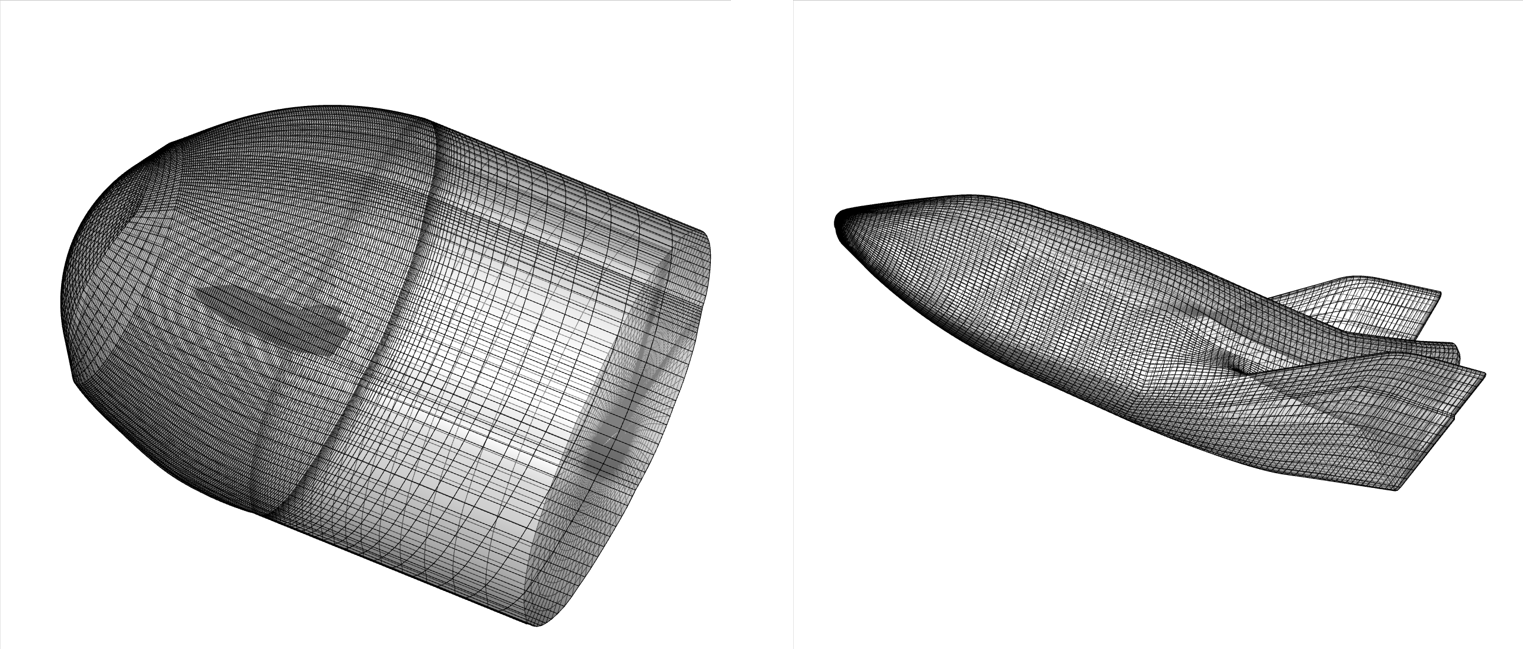}}
    \caption{Spatial discretization of the hypersonic flow around an X38-like space vehicle, where 961,080 hexahedral cells are generated in total: (left) global view of the simulation domain, (right) meshes on the wall surface. 
}
    \label{fig:x38_mesh}
\end{figure}

We consider the hypersonic flow around an X38-like space vehicle at $\text{Ma}=8$, see the back, top and side views in Ref.~\cite{wei2022unified}. To make a fair comparison with the DSMC \cite{li2021kinetic}, we use the same gas properties as those in Ref.~\cite{li2021kinetic}: $\mu_{ref}=1.7805\times 10^{-5}~\text{Pa}\cdot \text{s}$ at $T_{ref}=300$ K and viscosity index is $\omega=0.75$. The Knudsen number, which is determined in terms of the reference length $L_0=0.1$ m, free stream temperature $T_\infty=56$~K and density $\rho _\infty$, is chosen to be $0.00443, 0.0443,0.443,4.33$, respectively.  Also, two cases with $\text{AoA}=0^\circ$ and $20^\circ$ are simulated for each free stream condition.
There are $961,080$ hexahedral cells used in the spatial discretization, see Fig.~\ref{fig:x38_mesh}, and $8,002$ structure-unstructured hybrid cells in the discretization of velocity space at $\text{AoA}=0^\circ$ (similar to the Fig.~\ref{fig:3D_Ma5_AoA30_X_hybrid_Volume}),  and a similar velocity space with $8,531$ cells at $\text{AoA}=20^\circ$. The computational resources required in the simulations include $N_x\times N_v=512\times 1$ cores and 1.32 TB RAM.





\begin{figure}[htb]
    \centering
    \subfigure[]{
        \label{x38_Ma8_Kn0443_Density_AoA20}
        {\includegraphics[scale=0.3,clip = true]{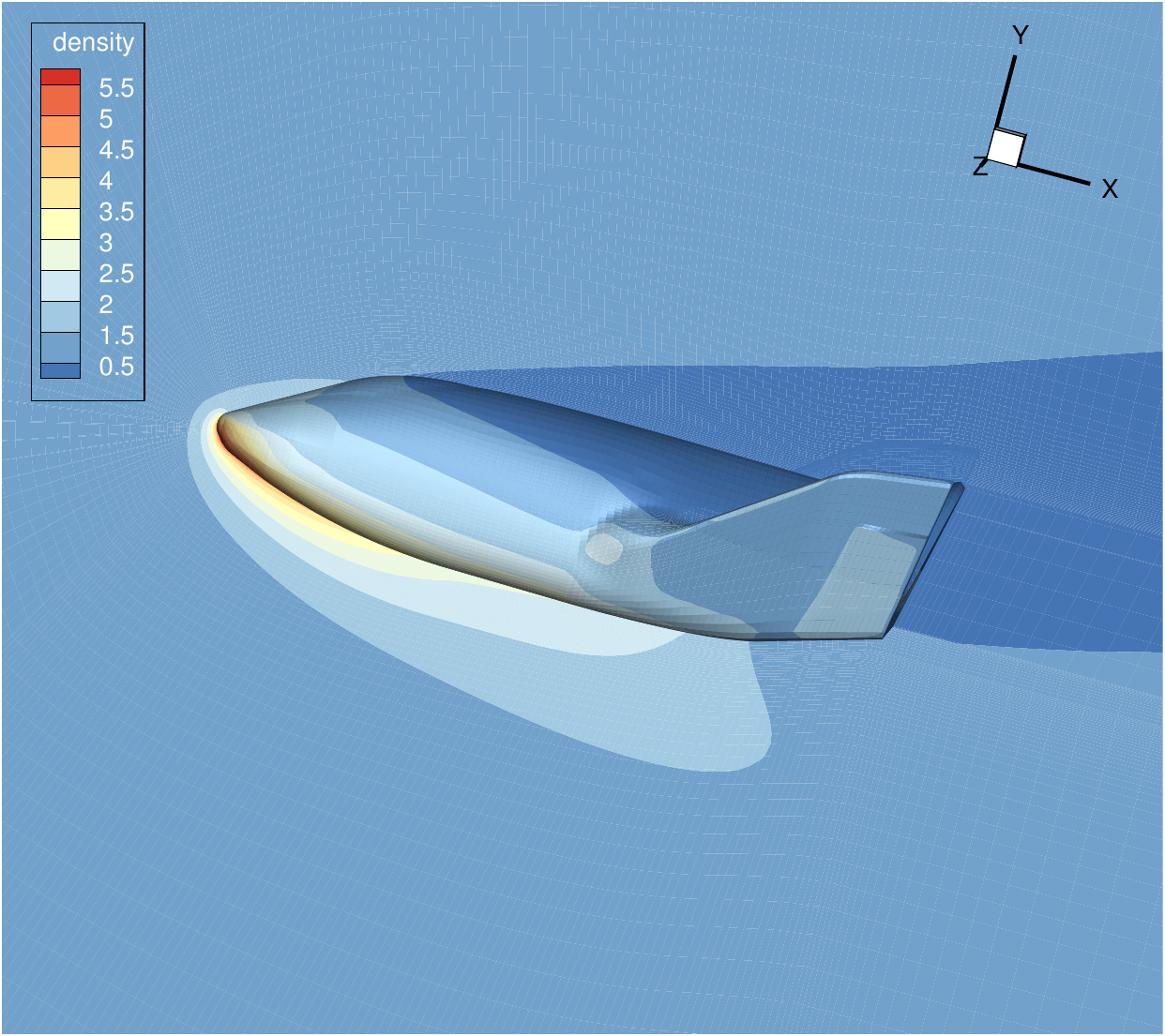}}}
    \subfigure[]{
        \label{x38_Ma8_Kn0443_Ma_AoA20}
        {\includegraphics[scale=0.3,clip = true]{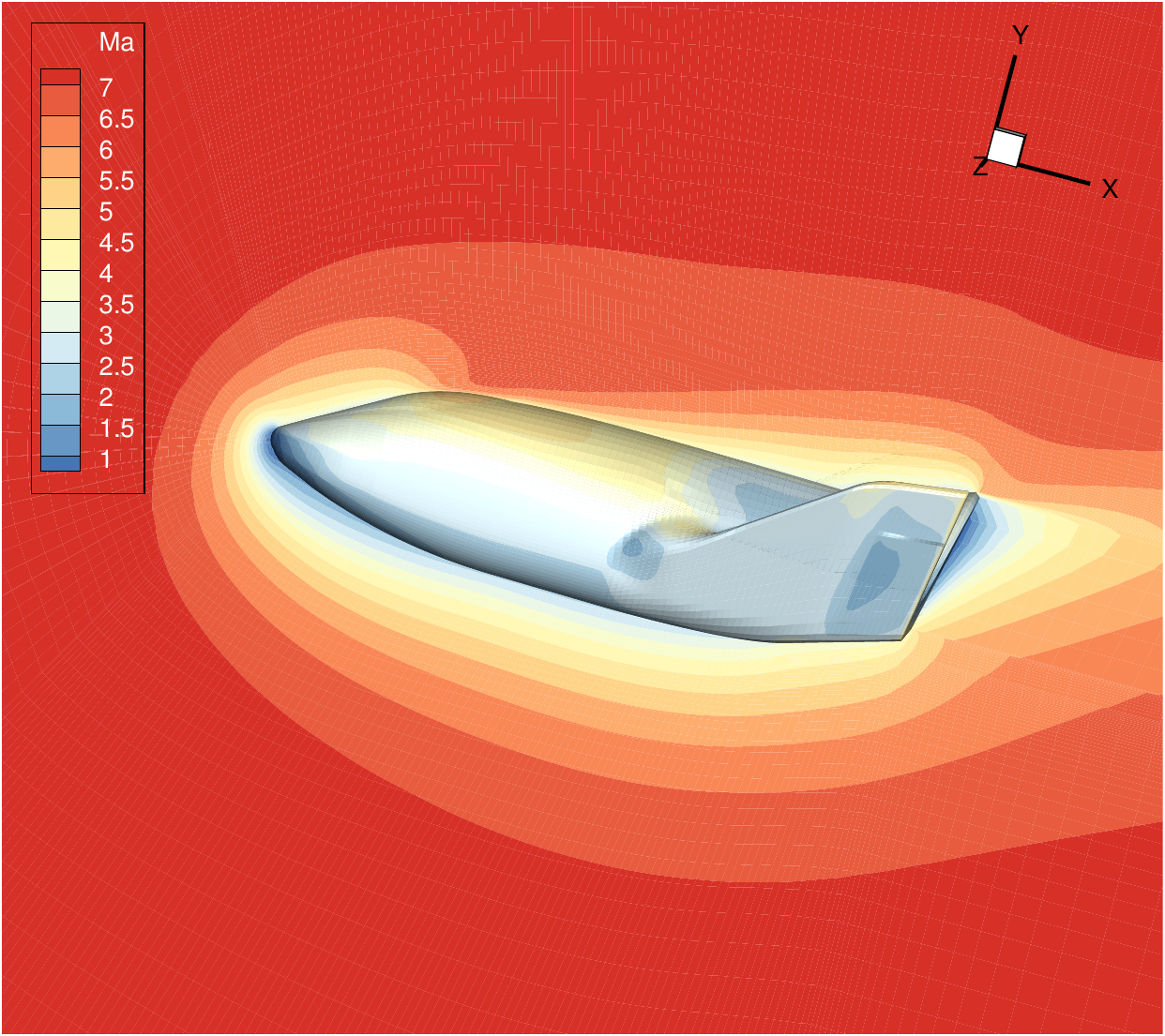}}}
    \\
    \subfigure[]{
        \label{x38_Ma8_Kn0443_Ttra_AoA20}
        {\includegraphics[scale=0.3,clip = true]{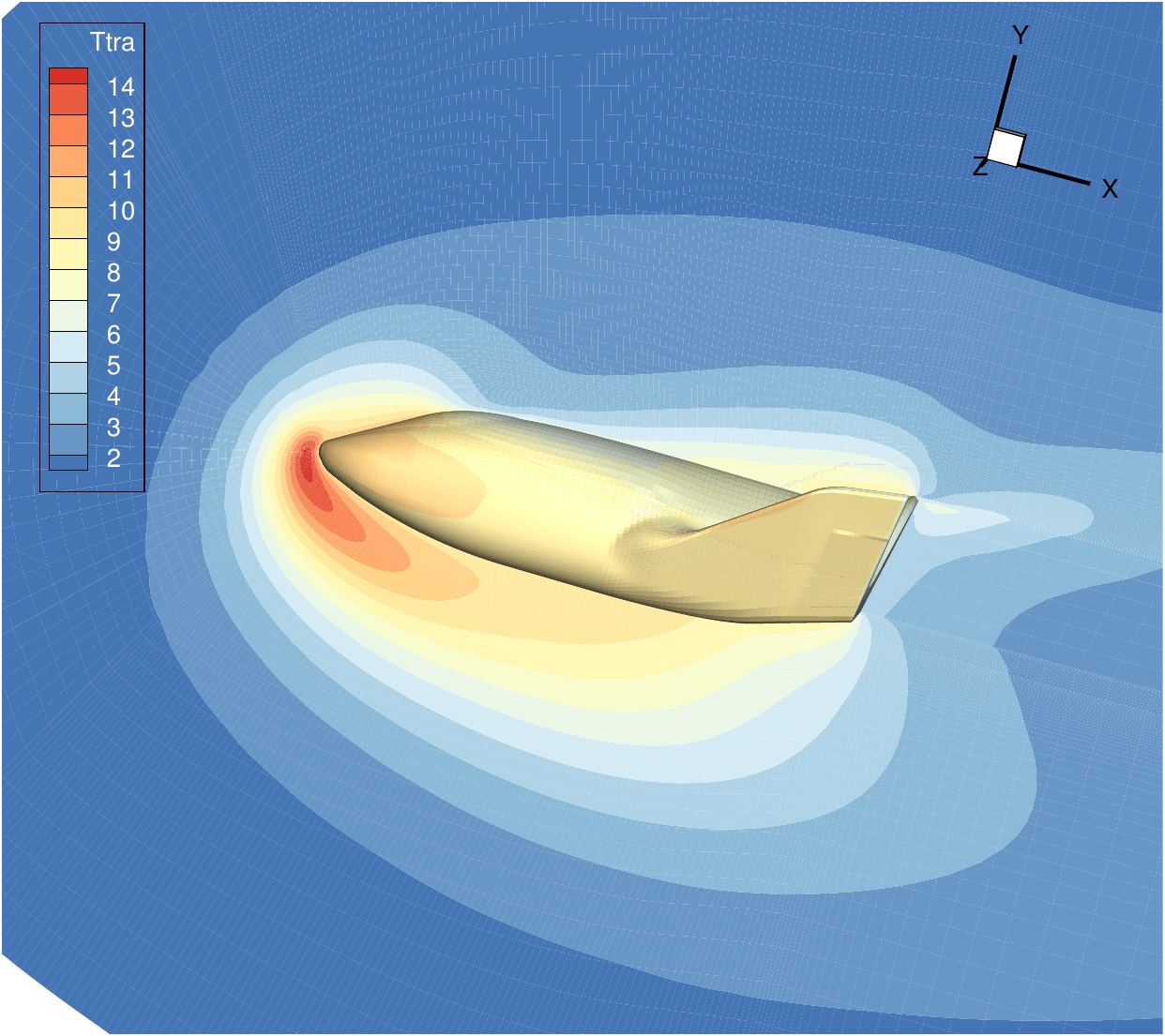}}}
    \subfigure[]{
        \label{x38_Ma8_Kn0443_Trot_AoA20}
        {\includegraphics[scale=0.3,clip = true]{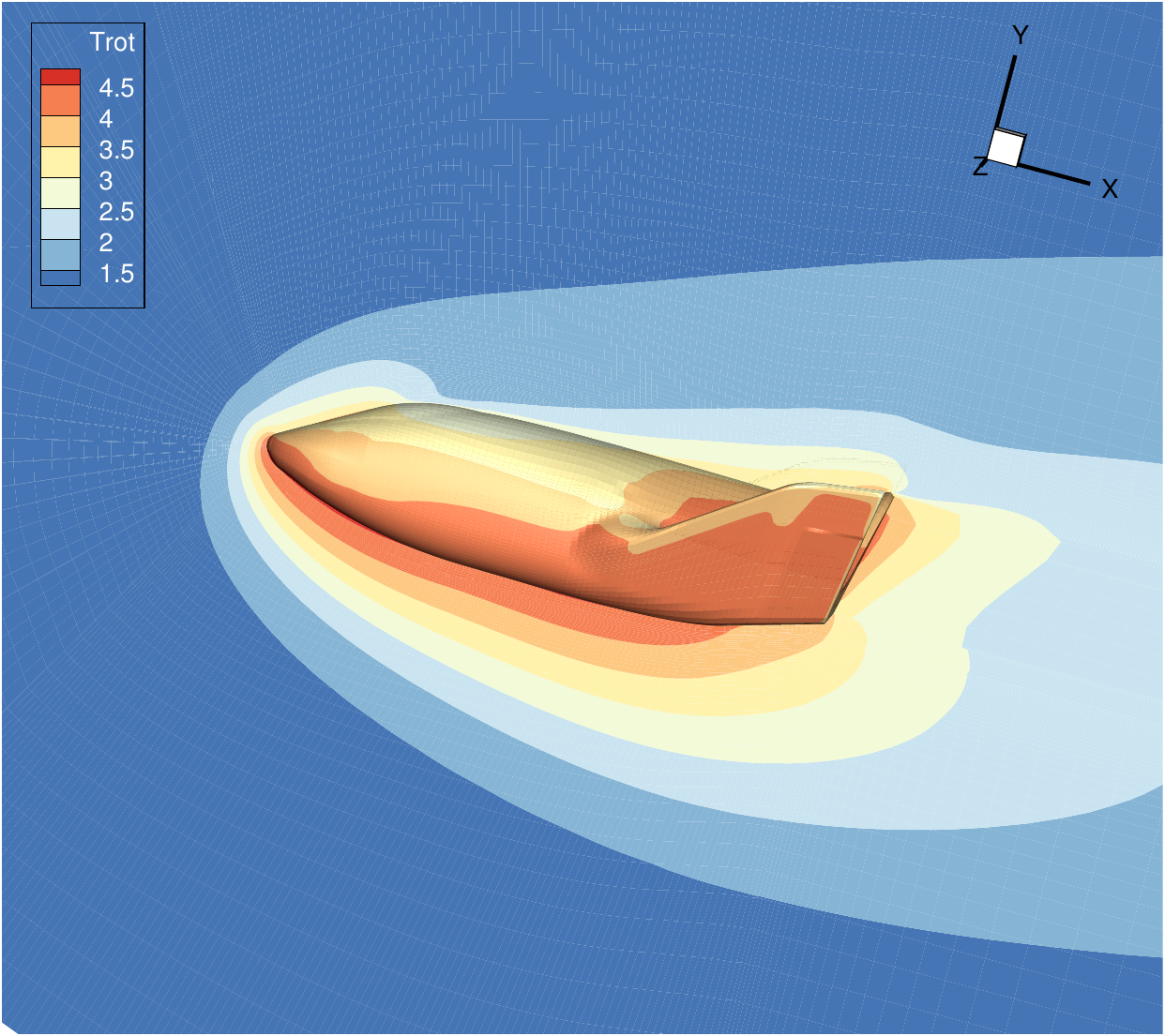}}}
    \caption{The distributions of dimensionless (a) density, (b) local Mach number, (c) translational and (d) rotational temperatures calculated by GSIS solver for the hypersonic flow passing an X38-like space vehicle, when $\text{Kn}=0.443$ and $\text{Ma}=8$, and $\text{AoA}=20^\circ$.}
    \label{fig:x38_Ma8_Kn0443macro_AoA20}
\end{figure}

\begin{figure}[htb]
    \centering
    \subfigure[]{
        \label{x38_Ma8_Kn00443_Density_AoA20}
        {\includegraphics[scale=0.3,clip = true]{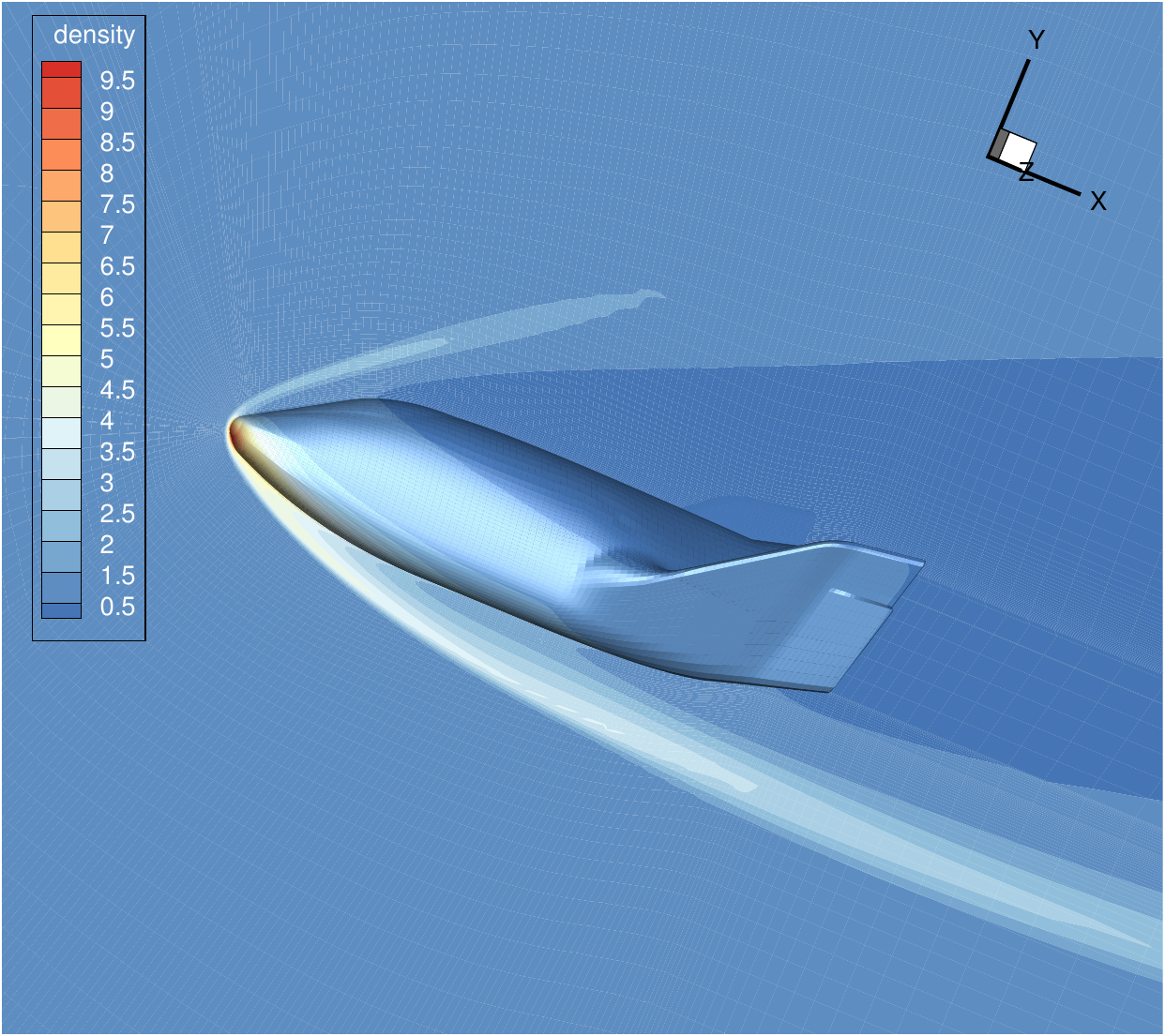}}}
    \subfigure[]{
        \label{x38_Ma8_Kn00443_Ma_AoA20}
        {\includegraphics[scale=0.3,clip = true]{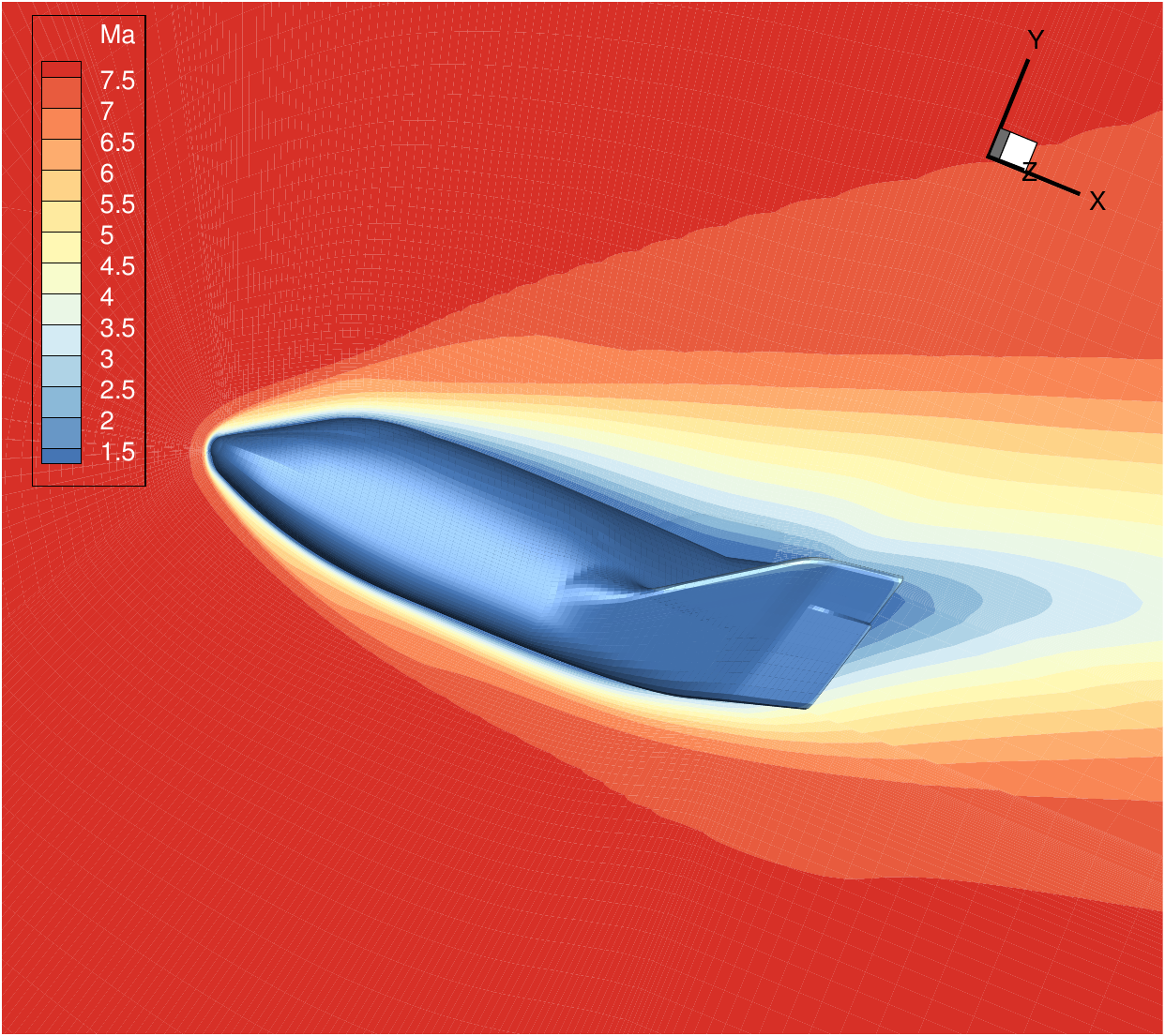}}}
    \\
    \subfigure[]{
        \label{x38_Ma8_Kn00443_Ttra_AoA20}
        {\includegraphics[scale=0.3,clip = true]{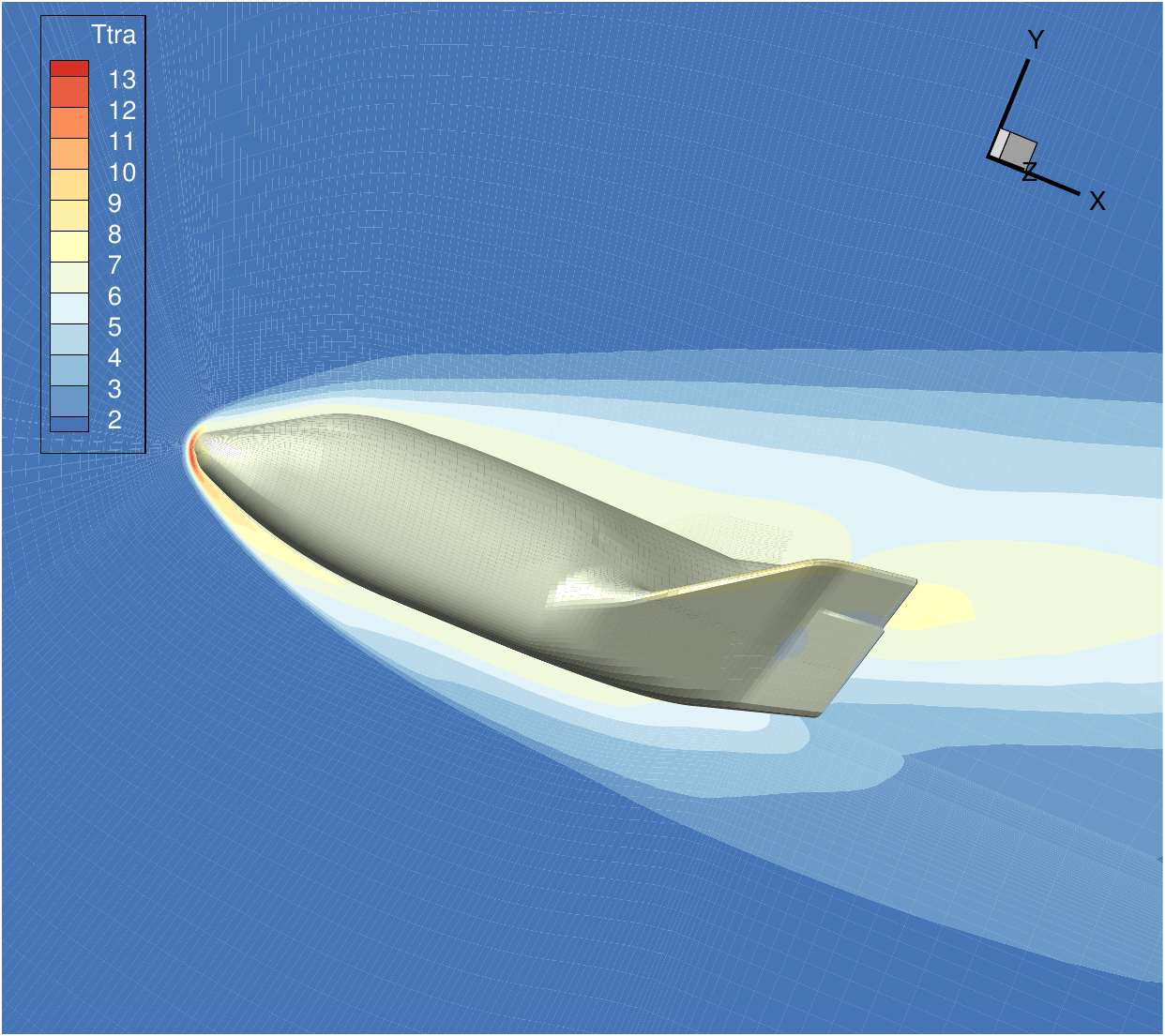}}}
    \subfigure[]{
        \label{x38_Ma8_Kn00443_Trot_AoA20}
        {\includegraphics[scale=0.3,clip = true]{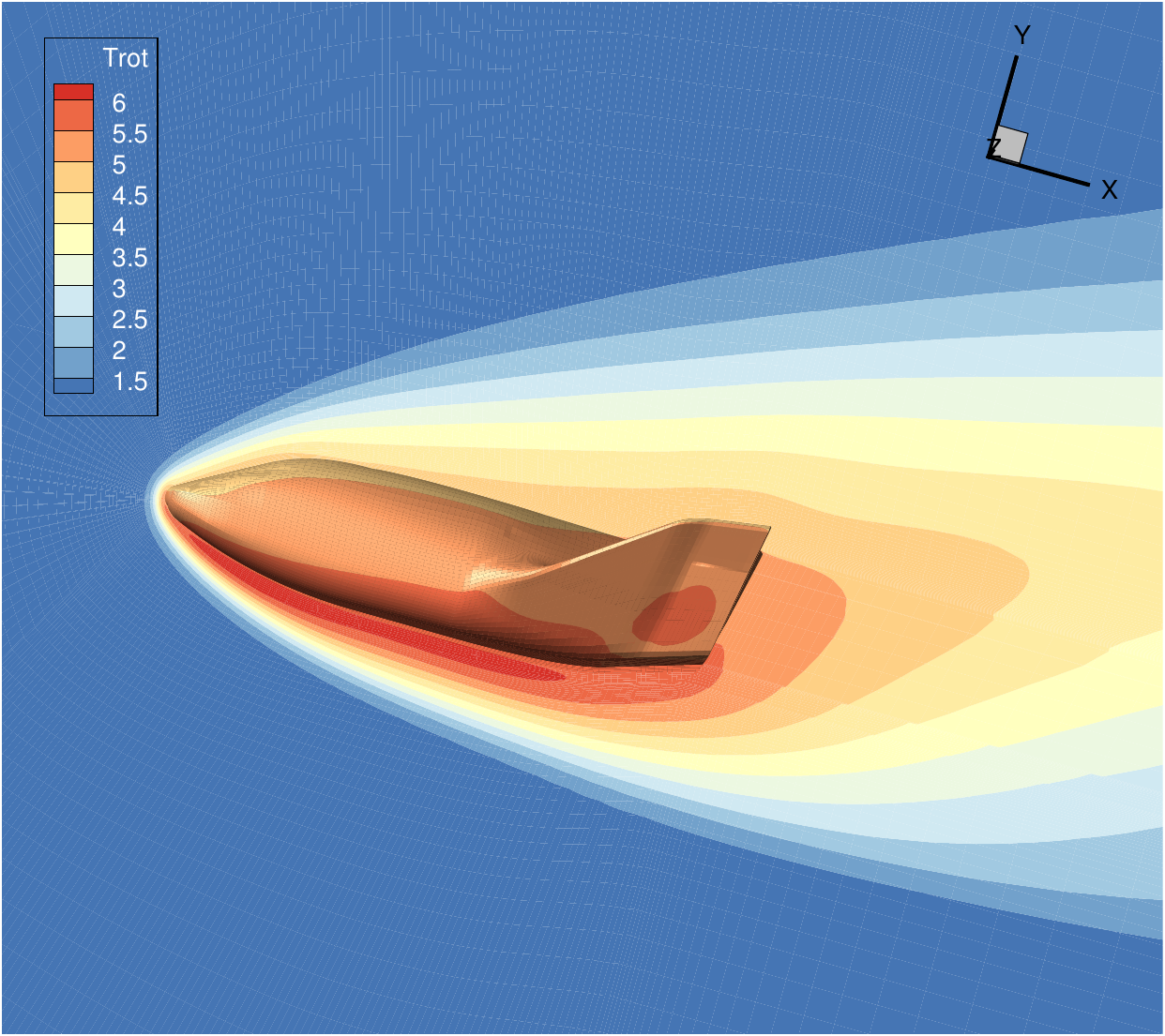}}}
    \caption{The distributions of dimensionless (a) density, (b) local Mach number, (c) translational and (d) rotational temperatures calculated by GSIS solver for the hypersonic flow passing an X38-like space vehicle, when $\text{Kn}=0.00443$ and $\text{Ma}=8$, and $\text{AoA}=20^\circ$.}
    \label{fig:x38_Ma8_Kn00443macro_AoA20}
\end{figure}

Figures~\ref{fig:x38_Ma8_Kn0443macro_AoA20} and \ref{fig:x38_Ma8_Kn00443macro_AoA20} plot the distributions of dimensionless density, local Mach number, translational and rotational temperatures, when $\text{Kn}=0.443$ and $\text{Kn}=0.00443$, respectively. The strong and sharp shock layer can be observed in the near continuum case when $\text{Kn}=0.00443$, while the non-equilibrium between translational and rotational temperature still significant. As the Knudsen number increases to 0.443, the bow shock in the windward region becomes much more diffuse, due to the rarefaction effects, e.g., the effective viscosity is larger.

\begin{figure}[tb]
    \centering
    \subfigure[]{
        \label{Cl_GSIS_DSMC}
        {\includegraphics[scale=0.5,clip = true]{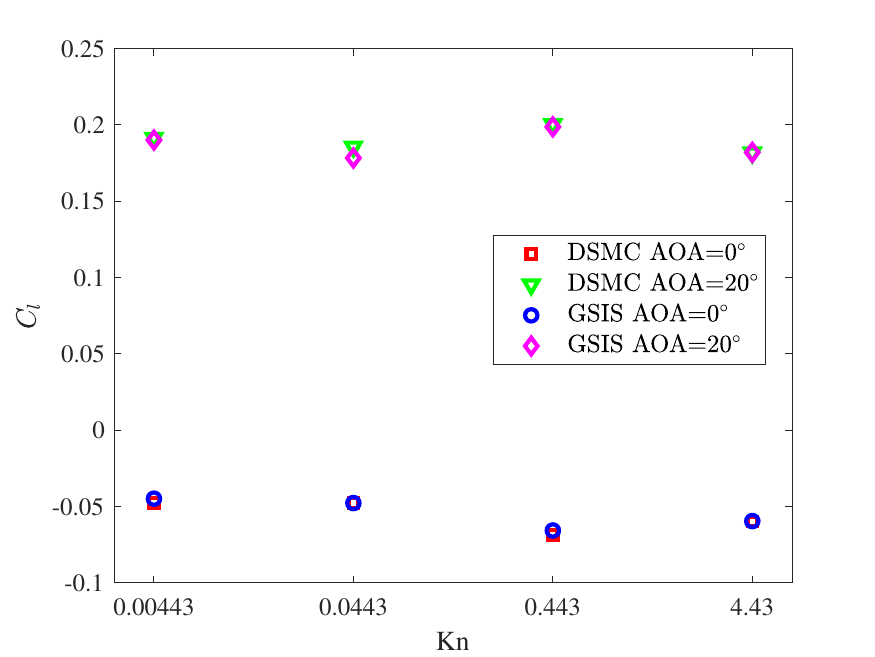}}}
    \subfigure[]{
        \label{Cd_GSIS_DSMC}
        {\includegraphics[scale=0.5,clip = true]{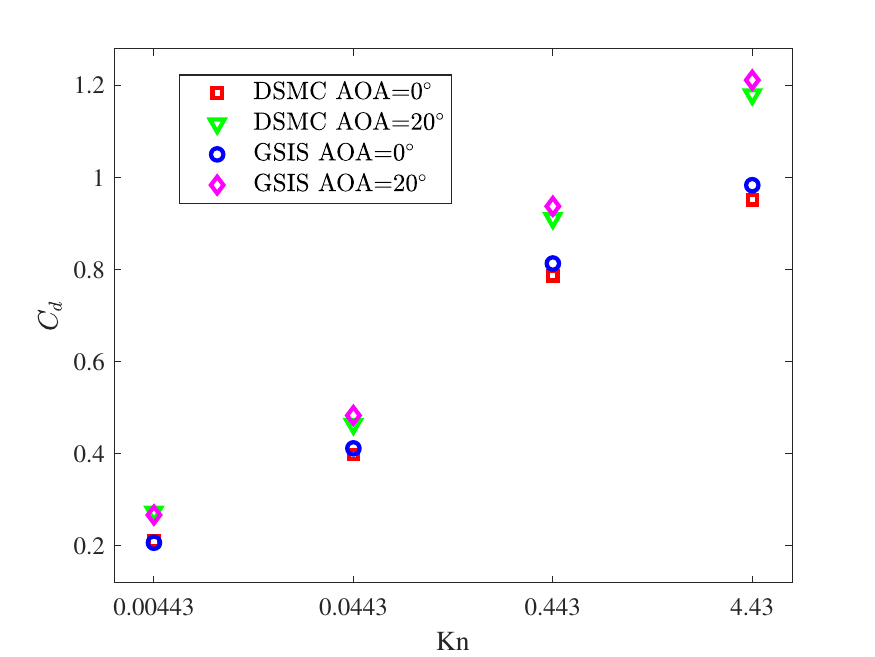}}}
    \caption{Comparison of the (a) lift and (b) drag coefficients for the hypersonic flow passing an X38-like space vehicle between GSIS solutions and DSMC results \cite{li2021kinetic}.}
    \label{fig:Cl_Cd_GSIS_DSMC_cmp}
\end{figure}

Figure~\ref{fig:Cl_Cd_GSIS_DSMC_cmp} compares the coefficients of aerodynamic lift and drag force calculated by GSIS and DSMC \cite{li2021kinetic}. Good agreement has been obtained. It is shown that the lift coefficient is not sensitive to the degree of rarefaction, while the drag coefficient increases significantly as the Knudsen number increases. 

\begin{table}[!t]
    \centering
    \caption{The computational costs in simulating the hypersonic flow passing an X38-like space vehicle ($\text{Ma}=8$). The gas employed in simulations by GSIS and UGKWP is nitrogen and argon, respectively. The number of spatial cells used is 961,080 in GSIS  and 560,593 (246,558 when $\text{Kn}=0.00443$) in UGKWP \cite{long2023nonequilibrium}.}
    \begin{threeparttable}
        \begin{tabular}{cc|cc|ccc|cccc}\hline
            \multirow{2}{*}{Kn} & \multirow{2}{*}{AoA} & \multicolumn{2}{c|}{UGKWP} & \multicolumn{3}{c|}{GSIS} & \multirow{2}{*}{Speedup ratio \tnote{$\star$}}              \\ \cline{3-4} \cline{5-7}
            ~            & ~ &       Cores & Wall time (h) &  Cores & Steps & Wall time (h) &    \\ \hline
            4.43                & $0^\circ$            & \multirow{8}{*}{640}       & 12.3                      & \multirow{8}{*}{512}    & 43       & 0.31         & 85.0  \\
            0.443               & $0^\circ$            &                            & 8.22                      &                         & 35       & 0.24         & 73.3  \\
            0.0443              & $0^\circ$            &                            & 15.1                      &                         & 46       & 0.33         & 98.0 \\
            0.00443             & $0^\circ$            &                            & 6.58                      &                         & 29       & 0.22         & 145.7 \\
            4.43                & $20^\circ$           &                            & 11.1                      &                         & 40       & 0.30         & 79.2  \\
            0.443               & $20^\circ$           &                            & 8.15                      &                         & 37       & 0.26         & 67.1   \\
            0.0443              & $20^\circ$           &                            & 13.6                      &                         & 45       & 0.32         & 91.0 \\
            0.00443             & $20^\circ$           &                            & 6.25                      &                         & 38       & 0.28         & 108.7 \\ \hline
        \end{tabular}
        \begin{tablenotes}
        \item [$\star$] The speedup ratio quantifies the computational efficiency of GSIS relative to UGKWP in these cases, where the computational time costs of the two schemes are normalized by the corresponding numbers of spatial cells.
    \end{tablenotes}
    \end{threeparttable}
    \label{tab:x38_gsis_Effency}
\end{table}

The computational costs for the simulations by GSIS at different Knudsen numbers are listed in Table~\ref{tab:x38_gsis_Effency}. For all cases, the converged solutions can be obtained within 1 hour on 512 cores, and particularly fast convergence is achieved in near-continuum flows. As a comparison, the corresponding cost by the UGKWP method \cite{long2023nonequilibrium} is also shown. The gas employed in simulations by GSIS and UGKWP is nitrogen and argon, respectively. Therefore, if the GSIS is applied to simulate the argon gas where only the translational motion is considered, the simulation time and storage will be reduced by half. 
Note that in UGKWP the number of spatial cells used is 246,558 for cases with $\text{Kn}=0.00443$ and 560,593 for the others, which is less than that used in our simulations. Assuming the linear scalability of the UGKWP method when the mesh size is increased to 961,080, it can be found that the GSIS can be faster than UGKWP by about one order of magnitude, see the last column in the table.

\subsection{Hypersonic flow passing a space station}



\begin{figure}[p]
    \centering
    {\includegraphics[width=0.45\textwidth,clip = true]{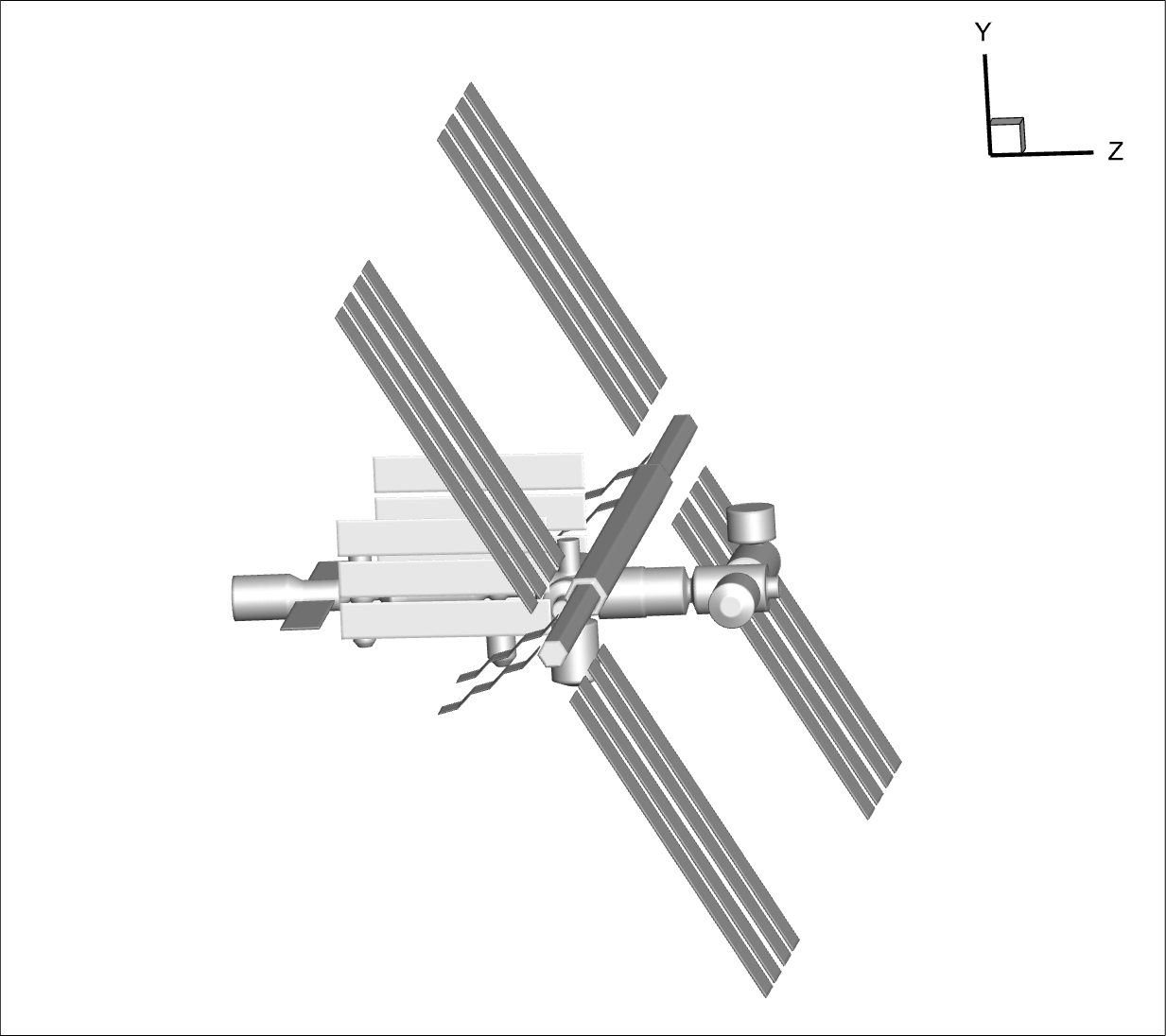}} 
    {\includegraphics[width=0.45\textwidth,clip = true]{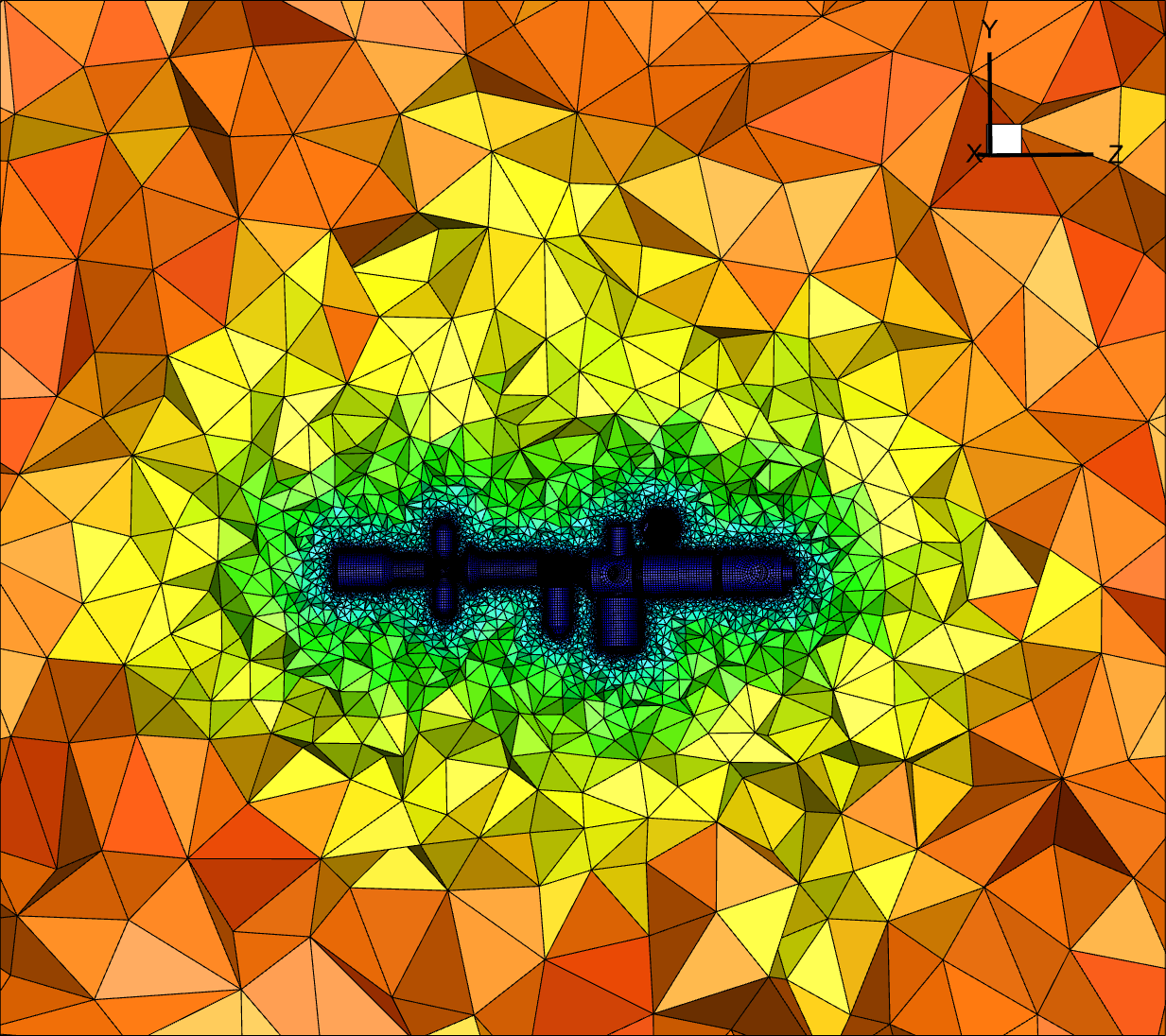}}
    \caption{Spatial domain discretization of the hypersonic flow around a space station, where 5,640,776 cells (composed of tetrahedron, pentahedron, triangular prism, and hexahedron) are generated in total. 
    }
    \label{fig:station_mesh}
\end{figure}

Note that the altitude of the space station is usually very high, so that the Knudsen number is large, and the traditional DSMC method is very efficient. However, recently scientists are interested in the falling and disintegration process of the out-of-control space station from outer space to earth as it reaches/exceeds its service life~\cite{li2019gas}. Therefore, as a test of the parallel performance and simulation capacity, the hypersonic flow passing a space station at $\text{Ma}=25$ is simulated for $\text{Kn}=0.01$, which are defined in terms of the reference length $L_0=0.01$ m and temperature $T_0=T_{\infty}=142.2$ K with $T_{\infty}$ being the free stream temperature. The direction of the incoming flow is the positive direction of the Z axis. The isothermal surface with $T_w=500$ K and fully diffuse gas-wall interaction is adopted. The configuration is shown in Fig.~\ref{fig:station_mesh}. The whole spatial domain is composed of tetrahedron, pentahedron, triangular prism, and hexahedron, with a total of 5,640,776 cells. The velocity domain is truncated to a sphere with diameter $42\sqrt{RT_0}$. Unstructured meshes with refinement around the stagnation and free stream velocity points are used, which result in 31,440 tetrahedral cells in the velocity domain discretization, which is similar to that in Fig.~\ref{fig:3D_Ma5_AoA30_X_unstructure_Volume}. 

\begin{figure}[p]
    \centering
    \subfigure[]{
        \label{Station_Ma25_Kn001_Density}
        {\includegraphics[width=0.45\textwidth,clip = true]{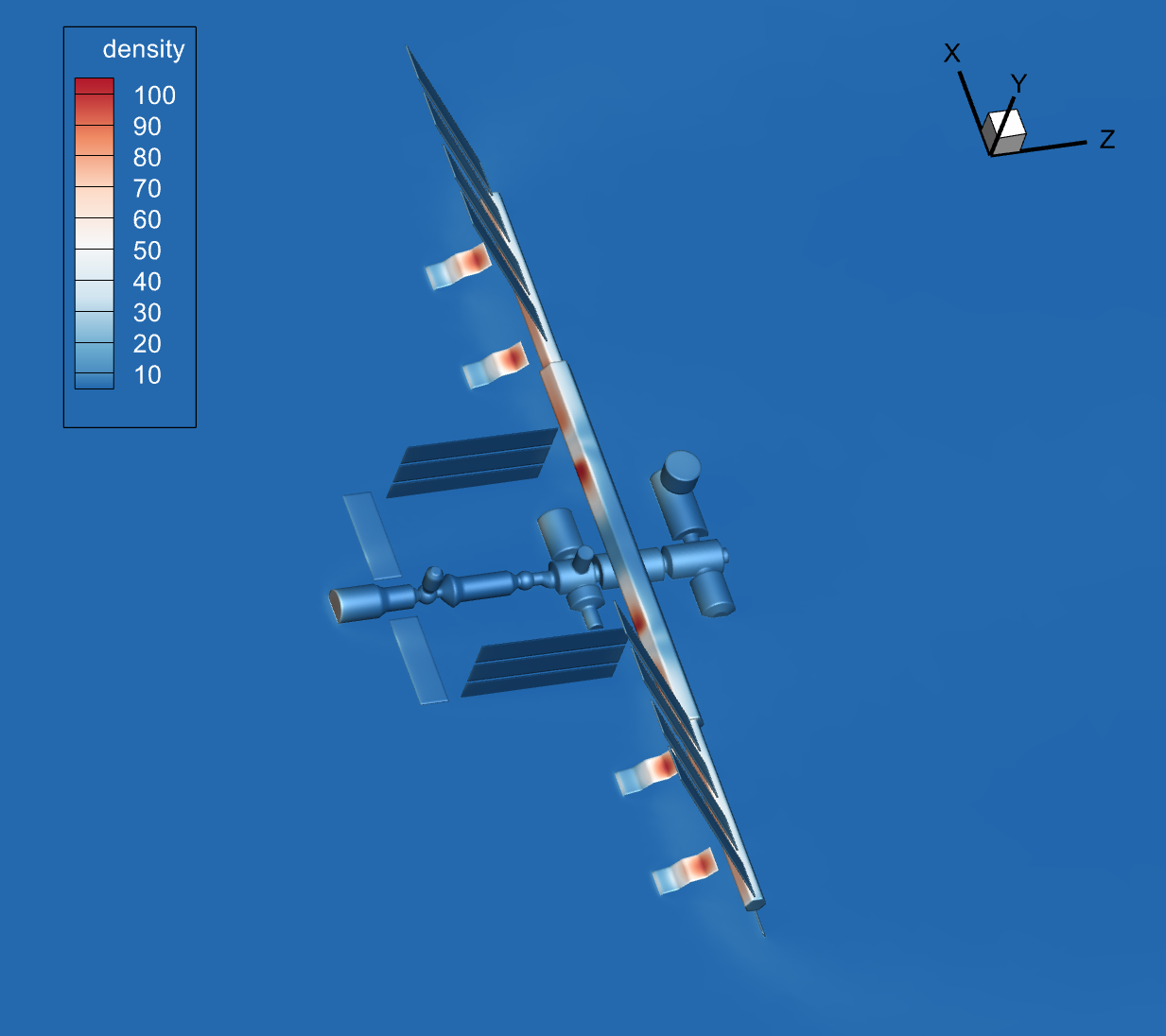}}}
    \subfigure[]{
        \label{Station_Ma25_Kn001_Ma}
        {\includegraphics[width=0.45\textwidth,clip = true]{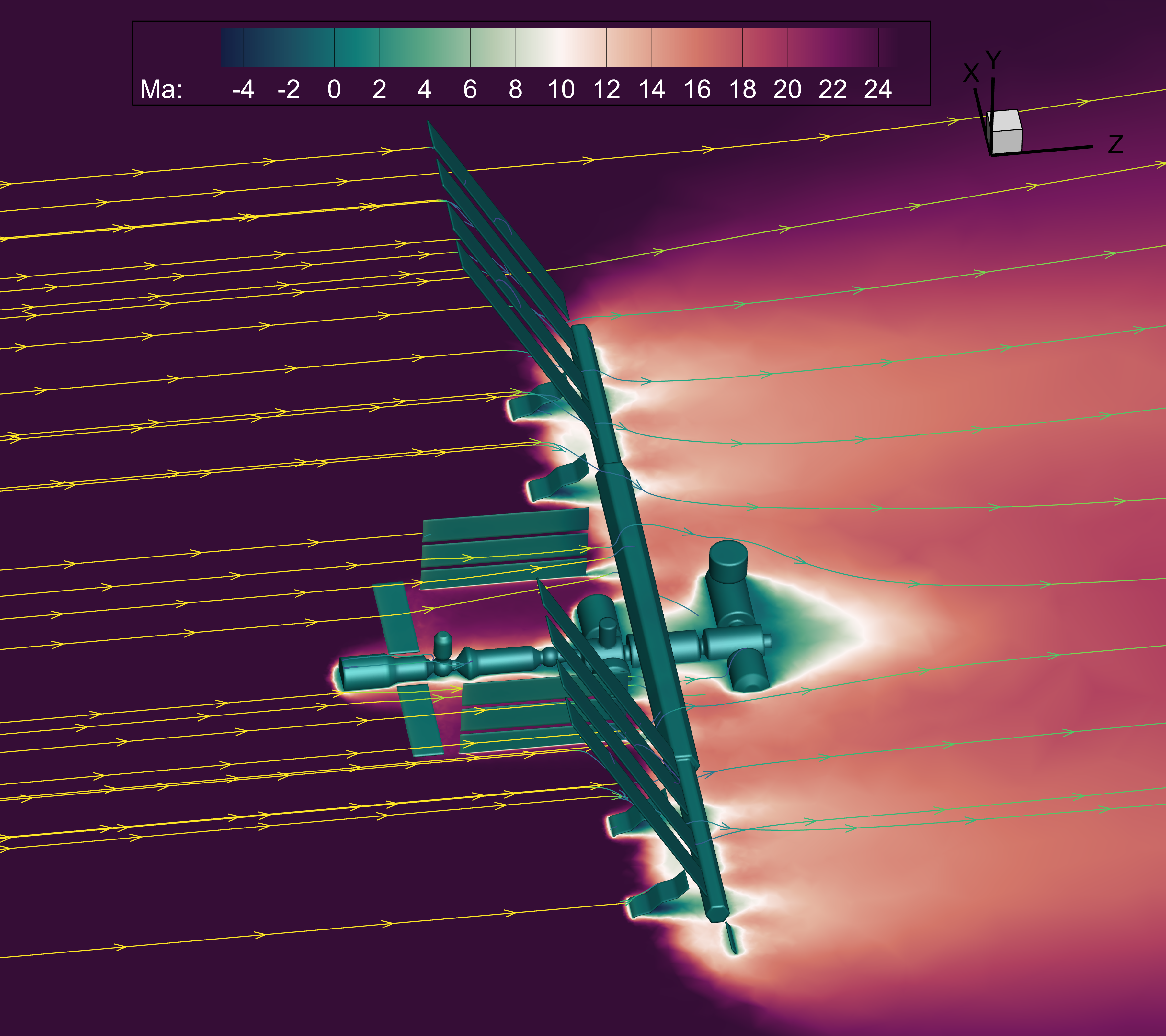}}}
    \\
    \subfigure[]{
        \label{Station_Ma25_Kn001_Ttra}
        {\includegraphics[width=0.45\textwidth,clip = true]{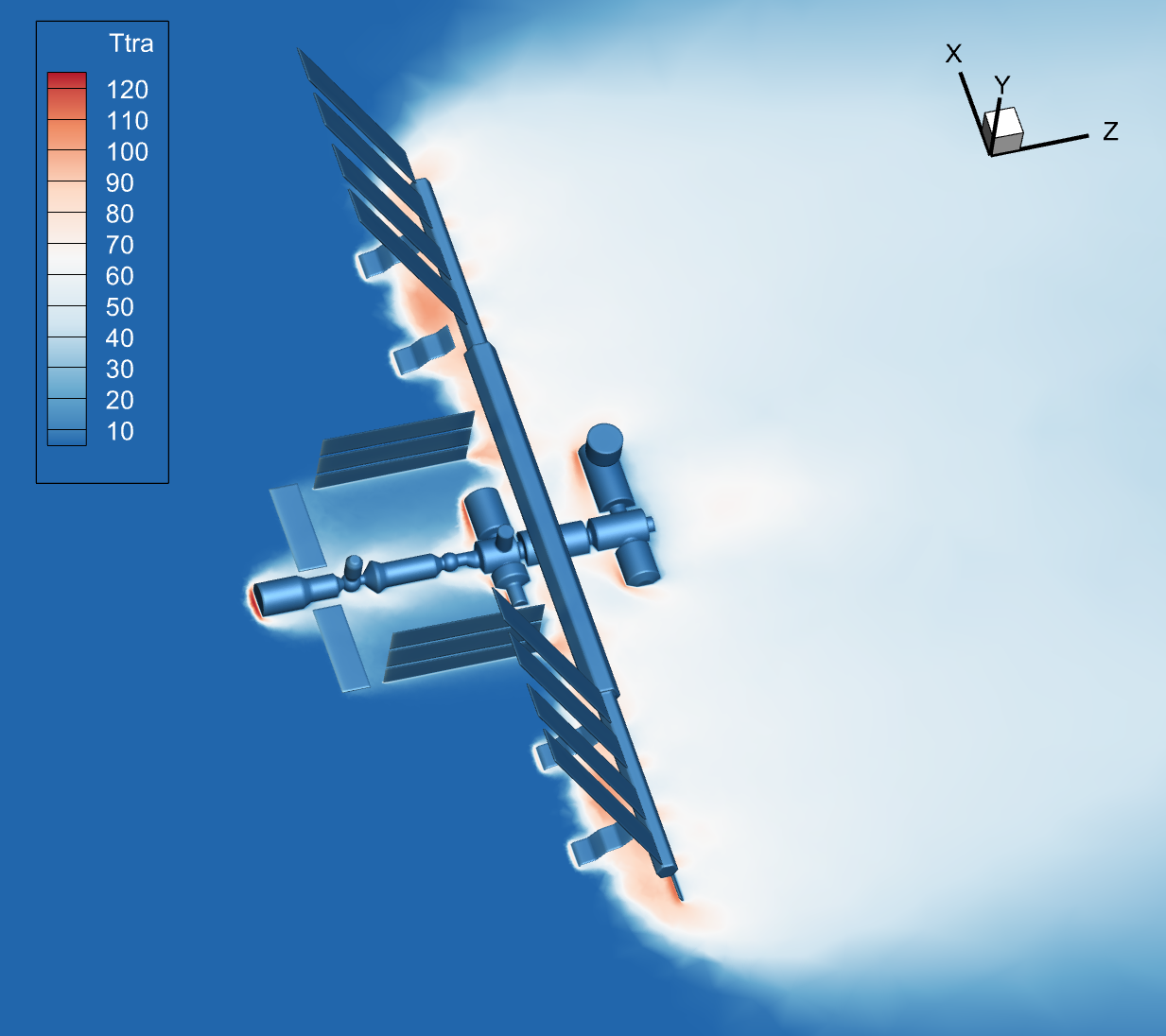}}}
    \subfigure[]{
        \label{Station_Ma25_Kn001_Trot}
        {\includegraphics[width=0.45\textwidth,clip = true]{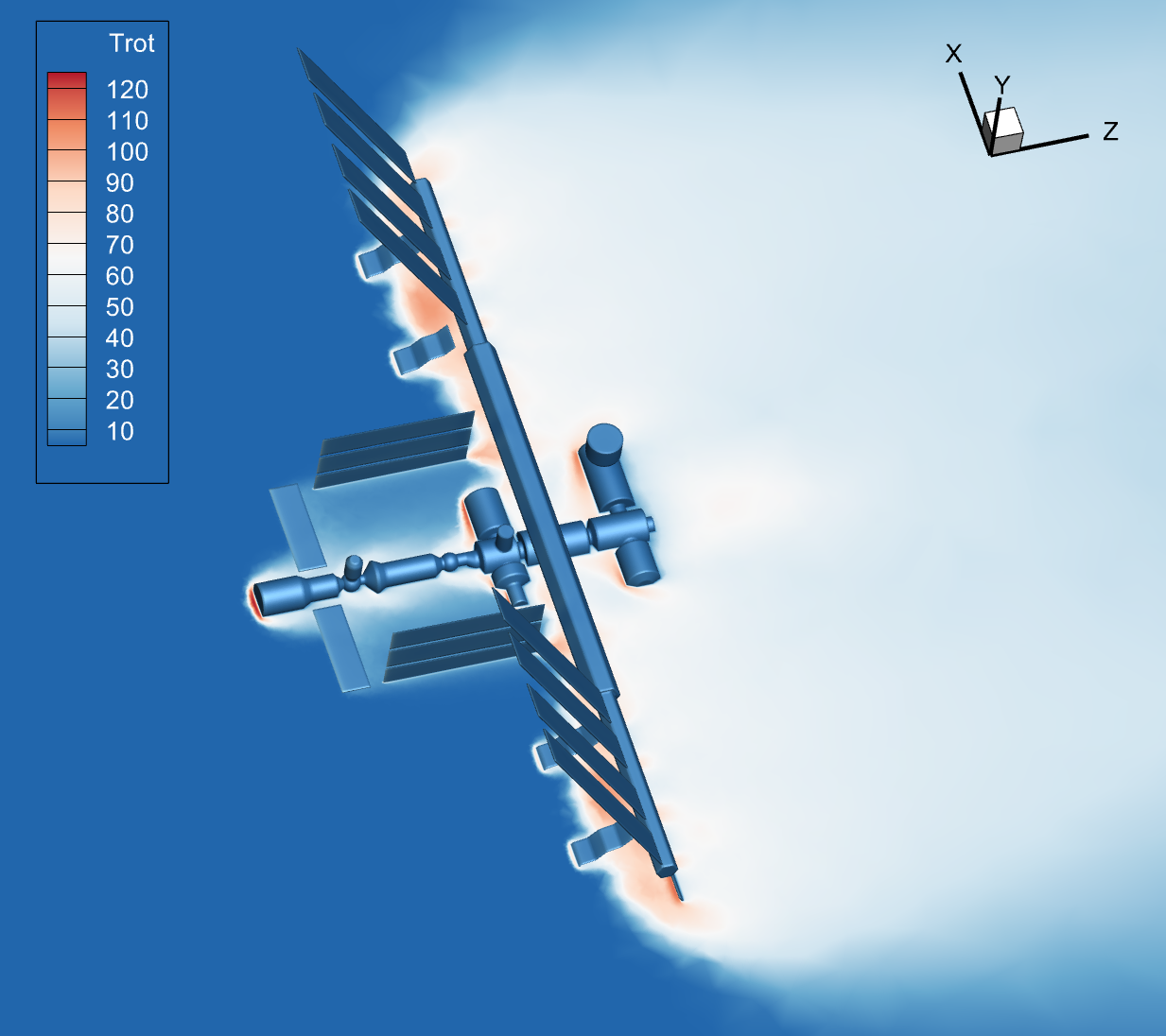}}}
    \caption{The distributions of dimensionless (a) density, (b) local Mach number, (c) translational and (d) rotational temperatures calculated by GSIS solver for the hypersonic flow passing a space station, when $\text{Kn}=0.01$ and $\text{Ma}=25$.    }
    \label{fig:Station_Ma25_Kn001_macro}
\end{figure}

Figure~\ref{fig:Station_Ma25_Kn001_macro} displays the dimensionless density, local Mach number, translational and rotational temperatures at $\text{Ma}=25$ and $\text{Kn}=0.01$.
Notice that in this case there is no significant difference between the translational and rotational temperatures. 
The computational resources required in the simulations include $N_x\times N_v=2304\times 4$ cores and 21.5 TB RAM. The initial field is calculated by 4000 steps of macroscopic solver with Euler constitutive relations and 10 steps of kinetic solver. After iterating GSIS for 52 steps, the error $\varepsilon$ reaches below $10^{-6}$, and the total computational time is 52 minutes.

 \section{Conclusions}\label{sec:conclusion}

In summary, we have developed an efficient parallel strategy to simulate the multiscale rarefied gas flow based on the gas kinetic equations. Due to the fast-converging property of GSIS, the iteration number of the kinetic equation, which is the most time-consuming part, is reduced to within 100 in the whole range of gas rarefaction. 
Eventually, the GSIS, which is a deterministic solver that uses a huge number of additional memory due to the discretization of velocity space, can be faster than the adaptive UGKWP method that combines the advantages of the stochastic and deterministic methods, in the simulation of high-speed multiscale flows. 


The GSIS framework is easy to implement, since the kinetic and the macroscopic equations can be solved efficiently by mature techniques in computational fluid dynamics. As a matter of fact, we believe that anyone who can write a program to solve the NSF equations can easily write the GSIS solver. Moreover, the GSIS solver is ready to be extended to time-dependent problems~\cite{Zeng2023CiCP}, where the two-body separation, fluid-solid interactions, and even the ablation can be incorporated. With these developments, we believe that the GSIS can become an indispensable tool in simulating large-scale three-dimensional hypersonic rarefied flows, e.g., in the simulation of the falling and disintegration process of out-of-control space stations.

\section*{Acknowledgments}

This work is supported by the National Natural Science Foundation of China (12172162) and the ``Climbing Program"  for Scientific and Technological Innovation in Guangdong (pdjh2023c10701).
Simulations are conducted in the Center for Computational Science and Engineering at the Southern University of Science and Technology. 
The authors thank Prof. Kun Xu in the Hong Kong University of Science and Technology for sharing the mesh of Apollo re-entry capsule and Dr. Liyan Luo in Southern University of Science and Technology for sharing the results of DSMC in 2D lid-driven flow in a square cavity.

%


\bibliographystyle{elsarticle-num}

\bibliography{ref}

\end{document}